\documentclass[10pt]{article}
\usepackage[utf8]{inputenc}
\usepackage[left=1in, right=1in, top=1in, bottom=1in]{geometry}
\usepackage{amsmath}
\usepackage{graphicx}
\usepackage{enumerate}
\usepackage{natbib}
\usepackage{url} 

\usepackage{bm}
\usepackage{amsfonts}   
\usepackage{amsmath} 
\usepackage{amsthm} 
\usepackage{mathrsfs} 
\usepackage[dvipsnames]{xcolor}
\usepackage{hyperref}
\usepackage{multirow}
\usepackage{enumitem}

\newcommand{\N}{\mathbb N} 
 
\newcommand{\PP}{\mathbb P} 
\newcommand{\pairP}{\mathcal{P}}
\newcommand{\C}{\mathcal{C}} 
\newcommand{\M}{\mathcal{M}}

\newcommand{\HH}{\mathcal{H}}
\newcommand{\bn}{\boldsymbol n}
\newcommand{\bh}{\boldsymbol h}
\newcommand{\bem}{\boldsymbol m}
\newcommand{\bu}{\boldsymbol u}
\newcommand{\bp}{\boldsymbol p}

\newcommand{\bd}{\boldsymbol d}
\newcommand{\balpha}{\boldsymbol \alpha}
\newcommand{\bgamma}{\boldsymbol \gamma}
\newcommand{\bX}{\boldsymbol X}
\newcommand{\bZ}{\boldsymbol Z}
\newcommand{\lfnm}{\lambda_{\scriptscriptstyle \mathrm{FNM}}}
\newcommand{\lfma}{\lambda_{\scriptscriptstyle \mathrm{FM1}}}
\newcommand{\lfmb}{\lambda_{\scriptscriptstyle \mathrm{FM2}}}

\def \serge#1{#1}
\def \mcio#1{#1}

\newtheorem{prop}{Proposition}
\newtheorem{prop2}{Proposition}
\newtheorem{defi}{Definition}

\title{Multifile Partitioning for Record Linkage and Duplicate Detection}
\author{Serge Aleshin-Guendel,  Mauricio Sadinle\thanks{
    Serge Aleshin-Guendel is a Ph.D. student, Department of Biostatistics, University of Washington, Seattle, WA 98195 (e-mail: aleshing@uw.edu); and Mauricio Sadinle is an Assistant Professor, Department of Biostatistics, University of Washington, Seattle, WA 98195 (e-mail: msadinle@uw.edu). This research was supported by the NSF under grant SES-1852841 and  by the NIH--NIDA under award R21DA051756.  The authors thank Jorge A. Restrepo for providing the Colombian homicide data.}\\ Department of Biostatistics, University of Washington, Seattle, Washington, U.S.A.}

\begin{document}

\maketitle
\begin{abstract}
Merging datafiles containing information on overlapping sets of entities is a challenging task in the absence of unique identifiers, and is further complicated when some entities are duplicated in the datafiles. Most approaches to this problem have focused on linking two files assumed to be free of duplicates, or on detecting which records in a single file are duplicates. However, it is common in practice to encounter scenarios that fit somewhere in between or beyond these two settings. We propose a Bayesian approach for the general setting of multifile record linkage and duplicate detection.  We use a novel partition representation to propose a structured prior for partitions that can incorporate prior information about the data collection processes of the datafiles in a flexible manner, and extend previous models for comparison data to accommodate the multifile setting.  We also introduce a family of loss functions to derive Bayes estimates of partitions that allow uncertain portions of the partitions to be left unresolved. The performance of our proposed methodology is explored through extensive simulations.
Code implementing the methodology is available at \url{https://github.com/aleshing/multilink}.
\end{abstract}

\section{Introduction}
\label{sec:intro}

When information on individuals is collected across multiple datafiles, it is natural to merge these  to harness all available information. This merging requires identifying \textit{coreferent} records, i.e., records that refer to the same entity, which is not trivial in the absence of unique identifiers. This problem arises in many fields, including public health \citep{Hof_2017}, 
official statistics \citep{Jaro_1989},
political science \citep{Enamorado_2019}, and human rights \citep{Sadinle_2014, Sadinle_2017, Ball_2019}.

Most approaches in this area have thus far focused on one of two settings. \textit{Record linkage} has traditionally referred to the setting where the goal is to find coreferent records across two datafiles, where the files are assumed to be free of duplicates. \textit{Duplicate detection} has traditionally referred to the setting where the goal is to find coreferent records within a single file. In practice, however, it is common to encounter problems that fit somewhere in between or beyond these two settings. For example, we could have multiple  datafiles that are all assumed to be free of duplicates, or we might have duplicates in some files but not in others. In these general settings, the data collection processes for the different datafiles possibly introduce different patterns of duplication, measurement error, and missingness into the records. Further, dependencies among these data collection processes determine which specific subsets of files contain records of the same entity. We refer to this general setting as \textit{multifile record linkage and duplicate detection}.

Traditional approaches to record linkage and duplicate detection have mainly followed the seminal work of \cite{Fellegi_1969}, by modeling comparisons of fields between pairs of records in a mixture model framework \citep{Winkler_1994, Jaro_1989, Larsen_2001}. These approaches work under, and take advantage of, the intuitive assumption that coreferent records will look similar, and non-coreferent records will look dissimilar.
However, these approaches output independent decisions for the coreference status of each pair of records, necessitating the use of ad hoc post-processing steps to reconcile incompatible decisions that ignore the logical constraints of the problem. 

Our approach to multifile record linkage and duplicate detection builds on previous Bayesian approaches where the parameter of interest is defined as a partition of the records. These Bayesian approaches have been carried out in two frameworks. In the \textit{direct-modeling} framework, one directly models the fields of information contained in the records \citep{Matsakis_2010, Tancredi_2011, Liseo_2011, Steorts_2015, Steorts_2016, Tancredi_2020, Marchant_2020, Enamorado_2020}, which requires a custom model for each type of field. 
While this framework can provide a plausible generative model for the records, it can be difficult to develop custom models for complicated fields like strings, so most approaches are limited to modeling categorical data, with some exceptions \citep{Liseo_2011, Steorts_2015}. 
In the \textit{comparison-based} framework, following the traditional approaches, one models comparisons of fields between pairs of records \citep{Fortini_2001, Larsen_2005, Sadinle_2014, Sadinle_2017}.  By modeling comparisons of fields, instead of the fields directly, a generic modeling approach can be taken for any field type, as long as there is a meaningful measure of similarity for that field type.

\cite{Sadinle_2013} generalized
\cite{Fellegi_1969} by linking $K>2$ files with no duplicates.
However, in addition to inheriting the issues of traditional approaches, their approach does not scale well in the number of files or the file sizes encountered in practice. \serge{\cite{Steorts_2016} presented a Bayesian approach in the direct-modeling framework for the general setting of multifile record linkage and duplicate detection, which has been extended by \cite{Steorts_2015} and \cite{Marchant_2020}. This approach uses a flat prior on arbitrary labels of partitions, which incorporates unintended prior information.}

In light of the shortcomings of existing approaches, we propose an extension of Bayesian comparison-based models 
that explicitly handles the setting of multifile record linkage and duplicate detection. We first present in Section \ref{sec:goals} a parameterization of partitions specific to the context of multifile record linkage and duplicate detection. Building on this parameterization, in Section \ref{sec:part_prior} we construct a structured prior for partitions that can incorporate prior information about the data collection processes of the files in a flexible manner. As a by-product, a family of priors for $K$-partite matchings is constructed. In Section \ref{sec:comp_model} we construct a likelihood function for comparisons of fields between pairs of records that accommodates possible differences in the datafile collection processes. In Section \ref{sec:bayesest} we present a family of loss functions that we use to derive Bayes estimates of partitions. These loss functions have an \textit{abstain option} which allow portions of the partition with large amounts of uncertainty to be left unresolved. Finally, we explore the performance of our proposed methodology through simulation studies in Section \ref{sec:sims}. In the Appendix we present an application of our proposed approach to link three Colombian homicide record systems.

\section{Multifile Partitioning}
\label{sec:goals}

\begin{figure*}[h]
\centering
\centerline{\includegraphics[width=0.93\linewidth]{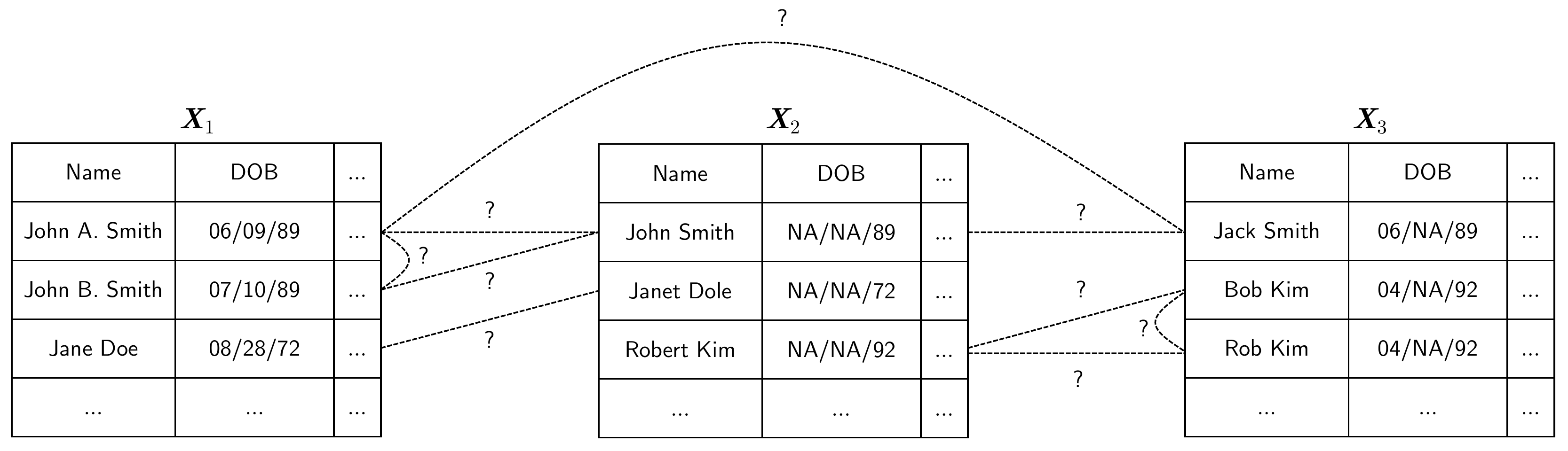}}
\begin{minipage}[b]{0.975\textwidth}
    \caption{A toy example of the multifile record linkage and duplicate detection problem.}
    \label{fig:toyex}
\end{minipage} 
\end{figure*}

Consider $K$ files $\bX_1, \cdots, \bX_K$, each containing information on possibly overlapping subsets of a population of entities. The goal of multifile record linkage and duplicate detection is to identify the sets of records in $\bX_1$, $\cdots$, $\bX_K$ that are coreferent, as illustrated in Figure \ref{fig:toyex}. Identifying coreferent records across datafiles represents the goal of record linkage, and identifying coreferent records within each file represents the goal of duplicate detection.

We denote the number of records contained in datafile $\bX_k$ as $r_k$, and the total number of records across all files as $r=\sum_{k=1}^K r_k$. We label the records in all datafiles in a consecutive order, that is, those in $\bX_1$ as $R_1=(1,\cdots,r_1)$, those in $\bX_2$ as $R_2=(r_1+1,\cdots,r_1+r_2)$, and so on, finally labeling the records in $\bX_K$ as $R_K=(\sum_{k=1}^{K-1}r_k+1,\cdots,r)$.  We denote $[r]=(1,\cdots,r)$, where it is clear that $[r]=(R_1,\cdots,R_K)$, which represents all the records coming from all datafiles. 

Formally, multifile record linkage and duplicate detection is a partitioning problem.  A partition of a set is a collection of disjoint subsets, called clusters, whose union is the original set.  In this context, the term \textit{coreference partition} refers to a partition $\C$ of all the records in $\bX_1$, $\cdots$, $\bX_K$, or equivalently a partition $\C$ of $[r]$, such that each cluster $c\in \C$ is exclusively composed of all the records generated by a single entity \citep{Matsakis_2010,Sadinle_2014}.  This implies that there is a one-to-one correspondence between the clusters in $\C$ and the entities represented in at least one of the datafiles.  Estimating $\C$ is the goal of multifile record linkage and duplicate detection.

\subsection{Multifile Coreference Partitions}
\label{sec:part_construct}

In the setting of multifile record linkage and duplicate detection, the datafiles are the product of $K$ data collection processes, which possibly introduce different patterns of duplication, measurement error, and missingness.  This indicates that records coming from different datafiles should be treated differently.   
To take this into account, we introduce the concept of a \emph{multifile coreference partition} by endowing a coreference partition $\C$ with additional structure to preserve the information on where  records come from.  Each cluster $c\in\C$ can be decomposed as $c=c^1\cup\cdots\cup c^k\cup\cdots\cup c^K$, where $c^{k}$ is the subset of records in cluster $c$ that belong to datafile $\bX_{k}$, which leads us to the following definition.

\begin{defi}
The \emph{multifile coreference partition} of datafiles $\bX_1, \cdots, \bX_K$ is obtained from the coreference partition $\C$ by expressing each cluster $c\in\C$ as a $K$-tuple $(c^1,\cdots, c^K)$, where $c^k$ represents the records of $c$ that come from datafile $\bX_k$.
\end{defi}

For simplicity we will continue using the notation $\C$ to denote a multifile coreference partition, although technically this new structure is richer and therefore different from a coreference partition that does not preserve the datafile membership of the records.  
The multifile representation of partitions is useful for decoupling the features that are important for within-file duplicate detection or for across-files record linkage. 

For duplicate detection, the goal is to identify coreferent records within each datafile. 
This can be phrased as estimating the \emph{within-file coreference partition} $\C_k$ of each datafile $\bX_k$.  Clearly, these $\C_k$ can be obtained from the multifile partition $\C$ by extracting the $k$th entry of each cluster $c=(c^1,\dots,c^K)\in\C$.  
Two useful summaries of a given within-file partition $\C_k$ are the number of within-file clusters $n_k=|\C_k|$, which is the number of unique entities represented in datafile $\bX_k$, and the within-file cluster sizes $\bd^k=\{|c^k| : c^k\in\C_k\}$, which represent the number of records associated with each entity in datafile $\bX_k$. 

On the other hand, in record linkage the goal is to identify coreferent records across datafiles.  Given the within-file partitions, $\C_1,\cdots,\C_K$, the goal can be phrased as identifying which clusters across these partitions represent the same entities.  This across-datafiles structure can be formally represented by a \textit{$K$-partite matching}.  Given $K$ sets $V_1,\cdots,V_K$, a $K$-partite matching $\M$ is a collection of subsets from $\cup_{k=1}^KV_k$ such that each $m\in\M$ contains maximum one element from each $V_k$.  If we think of each $V_k$ as the set of clusters $\C_k$ representing the entities in datafile $\bX_k$, then it is clear that the goal is to identify the $K$-partite matching $\M$ that puts together the clusters that refer to the same entities across datafiles.  This structure can be extracted from a multifile coreference partition $\C$, given that each element $c=(c^1,\dots,c^K)\in\C$ contains the coreferent clusters across all within-file partitions.  Indeed, a multifile coreference partition can be thought of as a $K$-partite matching of within-file coreference partitions.

A useful summary of the across-datafile structure is the amount of entity-overlap between datafiles, represented by the number of clusters $c=(c^1,\dots,c^K)\in\C$ with records in specific subsets of the files.  We can concisely summarize the entity-overlap of the datafiles through a contingency table. In particular, consider a $2^K$ contingency table with cells indexed by ${\bh}\in \{0,1\}^K$ and corresponding cell counts $n_{\bh}$. Here, $\bh$ represents a pattern of inclusion of an entity in the datafiles, where a 1 indicates inclusion and a 0 exclusion.  For instance,  
if $K=3$, $n_{011}$ is the number of clusters $c=(c^1, c^2, c^3)\in\C$ representing entities with records in datafiles $2$ and $3$ but without records in datafile $1$. We let $\HH=\{0,1\}^K \setminus \{0\}^K$ and denote the (incomplete) contingency table of counts as $\bn=\{n_{\bh}\}_{\bh\in\HH}$, which we refer to as the \textit{overlap table}. We ignore the cell $\{0\}^K$ which would represent entities that are not recorded in any of the $K$ files.  This cell is not of interest in this article, although it is the parameter of interest in population size estimation \citep[see e.g.][]{Bird_2018}.

\begin{figure*}[ht]
\centering
\centerline{\includegraphics[width=0.975\linewidth]{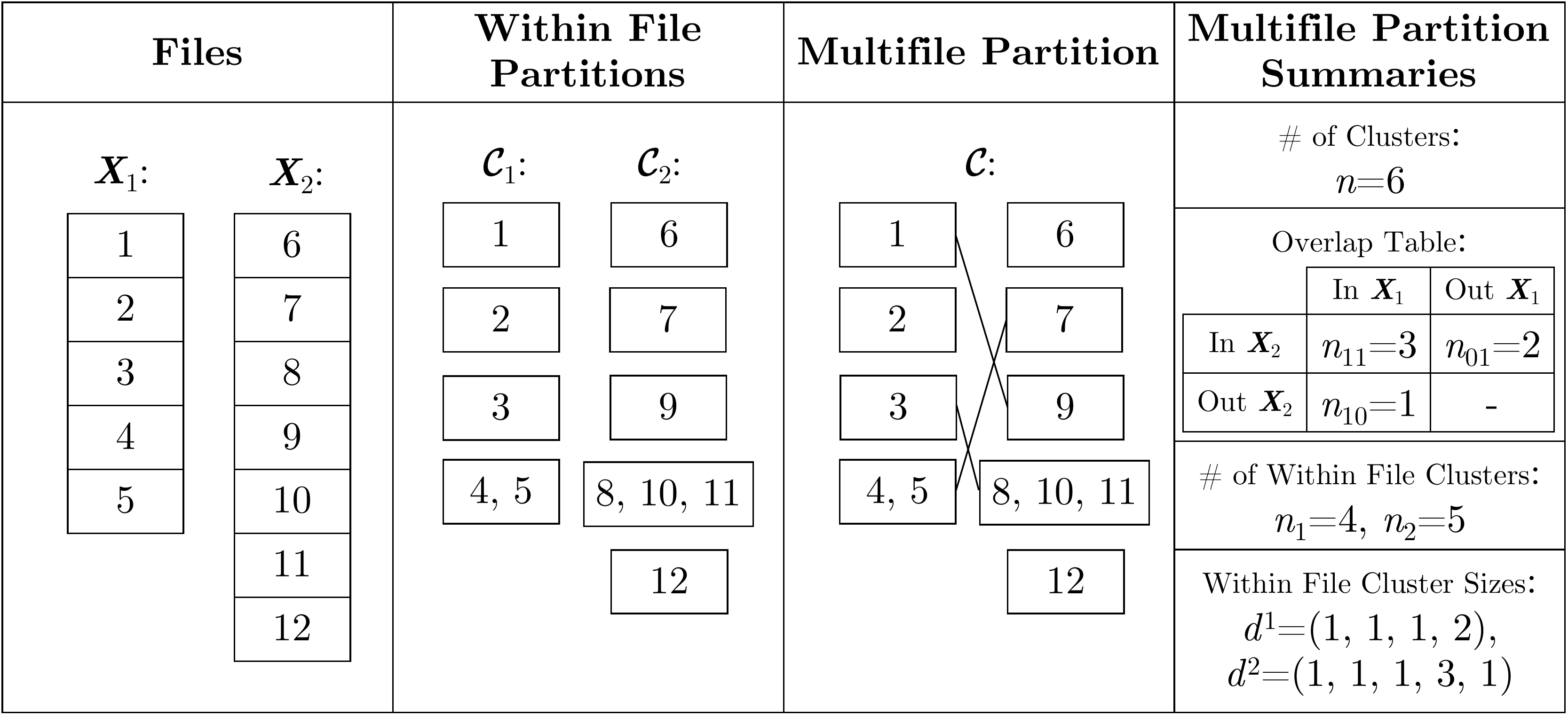}}
\begin{minipage}[b]{0.975\textwidth}
    \caption{An illustration of a multifile partition of $[12]$, where $\bX_1$ contains records $1-5$ and $\bX_2$ contains records $6-12$.}
    \label{fig:multifilepart}
\end{minipage} 
\end{figure*}

\emph{Example}.  To illustrate the concept of a multifile partition, consider two files with five and seven records respectively, so that $\bX_1$ contains records $1-5$ and $\bX_2$ contains records $6-12$. Suppose the coreference partition is $\{ \{1,9 \}, \{2 \},\{3,8,10, 11\},\{4,5,7 \},\{6 \},\{12\}\}$.  The corresponding multifile partition is $\C=\{ (\{1\}, \{9\}), (\{2 \}, \emptyset), (\{3\}, \{8,10,11 \}), (\{4,5\}, \{7\}), (\emptyset, \{6 \}), (\emptyset, \{12 \})\}$.  As illustrated in Figure \ref{fig:multifilepart}, the within-file partitions can be extracted as $\C_1=\{ \{1\}, \{2 \}, \{3\}, \{4,5\} \}$ and $\C_2=\{ \{6\}, \{7 \}, \{9 \}, \{8,10,11\}, \{12\} \}$, and the within-file cluster sizes are $\bd^1=(1,1,1,2)$ and $\bd^2=(1,1,1, 3,1)$.  The overlap table in this case is $\{n_{11},n_{10},n_{01}\}$, indicating that $n_{11}=3$ entities are represented in both datafiles, $n_{10}=1$ entity is represented only in the first datafile, and $n_{01}=2$ entities are represented only in the second datafile.  In total, there are $n_1=|\C_1|=n_{11}+n_{10}=4$ unique entities represented in $\bX_1$, $n_2=|\C_2|=n_{11}+n_{01}=5$ unique entities represented in $\bX_2$, and $n=|\C|=n_{11}+n_{10}+n_{01}=6$ entities among both datafiles.  

\section{A Structured Prior for Multifile Partitions}
\label{sec:part_prior}

Bayesian approaches to multifile record linkage and duplicate detection require prior distributions on multifile coreference partitions.  We present a generative process for multifile partitions, building on our representation introduced in Section \ref{sec:part_construct}. The idea is to generate a multifile partition by first generating summaries that characterize it, as follows:
\begin{enumerate}
    \itemsep-0.5em 
	\item Generate the number of unique entities $n$ represented in the datafiles, which also corresponds to the number of clusters of the multifile partition.
	\item Given $n$, generate an overlap table $\bn=\{n_{\bh}\}_{\bh\in\HH}$ so that $n=\sum_{{\bh}\in\HH} n_{\bh}$, where $\HH=\{0,1\}^K \setminus \{0\}^K$. From $\bn$ we can derive the number of entities in datafile ${\bX}_k$ as $n_k=\sum_{{\bh}\in\HH} {h}_k n_{\bh}$, where ${h}_k$ is the $k$th entry of $\bh$.
	\item For each $k=1,\cdots,K$, given $n_k$, independently generate a set of counts $\bd^{k}=\{d^k_i\}_{i=1}^{n_k}$, representing the number of records associated with each entity in file ${\bX}_k$. From $\bd^{k}$, we can derive the number of records in file ${\bX}_k$ as $r'_k=\sum_{i=1}^{n_k}d^{k}_i$.  Index the $r'_k$ records as $R'_k=(\sum_{l=1}^{k-1}r'_l+1,\cdots,\sum_{l=1}^{k}r'_l)$.
	\item For each $k=1,\cdots,K$, given $\bd^{k}$, induce a within-file partition $\C_k$ by randomly allocating $R'_k$ into $n_k$ clusters of sizes $d^{k}_1, \cdots, d^{k}_{n_k}$.
	\item Given the overlap table $\bn$ and within-file partitions $\{\C_1,\cdots, \C_K\}$, generate a $K$-partite matching of the within-file partitions by selecting uniformly at random from the set of all $K$-partite matchings with overlap table $\bn$. By definition, the result is a multifile coreference partition.
\end{enumerate}  
By parameterizing each step of this generative process, we can construct a prior distribution for multifile partitions, as we now show. 
\subsection{Parameterizing the Generative Process}\label{sec:param_gen}
\textit{Prior for the Number of Entities or Clusters}. 
In the absence of substantial prior information, we follow a simple choice for the prior on the number of clusters, by taking a uniform distribution over the integers less than some upper bound, $U$, i.e. $\PP(n)=U^{-1}I(n \in \{1,\cdots, U\})$. In practice, we set $U$ to be the actual number of records across all datafiles $r$, which is observed. \serge{More informative specifications are discussed in Appendix A.}

\noindent \textit{Prior for the Overlap Table}.
Conditional on $n$, we use a Dirichlet-multinomial distribution as our prior on the overlap table $\bn=\{n_{\bh}\}_{{\bh}\in\HH }$. Given a collection of positive hyperparameters for each cell of the overlap table, $\balpha=\{\alpha_{\bh}\}_{{\bh}\in\HH }$, and letting $\alpha_0=\sum_{{\bh}\in\HH }\alpha_{\bh}$, the prior for the overlap table under this choice is
$\PP(\bn\mid n)=
[(n!)\Gamma(\alpha_0)/\Gamma(n+\alpha_0)]\prod_{{\bh}\in\HH }[\Gamma(n_{\bh}+\alpha_{\bh})/(n_{\bh}!)\Gamma(\alpha_{\bh})].$
Due to conjugacy, 
$\balpha$ can be interpreted as prior cell counts, which can be used to incorporate prior information about the overlap between datafiles. \serge{In the absence of substantial prior information, when the number of files is not too large and the overlap table is not expected to be sparse, we recommend setting $\balpha=(1,\cdots,1)$. In Appendix A we discuss alternative specifications when the overlap table is expected to be sparse.} 

\noindent \textit{Prior for the Within-File Cluster Sizes}.
Given the number of entities in datafile $\bX_k$, $n_k=\sum_{{\bh}\in\HH} {h}_k n_{\bh}$, we generate the within-file cluster sizes $\bd^{k}=\{d^k_i\}_{i=1}^{n_k}$
assuming that $d^k_1,\cdots, d^k_{n_k}\mid n_k\stackrel{iid}{\sim} p_k(\cdot)$. Here $p_k(\cdot)$ represents the probability mass function of a distribution on the positive integers, so that $\PP(\bd^k\mid n_k)=\prod_{i=1}^{n_k}p_k(d^k_i)$.  We do not expect a-priori many duplicates per entity, and therefore we expect the counts in $\bd^{k}$ to be mostly ones or to be very small \citep{Miller_2015, Zanella_2016}. We therefore use a similar approach to \cite{Klami_2016}, and use distributions truncated to the range $\{1,\cdots, U^k\}$, where $U^k$ is a file-specific upper bound on cluster sizes.  We further use distributions where prior mass is concentrated at small values.  A default specification is to use a Poisson distribution with mean $1$ truncated to $\{1,\cdots, U^k\}$, i.e. $p_k(d^k_i)\propto (d^k_i!)^{-1}I(d^k_i\in\{1,\cdots, U^k\})$. \serge{
More informative options could be used for $p_k(\cdot)$ by using any distribution on $\{1,\cdots, U^k\}$, where this could vary from file to file if some files were known to have more or less duplication.}

\noindent \textit{Prior for the Within-File Partitions}. 
Given the within-file cluster sizes $\bd^{k}$, the number of ways of assigning $d^{k}_1,\cdots,d^{k}_{n_k}$ records to clusters $1,\cdots,n_k$, respectively, is given by the multinomial coefficient $r'_k!/\prod_{i=1}^{n_k}d^{k}_i!$, with $r'_k=\sum_{i=1}^{n_k}d^{k}_i$.  However, the ordering of the clusters is irrelevant for constructing the within-file partition $\C_k$ of $R'_k$.  There are $n_k!$ ways of ordering the $n_k$ clusters of $\C_k$, which leads to $r'_k!/(n_k! \prod_{i=1}^{n_k}d^{k}_i!)$ partitions of $R'_k$ into clusters of sizes $d^{k}_1,\cdots,d^{k}_{n_k}$.  
We then have $\PP(\C_k\mid \bd^k)=(n_k!/r'_k!)\prod_{i=1}^{n_k}d^{k}_i!$.

\noindent \textit{Prior for the \texorpdfstring{$K$}{K}-Partite Matching}.
Given the overlap table $\bn$ and the within-file partitions $\{\C_1,\cdots, \C_K\}$, our prior over $K$-partite matchings of the within-file partitions is uniform. Thus we just need to count the number of $K$-partite matchings with overlap table $\bn$. This is taken care of by Proposition \ref{prop:kpart}, proven in Appendix A.
\begin{prop}
    \label{prop:kpart}
	The number of $K$-partite matchings that have the same overlap table, $\bn=\{n_{\bh}\}_{{\bh}\in\HH }$, is  $\prod_{k=1}^{K}n_k!/\prod_{{\bh}\in\HH }n_{\bh}!$.
	Thus $\PP(\C\mid \{\C_k\}_{k=1}^{K}, \bn)=
\prod_{{\bh}\in\HH }n_{\bh}!/\prod_{k=1}^{K}n_k!.$
\end{prop}

\noindent \textit{The Structured Prior for Multifile Partitions}.
Letting quantities followed by $(\C)$ mean they are computable from $\C$, the density of our structured prior for multifile partitions is
\begin{align}
\label{eq:full_prior}
\PP(\C) &= \PP(n) \PP(\bn\mid n) \prod_{k=1}^{K}\left[\PP(\bd^k\mid n_k) \PP(\C_k\mid \bd^k)\right] \PP(\C\mid \{\C_k\}_{k=1}^{K}, \bn) \nonumber\\
&= \PP(n(\C)) \dfrac{n(\C)!~\Gamma(\alpha_0)}{\Gamma(n(\C)+\alpha_0)}\prod_{{\bh}\in\HH }\left[\dfrac{\Gamma(n_{\bh}(\C)+\alpha_{\bh})}{\Gamma(\alpha_{\bh})}\right] 
\prod_{k=1}^{K} \left[ \frac{1}{r'_k(\C)!}
\prod_{c_k\in\C_k}[|c_k|!~p_k(|c_k|)]\right].
\end{align}

\subsection{Comments and Related Literature}
\label{sec:prior_comments} 

The structured prior for multifile partitions allows us to incorporate prior information about the total number of clusters, the overlap between files, and the amount of duplication in each file. If we restrict the prior for the within-file cluster sizes to be $p_k(d^k_i)=I(d^k_i=1)$ for a given datafile $\bX_k$, then we enforce the assumption that there are no duplicates in that file.  Imposing this restriction for all datafiles leads to the special case of a prior for $K$-partite matchings, which is of independent interest, as we are not aware of any such constructions outside of the bipartite case \citep{Fortini_2001, Larsen_2005, Sadinle_2017}.

Our prior construction, where priors are first placed on interpretable summaries of a partition 
and then a uniform prior is placed on partitions which have those summaries,
mimics the construction of the priors on bipartite matchings of \cite{Fortini_2001, Larsen_2005} and \cite{Sadinle_2017}, 
and the Kolchin and allelic partition priors of \cite{Zanella_2016} and \cite{Betancourt_2020b}.
While the Kolchin and allelic partition priors could both be used as priors for multifile partitions, these do not incorporate the datafile membership of records. 
Using these priors in the multifile setting would imply 
that the sizes of clusters containing records from only one file have the same prior distribution as the sizes of clusters containing records from two files, which should not be true in general. 

\cite{Miller_2015} and \cite{Zanella_2016} proposed the \textit{microclustering property} as a desirable requirement for partition priors in the context of duplicate detection: 
denoting the size of the largest cluster in a partition of $[r]$ by $M_r$, a prior satisfies the microclustering property if $M_r/r\to0$ in probability as $r\to\infty$. 
A downside of priors with this property is that they can still allow the size of the largest cluster to go to $\infty$ as $r$
increases. For this reason \cite{Betancourt_2020b} 
introduced the stronger \textit{bounded microclustering property}, which we believe is more practically important: for any $r$, $M_r$ is finite with probability $1$. Our prior satisfies the bounded microclustering property as $M_r\leq \sum_{k=1}^K U^k$.

While our parameter of interest is a partition $\C$ of $r$ records, the prior developed in this section is a prior for a partition of a random number of records. In practice we condition on the file sizes, $\{r_k\}_{k=1}^K$, and use the prior 
$\PP(\C\mid \{r_k\}_{k=1}^K)\propto \PP(\C)I(r'_k(\C)=r_k(\C) \text{ for all } k )$, which alters the interpretation of the prior. A similar problem occurs for the 
Kolchin partition priors of \cite{Zanella_2016}. This motivated the exchangeable sequences of clusters priors of \cite{Betancourt_2020}, which are similar to Kolchin partition priors, but lead to a directly interpretable prior specification. It would be interesting in future work to see if an analogous prior could be developed for our structured prior for multifile partitions. Despite this limitation, we demonstrate in simulations in Section \ref{sec:sims} that incorporating strong prior information into our structured prior for multifile partitions can lead to 
improved frequentist performance over a default specification.

\section{A Model for Comparison Data}
\label{sec:comp_model}

We now introduce a comparison-based modeling approach to multifile record linkage and duplicate detection, building on the work of \cite{Fellegi_1969, Jaro_1989, Winkler_1994, Larsen_2001, Fortini_2001, Larsen_2005} and \cite{Sadinle_2014, Sadinle_2017}. Working under the intuitive assumption that coreferent records will look similar, and non-coreferent records will look dissimilar, these approaches construct statistical models for comparisons computed between each pair of records.

There are two implications of the  multifile setting described in Section \ref{sec:goals} that are important to consider when constructing a model for the comparison data. First, models for the comparison data should account for the fact that the distribution of the comparisons between record pairs might potentially change across different pairs of files. For example, if files $\bX_k$ and $\bX_{k'}$ are not accurate, whereas files $\bX_q$ and $\bX_{q'}$ are,
then the distribution of comparisons between $\bX_k$ and $\bX_{k'}$ will look very different compared with the distribution of comparisons between $\bX_q$ and $\bX_{q'}$. Second, the fields available for comparison will vary across pairs of files. For example, files $\bX_k$ and $\bX_{k'}$ may have collected information on a field that file $\bX_q$ did not. In this scenario, we would like a model that is able to utilize this extra field when linking $\bX_k$ and $\bX_{k'}$, even though it is not available in $\bX_q$.
In this section we introduce a Bayesian comparison-based model that explicitly 
handles the multifile setting by constructing a likelihood function that models comparisons of fields between different pairs of files separately. 
The separate models are able to adapt to the level of noise of each file pair, and the maximal number of fields are able to be compared for each file pair. 

\subsection{Comparison Data}

We construct comparison vectors for pairs of records  to provide evidence for whether they correspond to the same entity. For $k\leq k'$, let $\pairP_{kk'}=\{(i,j) : i<j, i\in {\bX}_k, j\in {\bX}_{k'} \}$ denote the set of all record pairs between files ${\bX}_k$ and ${\bX}_{k'}$, and let $F$ be the total number of different fields available from the $K$ files. For each file pair $(k, k')$, $k\leq k'$, and record pair $(i,j)\in \pairP_{kk'}$, we compare each field $f=1,\dots,F$ using a similarity measure $\mathcal{S}_f(i, j)$, which will depend on the data type of field $f$. For unstructured categorical fields such as race, $\mathcal{S}_f$ can be a binary comparison which checks for agreement. For more structured fields containing strings or numbers, $\mathcal{S}_f$ should be able to capture partial agreements. For example, string fields can be compared using a string metric like the Levenshtein edit distance \citep[see e.g.][]{Bilenko_2003}, and numeric fields can be compared using absolute differences. Comparison $\mathcal{S}_f(i, j)$ will be missing if field $f$ is not recorded in record $i$ or record $j$, which includes the case where field $f$ is not recorded in datafiles $\bX_k$ or $\bX_{k'}.$

While we could directly model the similarity measures $\mathcal{S}_f(i, j)$, this would require a custom model for each type of comparison, which inherits similar problems to the direct modeling of the fields themselves. Instead, we follow \cite{Winkler_1990} and \cite{Sadinle_2014, Sadinle_2017} in dividing the range of $\mathcal{S}_f$ into $L_f+1$ intervals $I_{f0}, I_{f1}, \cdots, I_{fL_f}$ that represent varying levels of agreement, with $I_{f0}$ representing the highest level of agreement, and $I_{fL_f}$ representing the lowest level of agreement. We then let $\gamma_{ij}^f=l$ if $\mathcal{S}_f(i, j)\in I_{fl}$, where larger values of $\gamma_{ij}^f$ represent larger disagreements between records $i$ and $j$ in field $f$.  
Finally, we form the comparison vector $\bgamma_{ij}=(\gamma_{ij}^1, \cdots, \gamma_{ij}^F)$. Constructing the comparison data this way allows us to build a generic modeling approach.
In particular, 
extending \cite{Fortini_2001, Larsen_2005} and  \cite{Sadinle_2014, Sadinle_2017}, our model for the comparison data is
\begin{align*}
\bgamma_{ij}\mid \C(i)=\C(j), (i,j)\in \pairP_{kk'} &\stackrel{iid}{\sim}\textsf{M}_{kk'}(\bem_{kk'}), \nonumber \\
\bgamma_{ij}\mid \C(i)\neq \C(j), (i,j)\in \pairP_{kk'} &\stackrel{iid}{\sim}\textsf{U}_{kk'}(\bu_{kk'}), \nonumber
\end{align*}
where $\C$ is a multifile partition, $\C(i)$ denotes record $i$'s cluster in $\C$, $\C(i)=\C(j)$ indicates that records $i$ and $j$ are coreferent, $\textsf{M}_{kk'}(\bem_{kk'})$ is a model for the comparison data among coreferent record pairs from the file pair ${\bX}_k$ and ${\bX}_{k'}$, 
$\textsf{U}_{kk'}(\bu_{kk'})$ is a model for the comparison data among non-coreferent record pairs from datafile pair ${\bX}_k$ and ${\bX}_{k'}$, and $\bem_{kk'}$ and $\bu_{kk'}$ are vectors of parameters. 

\serge{In the next section we make two further assumptions that simplify the model parameterization. Before doing so, we note a few 
limitations of our comparison-based model.
First, 
computing comparison vectors scales quadratically in the number of records. 
Second, comparison vectors for different record pairs are not actually independent conditional on the partition \citep[see Section 2 of][]{Tancredi_2011}. 
Third, 
 modeling discretized comparisons of record fields represents a loss of information. While the first limitation is computational and unavoidable in the absence of blocking (see Appendix B), the other two limitations are inferential. Despite these  limitations, we find in Section \ref{sec:sims} that the combination of our structured prior for multifile partitions and our comparison-based model can produce linkage estimates with satisfactory frequentist performance.}

\subsection{Conditional Independence and Missing Data}

Under the assumptions that the fields in the comparison vectors are conditionally independent given the multifile partition of the records and that missing comparisons are ignorable \citep{Sadinle_2014, Sadinle_2017}, 
the likelihood of the observed comparison data, $\bgamma^{obs}$, becomes 
\begin{align}\label{eq:lhood}
\mathcal{L}(\C, \Phi\mid \bgamma^{obs})&=
\prod_{k\leq k'}\prod_{f=1}^F\prod_{l=0}^{L_f} (m_{kk'}^{fl})^{a_{kk'}^{fl}(\C)} (u_{kk'}^{fl})^{b_{kk'}^{fl}(\C)}.
\end{align}
Here $m_{kk'}^{fl}=\PP(\gamma_{ij}^f=l\mid \C(i)=\C(j), (i,j)\in \pairP_{kk'})$,
$u_{kk'}^{fl}=\PP(\gamma_{ij}^f=l\mid \C(i)\neq\C(j), (i,j)\in \pairP_{kk'})$, $a_{kk'}^{fl}(\C) = \sum_{(i,j) \in \pairP_{kk'}} I_{obs}(\gamma_{ij}^f) I(\gamma_{ij}^f=l) I(\C(i)=\C(j)),$ $b_{kk'}^{fl}(\C) = \sum_{(i,j) \in \pairP_{kk'}} I_{obs}(\gamma_{ij}^f) I(\gamma_{ij}^f=l) I(\C(i)\neq\C(j)),$
$I_{obs}(\cdot)$ is an indicator of whether its argument was observed,   and $\Phi=(\bem,\bu)$ where $\bem$ collects all of the $\bem_{kk'}^{f}=(m_{kk'}^{f0},\cdots, m_{kk'}^{fL_f})$ and $\bu$ collects all of the $\bu_{kk'}^{f}=(u_{kk'}^{f0},\cdots, u_{kk'}^{fL_f})$. For a given multifile partition $\C$, $a_{kk'}^{fl}(\C)$ represents the number of record pairs in $\pairP_{kk'}$ that belong to the same cluster with observed agreement at level $l$ in field $f$, and $b_{kk'}^{fl}(\C)$ represents the number of record pairs in $\pairP_{kk'}$ that do not belong to the same cluster with observed agreement at level $l$ in field $f$.

\section{Bayesian Estimation of Multifile Partitions}
\label{sec:bayesest}

Bayesian estimation of the multifile coreference partition $\C$ is based on the posterior distribution $p(\C, \Phi \mid \bgamma^{obs})\propto  \PP(\C)p(\Phi) \mathcal{L}(\C, \Phi\mid \bgamma^{obs})$, where $\PP(\C)$ is our structured prior for multifile partitions \eqref{eq:full_prior}, $\mathcal{L}(\C, \Phi\mid \bgamma^{obs})$ is the likelihood from our model for comparison data \eqref{eq:lhood}, and $p(\Phi)$ represents a prior distribution for the $\Phi=(\bem,\bu)$ model parameters.  We now specify this prior $p(\Phi)$, outline a Gibbs sampler to sample from $p(\C, \Phi \mid \bgamma^{obs})$, and present a strategy to obtain point estimates of the multifile partition $\C$.

\subsection{Priors for \texorpdfstring{$\bem$}{m} and \texorpdfstring{$\bu$}{u}} \label{sec:m_u_prior}

We will use independent, conditionally conjugate priors for $\bem_{kk'}^{f}$ and $\bu_{kk'}^{f}$, namely $\bem_{kk'}^{f}\sim\textsf{Dirichlet}(\mu_{kk'}^{f0},\cdots, \mu_{kk'}^{fL_f})$ and $\bu_{kk'}^{f}\sim\textsf{Dirichlet}(\nu_{kk'}^{f0},\cdots, \nu_{kk'}^{fL_f})$.
\serge{In this article we will use a default specification of $(\mu_{kk'}^{f0},\cdots, \mu_{kk'}^{fL_f})=(\nu_{kk'}^{f0},\cdots, \nu_{kk'}^{fL_f})=(1,\cdots, 1)$. We believe this prior specification is sensible for the $\bu$ parameters, following the discussion in Section 3.2 of \cite{Sadinle_2014}, as comparisons amongst non-coreferent records are likely to be highly variable and it is more likely than not that eliciting meaningful priors for them is too difficult. For the $\bem$ parameters, it might be desirable in certain applications to introduce more information into the prior. For example, one could set $\mu_{kk'}^{f0}>\cdots> \mu_{kk'}^{fL_f}$ to incorporate the prior belief that higher levels of agreement should have larger prior probability than lower levels of agreement. Another route would be to use the sequential parameterization of the $\bem$ parameters, and the associated prior recommendations, described in \cite{Sadinle_2014}.}

\subsection{Posterior Sampling}

In Appendix B we outline a Gibbs sampler that produces a sequence of samples $\{\C^{[t]},\Phi^{[t]}\}_{t=1}^T$ from the posterior distribution $p(\C, \Phi \mid \bgamma^{obs})$, which we will use to obtain Monte Carlo approximations of posterior expectations involved in the derivation of point estimates $\hat{\C}$, as presented in the next section. 
\serge{In Appendix B, we discuss the computational complexity of the Gibbs sampler, how computational performance can be improved through the usage of indexing techniques, and the initialization of the Gibbs sampler.}

\subsection{Point Estimation}

In a Bayesian setting, one can obtain a point estimate $\hat{\C}$ of the multifile partition using the 
posterior $\PP(\C\mid\bgamma^{obs})=\int p(\C, \Phi \mid \bgamma^{obs})d\Phi$ and a loss function $L(\C, \hat{\C})$. The Bayes estimate is the multifile partition $\hat{\C}$ that minimizes the expected posterior loss $\mathbb{E}[L(\C, \hat{\C})\mid\bgamma^{obs}]=\sum_{\C}L(\C, \hat{\C})\PP(\C\mid\bgamma^{obs})$, although in practice such expectations are approximated using posterior samples. \serge{Previous examples of loss functions for partitions included Binder's loss \citep{Binder_1978} and the variation of information \citep{Meila_2007}, both recently surveyed in \cite{Wade_2018}. \mcio{The quadratic and absolute losses presented in \cite{Tancredi_2011} are special cases of Binder's loss, and \cite{Steorts_2016} drew connections between their proposed maximal matching sets and the losses of \cite{Tancredi_2011}.}}

In many applications there may be much uncertainty on  the linkage decision for some records in the datafiles. For example, in Figure \ref{fig:toyex} it is unclear which of the records with last name ``Smith" are coreferent. It is thus desirable to leave decisions for some records unresolved, so that the records can be hand-checked during a clerical review, which is common in practice \citep[see e.g.][]{Ball_2019}.  In the classification literature, leaving some decisions unresolved is done through a \textit{reject option} \cite[see e.g.][]{Herbei_2006}, which here we will refer to as an \textit{abstain option}.  We will refer to point estimates with and without abstain option as \textit{partial estimates} and \textit{full estimates}, respectively.
We now present a family of loss functions for multifile partitions which incorporate an abstain option, building upon the family of loss functions for bipartite matchings presented in \cite{Sadinle_2017}.

\subsubsection{A Family of Loss Functions with an Abstain Option}
\label{sec:loss}

For the purpose of this section we will represent a multifile partition $\C$ as a vector $\bZ=(Z_1,\cdots, Z_r)$ of labels, where $Z_i\in\{1,\dots,r\}$, such that $Z_i=Z_j$ if $\C(i)=\C(j)$. We represent a Bayes estimate here as a vector $\hat{\bZ} = (\hat{Z_{1}}, \dots, \hat{Z_{r}})$, where $\hat{Z_i}\in\{1, \dots, r, A\}$, with $A$ representing an abstain option intended for records whose linkage decisions are not clear and need further review. \mcio{We assign different losses to using the abstain option and to different types of matching errors.} We propose to compute the overall loss additively, as 
$L(\bZ,\hat\bZ) = \sum_{i=1}^{r} L_i(\bZ,\hat{\bZ})$.  To introduce the expression for the $i$th-record-specific loss $L_i(\bZ,\hat{\bZ})$, we use the notation $\Delta_{ij}=I(Z_i=Z_{j})$, and likewise $\hat \Delta_{ij}=I(\hat Z_i= \hat Z_{j})$. 

The proposed individual loss for record $i$ is
\begin{equation}\label{eq:LossR_j}
L_i(\bZ,\hat{\bZ})=\left\{
  \begin{array}{ll}
    \lambda_{A}, & \hbox{ if } \hat Z_i= A, \\
    0, & \hbox{ if } \Delta_{ij}=\hat \Delta_{ij} \text{ for all } j \text{ where } \hat Z_{j}\neq A, \\
    \lfnm, & \hbox{ if } \hat Z_i\neq A, ~~ \sum_{j\neq i}\hat \Delta_{ij}=0, ~~ \sum_{j\neq i} \Delta_{ij}>0, \\
		\lfma, & \hbox{ if } \hat Z_i\neq A, ~~ \sum_{j\neq i}\hat \Delta_{ij}>0, ~~ \sum_{j\neq i} \Delta_{ij}=0, \\
		\lfmb, & \hbox{ if } \hat Z_i\neq A, ~~ \sum_{j\neq i} \hat\Delta_{ij}>0, ~~ \sum_{j\neq i}(1-\hat\Delta_{ij})\Delta_{ij}>0.\\
  \end{array}
\right.
\end{equation}
That is, $\lambda_{A}$ represents the loss from abstaining from making a decision; $\lfnm$ is the loss from a false non-match (FNM) decision, that is, deciding that record $i$ does not match any other record ($\sum_{j\neq i}\hat \Delta_{ij}=0$) when in fact it does ($\sum_{j\neq i} \Delta_{ij}>0$); $\lfma$ is the loss from a type 1 false match (FM1) decision, that is, deciding that record $i$ matches other records ($\sum_{j\neq i}\hat \Delta_{ij}>0$) when it does not actually match any other record ($\sum_{j\neq i} \Delta_{ij}=0$); and $\lfmb$ is the loss from a type 2 false match (FM2), that is, a false match decision when record $i$ is matched to other records ($\sum_{j\neq i} \hat\Delta_{ij}>0$) but it does not match all of the records it should be matching ($\sum_{j\neq i}(1-\hat\Delta_{ij})\Delta_{ij}>0$).  

The posterior expected loss is 
$ \mathbb{R}(\hat\bZ)
= \sum_{i=1}^{r}\mathbb{E}[L_i(\bZ,\hat{\bZ})\mid \bgamma^{obs}],$
where
\begin{flalign}\label{eq:PostExpLoss_j}
\hspace{-.2cm}\mathbb{E}[L_i(\bZ,\hat{\bZ})\mid \bgamma^{obs}]&=\left\{
  \begin{array}{ll}
    \lambda_{A}, & \hspace{-.3cm}\hbox{if } \hat Z_i=A, \\		
		\lfnm ~ \PP(\sum_{j\neq i} \Delta_{ij}>0\mid \bgamma^{obs}), & \hspace{-.3cm}\hbox{if } \hat Z_i\neq A, \hspace{-.05cm}\sum_{j\neq i}\hat \Delta_{ij}=0, \\
		\lfma ~ \PP(\sum_{j\neq i} \Delta_{ij}=0\mid \bgamma^{obs}) ~ +&\\
		\lfmb ~ \PP(\sum_{j\neq i}(1-\hat\Delta_{ij})\Delta_{ij}>0\mid \bgamma^{obs}), & \hspace{-.3cm}\hbox{if } \hat Z_i\neq A, \hspace{-.05cm} \sum_{j\neq i}\hat \Delta_{ij}>0,
  \end{array}
\right.
\end{flalign}
and quantities computed with respect to the posterior distribution, $\PP(\bZ\mid \bgamma^{obs})$, can all be approximated using posterior samples.  While this presentation is for general positive losses $\lfnm,\lfma,\lfmb$ and $\lambda_A$, these only have to be specified up to a proportionality constant \citep{Sadinle_2017}.  If we do not want to allow the abstain option, then we can set $\lambda_{A}=\infty$ and the derived full estimate $\hat\bZ$ will have a linkage decision for all records.  Although we have been using partition labelings $\bZ$, the expressions in \eqref{eq:LossR_j} and \eqref{eq:PostExpLoss_j} are invariant to different labelings of the same partition.  In the two-file case, \cite{Sadinle_2017} provided guidance on how to specify the individual losses $\lfnm,\lfma,\lfmb$ and $\lambda_A$ in cases where there is a notion of false matches being worse than false non-matches or vise versa.
\cite{Sadinle_2017} also gave recommendations for default values of these losses that lead to good frequentist performance in terms not over- or under-matching across repeated samples. In Appendix C, we discuss how our proposed loss function differs from the loss function of \cite{Sadinle_2017} and propose a strategy for approximating the Bayes estimate.

\section{Simulation Studies} \label{sec:sims}

To explore the performance of our proposed approach for linking three duplicate-free files, as in the application to the Colombian homicide record systems of Appendix E, we present two simulation studies under varying scenarios of measurement error and datafile overlap. The two studies correspond to scenarios with equal and unequal measurement error across files, respectively. Both studies present results based on full estimates. In Appendix D we further explore the performance of our proposed approach for linking three files with duplicates, with results based on full and partial estimates.

\subsection{General Setup} \label{sec:sim_setup}

We start by describing the general characteristics of the simulations. For each of the simulation scenarios we conduct 100 replications, for each of which we generate three files as follows.  For each of $n=500$ entities, $\bh\in\HH$ is drawn from a categorical distribution with probabilities $\{p_{\bh}\}_{\bh\in\HH}$, where $\bh$ represents the subset of files the entity appears in, and so we change the values of $\{p_{\bh}\}_{\bh\in\HH}$ across simulation scenarios to represent varying amounts of file overlap.  Files are then created by generating the implied number of records for each entity. In the additional simulations considered in Appendix D, the generated number of records for each entity depends not only on $\bh$, but also on the duplication mechanism. 

All records are generated using a synthetic data generator developed in \cite{Tran_2013}, which allows for the incorporation of different forms of measurement error in individual fields, along with dependencies between fields we would expect in applications.  The data generator first generates clean records before distorting them to create the observed records. In particular, each observed record will have a fixed number of erroneous fields, where errors selected uniformly at random from a set of field dependent errors displayed in Table 3 of \cite{Sadinle_2014} (reproduced in Appendix D), with a maximum of two errors per field.  We generate records with seven fields of information: sex, given name, family name, age, occupation, postal code, and phone number.  

For each simulation replicate, we construct comparison vectors as given in Table 4 of \cite{Sadinle_2014} (reproduced in Appendix D).
We use the model for comparison data proposed in Section \ref{sec:comp_model} with flat priors on $\bem$ and $\bu$ as discussed in Section \ref{sec:m_u_prior}, and the structured prior proposed in Section \ref{sec:part_prior} with a uniform prior on the number of clusters and $\balpha=(1,\cdots,1)$ as described in Section \ref{sec:param_gen}. Using the Gibbs sampler presented in Appendix B we obtain $1,000$ samples from the posterior distribution of multifile partitions, and discard the first $100$ as burn-in. \serge{In Appendix D we discuss convergence of the Gibbs sampler, present running times of the proposed approach, and present an extra simulation exploring the running time of the approach with a larger number of records.} 
We then approximate the Bayes estimate $\hat{\bZ}$ for multifile partitions using the loss function described in Section \ref{sec:loss} as described in Appendix C. For full estimates, we use the default values of $\lfnm=\lfma=1$ and $\lfmb=2$ recommended by \cite{Sadinle_2017}. \serge{ In Appendix D we explore alternative specifications of the loss function.}

We will assess the performance of the Bayes estimate using 
\textit{precision} and \textit{recall} with respect to the true coreference partition $\bZ$. Let $\pairP$ be the set of all record pairs. Using notation from Section \ref{sec:loss}, let
$TM(\bZ, \hat{\bZ})=\sum_{(i,j)\in \pairP}\Delta_{ij}\hat{\Delta}_{ij}$ be the number of true matches (record pairs correctly declared coreferent),
$FM(\bZ, \hat{\bZ})=\sum_{(i,j)\in \pairP}(1-\Delta_{ij})\hat{\Delta}_{ij}$ be 
the number of false matches (record pairs incorrectly declared coreferent),
and $FNM(\bZ, \hat{\bZ})=\sum_{(i,j)\in \pairP}\Delta_{ij}(1-\hat{\Delta}_{ij})$ the number of false non-matches (record pairs incorrectly declared non-coreferent).
Then \textit{precision} is $TM(\bZ, \hat{\bZ}) /[TM(\bZ, \hat{\bZ})+FM(\bZ, \hat{\bZ})]$, the proportion of record pairs declared as coreferent that were truly coreferent, and \textit{recall} is $TM(\bZ, \hat{\bZ})/[TM(\bZ, \hat{\bZ})+FNM(\bZ, \hat{\bZ})]$, the proportion of record pairs that were truly coreferent that were correctly declared as coreferent. Perfect performance corresponds to precision and recall both being 1. 
In the simulations, we computed the median, $2$nd, and $98$th percentiles of these measures over the 100 replicate data sets.
{Additionally, in Appendix D, we assess the performance of the Bayes estimate when estimating the number of entities, $n$.

\subsection{Duplicate-Free Files, Equal Errors Across Files}
\label{sec:sim1}

In this simulation study we explore the performance of our methodology by varying the} number of erroneous fields per record over $\{1,2,3,5\}$, and also varying $\{p_{\bh}\}_{\bh\in\HH}$, which determines the amount of overlap, over four scenarios:
\begin{itemize}
    \itemsep0em 
    \item High Overlap: $p_{001}=p_{010}=p_{100}=0.4/3, p_{011}=p_{101}=p_{110}=0.15, p_{111}=0.15$,
    \item Medium Overlap: $p_{001}=p_{010}=p_{100}=0.7/3,
    p_{011}=p_{101}=p_{110}=0.05, p_{111}=0.15$,
    \item Low Overlap: 
    $p_{001}=p_{010}=p_{100}=0.8/3,
    p_{011}=p_{101}=p_{110}=0.05/3, p_{111}=0.15$,
    \item No-Three-File Overlap: $p_{001}=p_{010}=p_{100}=0.55/3,
    p_{011}=p_{101}=p_{110}=0.15, p_{111}=~0$.
\end{itemize}
These are intended to represent a range of scenarios that could occur in practice. In the high overlap scenario 60\% of the entities are expected to be in more than one datafile, in the low overlap scenario 80\% of the entities are expected to be represented in a single datafile, and in the no-three-file overlap scenario no entities are represented in all datafiles.

To implement our methodology, in addition to the general set-up described in Section \ref{sec:sim_setup}, we restrict the prior for the within-file cluster sizes so that they have size one with probability one, incorporating the assumption of no-duplication within files (see Section \ref{sec:prior_comments}). Imposing this restriction for all datafiles leads to a prior for tripartite matchings.
To illustrate the impact of using our structured prior, we compare with the results obtained using our model for comparison data with a flat prior on tripartite matchings.

\begin{figure}[h]
\centering
\includegraphics[width=0.9\linewidth]{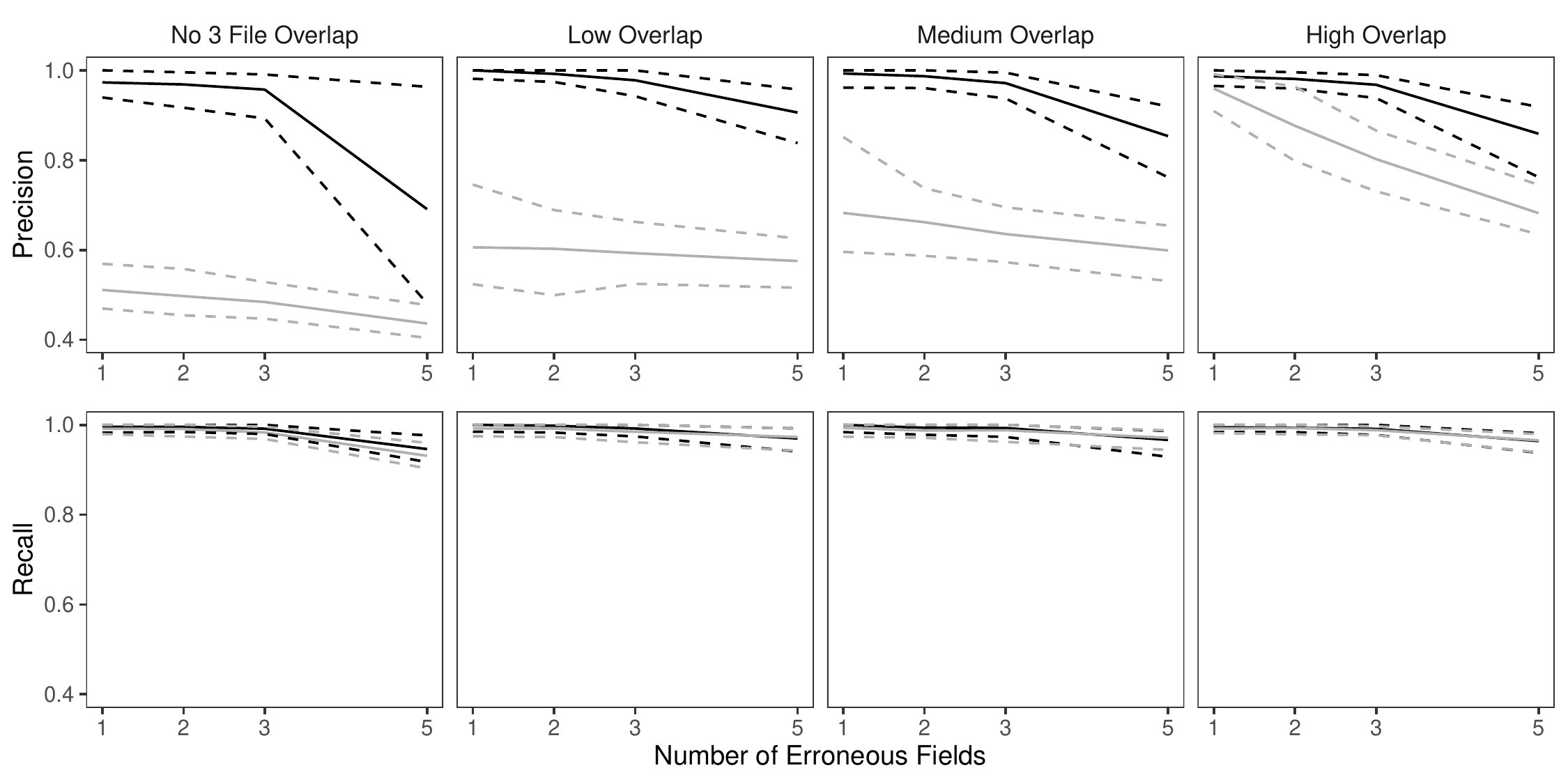}
\begin{minipage}{0.95\textwidth}
  \caption{Performance comparison for simulation with equal measurement error across files. Black lines refer to results under our structured prior, grey lines to results under the flat prior, solid lines show medians, and dashed lines show 2nd and 98th percentiles.} 
  \label{fig:no_dup_sim}
\end{minipage} 
\end{figure}

The results of the simulation are seen in Figure \ref{fig:no_dup_sim}. We see that our proposed approach performs consistently well across different settings, with the exception of the no-three-file overlap setting under high measurement error, where the precision decreases dramatically.  The approach using a flat prior on tripartite matchings has poor precision in comparison, and it is particularly low when the amount of overlap is low. This suggests that our structured prior improves upon the flat prior by protecting against over-matching (declaring noncoreferent record pairs as coreferent). \serge{In Appendix D we demonstrate how the performance in the no-three-file overlap setting can be improved through the incorporation of an informative prior for the overlap table through $\balpha$.}

\subsection{Duplicate-Free Files, Unequal Errors Across Files}
\label{sec:sim4}

In this simulation study we have different patterns of measurement error across the three files. Rather than each field in each record having a chance of being erroneous according to Table 3 of \cite{Sadinle_2014}, we will use the following measurement error mechanism to generate the data. For each record in the first file, age is missing, given name has up to seven errors, and all other fields are error free.
For each record in the second file, sex and occupation are missing, last name has up to seven errors, and all other fields are error free. For each record in the third file, phone number and postal code have up to seven errors and all other fields are error free. Under this measurement error mechanism, there is enough information in the error free fields to inform pairwise linkage of the files. We further vary $\{p_{\bh}\}_{\bh\in\HH}$ over the no-three-file and high overlap settings from Section \ref{sec:sim1}.

Our goal in this study is to demonstrate that having both the structured prior for partitions and the separate models for comparison data from each file-pair can lead to better performance than not having these components. We will compare our model as described in Section \ref{sec:sim1} to both our model for comparison data with a flat prior on tripartite matchings (as in Section \ref{sec:sim1}) and a simplification of our model for comparison data using a single model for all file-pairs but with our structured prior for partitions.

\begin{figure}[!h]
\centering
\includegraphics[width=0.95\linewidth]{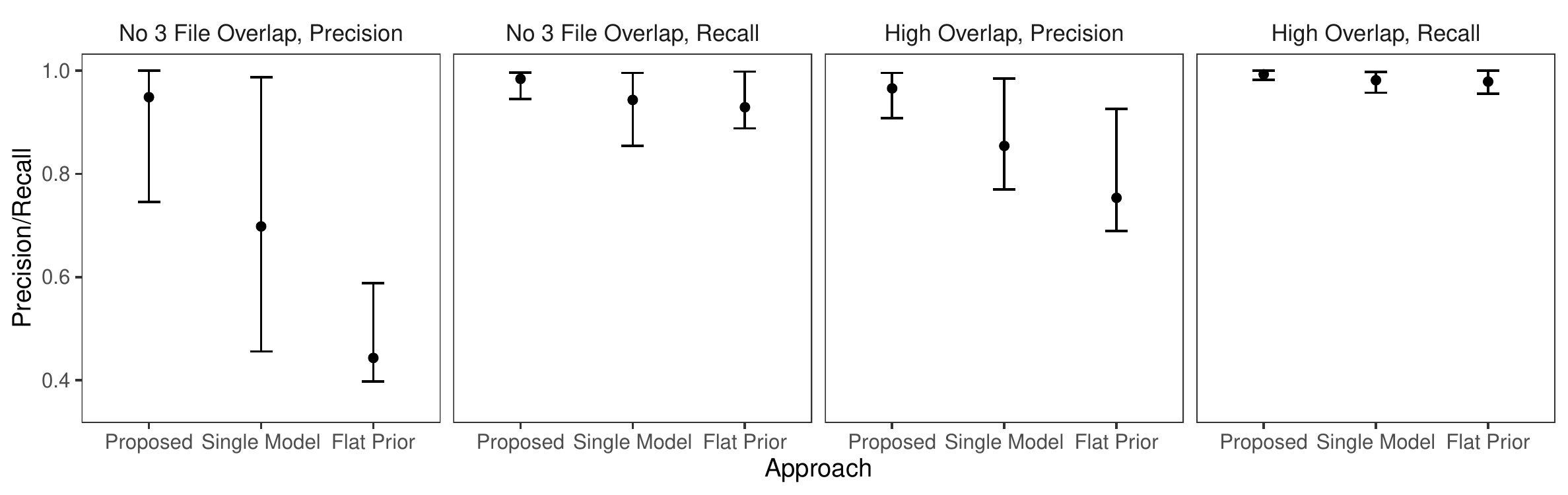}
\begin{minipage}[b]{0.975\textwidth}
    \caption{Performance comparison for simulation with unequal measurement error across files. ``Proposed" refers to our proposed approach, ``Single Model" refers to the approach using a single model for all file-pairs and our structured prior for partitions, and ``Flat Prior" refers to the approach using our model for comparison data with a flat prior on tripartite matchings. Dots show medians, and bars show 2nd and 98th percentiles.}
    \label{fig:no_dup_error_sim}
\end{minipage} 
\end{figure}

The results of the simulation are given in Figure \ref{fig:no_dup_error_sim}. We see that our proposed approach outperforms both alternative approaches in both precision and recall in both overlap settings. This suggests that both the structured prior for tripartite matchings and the separate models for comparison data from each file-pair can help improve performance over alternative approaches. We note that in the no-three-file (high) overlap setting the precision of the proposed approach is greater than or equal to the precision of the approach using a single model for all file-pairs in $98$ ($100$) of the $100$ replications. 

\section{Discussion and Future Work}
\label{sec:disc}

The methodology proposed in this article makes three contributions. First, the multifile partition parameterization, specific to the context of multifile record linkage and duplicate detection, allows for the construction of our structured prior for partitions, which provides a flexible mechanism for incorporating prior information about the data collection processes of the files. This prior is applicable to any Bayesian approach which requires a prior on partitions, including direct-modeling approaches such as \cite{Steorts_2016}.  We are not aware of any priors for $K$-partite matchings when $K>2$, so we hope our construction will lead to more development in this area. The second contribution is an extension of previous comparison-based models that explicitly handles the multifile setting.  Allowing separate models for comparison data from each file pair leads to higher quality linkage.  The third is a novel loss function for multifile partitions which can be used to derive Bayes estimates with good frequentist properties. Importantly, the loss function allows for linkage decisions to be left unresolved for records with large matching uncertainty. \serge{As with our structured prior on partitions, the loss function is applicable to any Bayesian approach which requires point estimates of partitions, including direct-modeling approaches.}

There are a number of directions for future work. One direction is the modeling of dependencies between the comparison fields \citep[see e.g.][]{Larsen_2001}, which should further improve the quality of the linkage. Another direction is the development of approaches to jointly link records and perform a downstream analysis, thereby propagating the uncertainty from the linkage. \serge{See Section 7.2 of \cite{Binette_2020} for a recent review of such joint models. In this direction, a natural task to consider next is population size estimation, where the linkage of the datafiles plays a central role \citep{Tancredi_2011, Tancredi_2020}.}

\newpage
\begin{center}
    \title{\Large \textbf{Supplementary Appendices for Multifile Partitioning for Record Linkage and Duplicate Detection}}
\end{center}
\appendix


\section{Structured Prior Appendix}
In this appendix, we prove Proposition 1 from the main text and provide additional guidance for the specification of the structured prior for multifile partitions.

\subsection{Proof of Proposition 1}
In this section, we restate and prove Proposition 1 from the main text.
\begin{prop2}
	The number of $K$-partite matchings that have the same overlap table, $\bn=\{n_{\bh}\}_{{\bh}\in\HH }$, is  $\prod_{k=1}^{K}n_k!/\prod_{{\bh}\in\HH }n_{\bh}!$ , where $n_k=\sum_{{\bh}\in\HH} {h}_k n_{\bh}$ is the number of entities in datafile ${\bX}_k$. 
	Thus $\PP(\C\mid \{\C_k\}_{k=1}^{K}, \bn)=
\prod_{{\bh}\in\HH }n_{\bh}!/\prod_{k=1}^{K}n_k!.$
\end{prop2}
\begin{proof}
	Let us first count all of the ways that we can place the clusters in file ${\bX}_k$ into the overlap table cells that ${\bX}_k$ is included in, $\HH _k=\{{\bh}\in\HH : {h}_k=1\}$.  This is just a multinomial coefficient, $n_k!/(\prod_{{\bh}\in\HH _k}n_{\bh}!)$.
	Thus the number of ways we can place all of the clusters from all of the files into the cells in $\HH $ is $\prod_{k=1}^{K}n_k!/(\prod_{{\bh}\in\HH _k}n_{\bh}!)=(\prod_{k=1}^{K}n_k!)/[\prod_{h\in\HH }(n_{\bh}!)^{\sum_{k=1}^{K}{h}_k}].$
	Given that there are $n_{\bh}$ clusters from each file with ${h}_k=1$ in cell ${\bh}$, now all we have to count is how many distinct complete matchings are possible between them, which is just $(n_{\bh}!)^{(\sum_{k=1}^{K}{h}_k)-1}$. Thus the number of $K$-partite matchings is $[(\prod_{k=1}^{K}n_k!)/(\prod_{{\bh}\in\HH }(n_{\bh}!)^{\sum_{k=1}^{K}{h}_k})][\prod_{{\bh}\in\HH }(n_{\bh}!)^{\sum_{k=1}^{K}{h}_k}/n_{\bh}!]=	\prod_{k=1}^{K}n_k!/\prod_{{\bh}\in\HH }n_{\bh}!.$
\end{proof}

\subsection{Prior Specification Guidance}
\label{ref:prior_guide}
In this section we provide additional guidance for the specification of the structured prior for multifile partitions described in Section 3 of the main text. In particular, we further discuss the priors for the number of clusters, the overlap tables, and the within-file cluster sizes.

\noindent \textit{Prior for the Number of Entities or Clusters}. In the main text we recommended, in the absence of substantial prior information, to use a uniform prior on $\{1,\cdots,U\}$ for the number of clusters, for some upper bound $U$. Our default recommendation was to set $U=r$, i.e. the total number of records. If one has substantive prior information about the number of clusters, this could instead be incorporated using  other distributions on the positive integers. \cite{Miller_2015} and \cite{Zanella_2016} both suggest to use a negative-binomial distribution with parameters $a>0$ and $q\in(0,1)$ truncated to the positive integers, i.e. 
$\PP(n)\propto \frac{\Gamma(n + a)}{(n!)\Gamma(a)}(1-q)^aq^nI(n\in\N)$. \cite{Zanella_2016} further suggest a weakly informative specification for this Negative-binomial prior where $a$ and $q$ are selected such that $\text{E}[n]=\sqrt{\text{Var}(n)}=r/2$. We follow \cite{Miller_2015} and \cite{Zanella_2016} and suggest a negative-binomial prior for the number of clusters $n$ when incorporating substantive prior information.

\noindent \textit{Prior for the Overlap Table}. In the main text we recommended using a Dirichlet-multinomial prior for the overlap table, specified by a collection of positive hyperparameters. In the absence of substantial prior information we recommended setting $\balpha=(1,\cdots,1)$. Due to conjugacy of the Dirichlet distribution with the multinomial, $\balpha$ can be interpreted as prior cell counts, and thus our recommendation amounts to incorporating a prior count of $1$ to each cell and an overall prior sample size of $2^K-1$.

In our simulations and application, we found across a variety of overlap settings that this default prior performed satisfactorily. However, in the no-three-file-overlap setting, our approach struggled when there was a large amount of measurement error. We find in Appendix D.2 that we can improve performance in this setting by using informative prior cell counts, rather than using our default specification.
What sets this no-three-file-overlap simulation setting apart from the other settings is that it is sparse, i.e. in this setting the count for the three-file-overlap cell of the overlap table is truly $0$. 

When linking a large number of files $K$, the size of the overlap table, $2^K-1$, becomes large very quickly, which makes it likely that the true overlap table is sparse. Our default prior specification may not be appropriate in these settings as using a prior cell count of $1$ for each cell may be incorporating prior information that is too strong, as illustrated in Example 1.4 of \cite{Berger_2015}. One possible alternative as a default specification when the overlap table is potentially sparse, would be to set $\balpha=(1/(2^K-1),\cdots, 1/(2^K-1))$ (see Section 3.2 of \cite{Berger_2015} for justification). If one has prior information concerning which cells of the overlap table are likely to be sparse, based on the results in Appendix D.2, we recommend attempting to incorporate this information into the prior. For example, if it is believed that some combination of files are likely not to have collected information on the same set of entities, one can incorporate this information by making the corresponding prior cell counts close to $0$.

Another route one could take would be to replace the Dirichlet-multinomial prior on $\bn$ with a multinomial prior on $\bn$, with a non-Dirichlet prior on the multinomial cell probabilities. For example, one could use the tensor-factorization priors of \cite{Dunson_2009} and \cite{Zhou_2015}, which have been shown to have lead to estimates of cell probabilities with good performance in large sparse contingency tables.

\noindent \textit{Prior for the Within-File Cluster Sizes}. Given that in a joint record linkage and duplicate detection scenario we do not expect there to be many duplicates per entity in any given file, in the main text we recommended specifying i.i.d. priors for the sizes of the within-file clusters, i.e. $d^k_1,\cdots, d^k_{n_k}\mid n_k\stackrel{iid}{\sim} p_k(\cdot)$ for a given file $\bX_k$. Here $p_k(\cdot)$ represents the probability mass function of a distribution on $\{1,\cdots, U^k\}$, where $U^k$ is a file-specific upper bound on cluster sizes.

When a given file $\bX_k$ is assumed to have no duplicates, in the main text we recommended enforcing this restriction that there are no duplicates in that file by setting $U^k=1$ and $p_k(d^k_i)=I(d^k_i=1)$. When a given file $\bX_k$ is assumed to have  duplicates, in the main text we recommended a Poisson distribution with parameter $\lambda=1$, i.e. $p_k(d^k_i)\propto (d^k_i!)^{-1}I(d^k_i\in\{1,\cdots, U^k\})$. This prior places most of the prior mass close to $1$, where various properties such as prior mean and standard deviation can be computed numerically (e.g. when $U^k=10$ the prior mean is $1.58$). If one has information on the average amount of duplication in file $k$, given a specified upper bound $U^k$, one could specify the parameter $\lambda$ of the Poisson prior such that the prior mean is equal to the average amount of duplication. Alternatively, one could place a hyperprior on $\lambda$. This is similar approach to \cite{Zanella_2016},  who used a negative-binomial distribution with parameters $r>0$ and $p\in(0,1)$ truncated to the positive integers, with a gamma hyperprior for $r$ and a beta hyperprior for $p$. Indeed, the negative-binomial prior of \cite{Zanella_2016} can be seen as a generalization of our Poisson prior, based on well-known connections between the Poisson and negative-binomial, and could be used instead of our Poisson recommendation if desired.

In the simulations presented in Appendix D.3 we explore using a Poisson prior with parameter $\lambda$ varying over $\{0.1,1,2\}$, when the within-file cluster sizes are generated from a Poisson with parameter $\lambda$ varying over $\{0.1,1,2\}$. We find in these simulations that when there is medium or high duplication (i.e. the within-file cluster sizes are generated from a Poisson with mean in $\{1,2\}$), the results are not sensitive to $\lambda$, whereas when there is low duplication (i.e. the within-file cluster sizes are generated from a Poisson with mean $0.1$), the results are sensitive to $\lambda$. This suggests that model performance is more sensitive to the specification of the prior distribution for within-file cluster sizes when there is a low amount of duplication, and that care should be taken when specifying the prior for the within-file cluster sizes for files which are expected to have very little duplication.


\section{Posterior Inference Appendix}
In this appendix, we first derive full conditional distributions of our structured prior for partitions, and then use the full conditional distributions to derive a Gibbs sampler for posterior inference in our model. We then discuss the computational complexity of our approach to posterior inference, how computational performance can be improved through the usage of indexing techniques, and the initialization of the Gibbs sampler.

\subsection{Conditional Assignment Probabilities}
\label{sec:fullc}
In this section we use the form of the prior distribution in 
Equation (1) of the main text
to derive the conditional probability for assigning a record $j$ from file $\bX_k$ to a given cluster of an existing multifile partition $\C_{-j}$ of the other records. 
Specifically, we derive $\PP(j\to c\mid \C_{-j})$, where $j\to c$ denotes adding record $j$ to a cluster $c\in\C_{-j}$ or to an empty cluster. 
Let a quantity followed by $(\C_{-j})$ denote that it is derived from $\C_{-j}$ analogously to in Section 
3.1 of the main text.
Let $\boldsymbol{1}_{k}$ denote the inclusion pattern indicating inclusion only in file ${\bX}_k$, that is, $\boldsymbol{1}_{k}$ is a vector of zeroes except for its $k$th entry which equals 1. Further, let $\boldsymbol{1}_{c}$ denote the  inclusion pattern of the cluster $c\in \C_{-j}$, that is, the $l$th entry of $\boldsymbol{1}_{c}$ is $I(c^{l}\neq \emptyset)$.  Finally, let $\boldsymbol{1}_{c\cup j}$ denote the inclusion pattern of the cluster $c\in \C_{-j}$ after adding record $j$ to it.
Then the conditional assignment probability is

\begin{equation*}
\begin{array}{l}
\PP(j\to c\mid \C_{-j})\propto \\
\left\{
\begin{array}{ll}
\left[ \dfrac{\PP(n(\C_{-j})+1)}{\PP(n(\C_{-j}))} \right] 
\left[ \dfrac{(n(\C_{-j})+1 )(n_{\boldsymbol{1}_{k}}(\C_{-j}) + \alpha_{\boldsymbol{1}_{k}})}{n(\C_{-j})+\alpha_0} \right] 
p_k(1)
&,\ \text{if $c=(\emptyset,\cdots,\emptyset)$} \\ 
\left[ \dfrac{n_{\boldsymbol{1}_{c\cup j}}(\C_{-j})+\alpha_{\boldsymbol{1}_{c\cup j}}}{n_{\boldsymbol{1}_{c}}(\C_{-j})+\alpha_{\boldsymbol{1}_{c}}-1} \right] 
p_k(1) 
&,\ \text{if $c\neq(\emptyset,\cdots,\emptyset), |c^k|=0$} \\ 
(|c^k|+1)\left[\dfrac{p_k(|c^k|+1)}{p_k(|c^k|)} \right]
&,\ \text{if $|c^k|>0$} .\\ 
\end{array}
\right.
\end{array}
\end{equation*} 

\subsection{Gibbs Sampler}
\label{sec:gibbs}
We will now derive a Gibbs sampler to explore the posterior of $\Phi$ and $\C$. Suppose we are at iteration $t+1$ of the sampler, with current samples $\Phi^{[t]}=(\bem^{[t]},\bu^{[t]})$ 
and $\C^{[t]}$. Then we obtain the samples for iteration $t+1$ through the following steps:
\begin{enumerate}
	\item For $k\leq k'$ and $f\in \{1,\dots,F\}$, sample
	\begin{equation*}
	\bem_{kk'}^{f[t+1]}\mid \ \C^{[t]}, \bgamma^{obs} \sim \textsf{Dirichlet}(a_{kk'}^{f0}(\C^{[t]})+\mu_{kk'}^{f0},\cdots, a_{kk'}^{fL_f}(\C^{[t]})+\mu_{kk'}^{fL_f})
	\end{equation*}
	and 
	\begin{equation*}
	\bu_{kk'}^{f[t+1]}\mid \ \C^{[t]}, \bgamma^{obs} \sim \textsf{Dirichlet}(b_{kk'}^{f0}(\C^{[t]})+\nu_{kk'}^{f0},\cdots, b_{kk'}^{fL_f}(\C^{[t]})+\nu_{kk'}^{fL_f}).
	\end{equation*}
	Call these samples $\Phi^{[t+1]}$.
	
	\item We now sample the cluster assignment for each record $j\in[r]$ sequentially. Suppose we have sampled the first $j-1$ records, and are sampling the cluster assignment for record $j$ from file ${\bX}_k$. Let $\C^{[t]}_{-j}$ denote the current partition of $[r]$, without record $j$, after sampling the first $j-1$ records. Then we sample the cluster assignment for record $j$ according to the following probabilities:
	
	\begin{equation*}
	\begin{array}{l}
	\PP(j\to c\mid \C^{[t]}_{-j}, \Phi^{[t+1]}, \bgamma^{obs})\propto\\
	\left\{
	\begin{array}{ll}
	\left[ \dfrac{\PP(n(\C^{[t]}_{-j})+1)}{\PP(n(\C^{[t]}_{-j}))} \right]
	\left[ \dfrac{(n(\C^{[t]}_{-j})+1)(n_{\boldsymbol{1}_{k}}(\C^{[t]}_{-j}) + \alpha_{\boldsymbol{1}_{k}})}{n(\C^{[t]}_{-j})+\alpha_0} \right]
	p_k(1)
	&,\ \text{if $c=(\emptyset,\cdots,\emptyset)$} \\ 
	\left[\prod_{k'=1}^K\prod_{i\in c^{k'}}\mathcal{L}_{ij}^{[t+1]}\right]
	\left[ \dfrac{n_{\boldsymbol{1}_{c\cup j}}(\C^{[t]}_{-j})+\alpha_{\boldsymbol{1}_{c\cup j}}}{n_{\boldsymbol{1}_{c}}(\C^{[t]}_{-j})+\alpha_{\boldsymbol{1}_{c}}-1} \right]
	p_k(1)
	&,\ \text{if $|c^k|=0, c\neq(\emptyset,\cdots,\emptyset)$} \\ 
	\left[\prod_{k'=1}^K\prod_{i\in c^{k'}}\mathcal{L}_{ij}^{[t+1]}\right](|c^k|+1)
	\left[\dfrac{p_k(|c^k|+1)}{p_k(|c^k|)} \right]
	&,\ \text{if $|c^k|>0$} ,\\ 
	\end{array}
	\right.
	\end{array}
	\end{equation*}
	where, letting $k'$ denote the file that record $i$ is in,
	\begin{align*}
	\mathcal{L}_{ij}^{[t+1]}&= \prod_{f=1}^F\left[\prod_{l=0}^{L_f} \left(\dfrac{m_{kk'}^{fl[t+1]}}{u_{kk'}^{fl[t+1]}}\right)^{I(\gamma_{ij}^f=l)} \right]^{I_{obs}(\gamma_{ij}^f)} 
	\nonumber\\
	&=\exp\left[\sum_{f=1}^F I_{obs}(\gamma_{ij}^f) \sum_{l=0}^{L_f}\log\left(\dfrac{m_{kk'}^{fl[t+1]}}{u_{kk'}^{fl[t+1]}}\right)I(\gamma_{ij}^f=l)\right].
	\end{align*}
\end{enumerate}

\subsection{Computational Complexity}
The computational complexity of posterior inference in our proposed approach can be broken up into the complexity of pre-computing comparison vectors, and the complexity of individual steps of the Gibbs sampler presented in Appendix \ref{sec:gibbs}.
\begin{itemize}
    \item The computational complexity of pre-computing comparison vectors is $\mathcal{O}(\texttt{rp}*F)$, where $\texttt{rp}$ is the number of valid record pairs. To be more specific, when we assume there are duplicates in every file, $\texttt{rp}=r(r-1)/2$, and when we assume there are no duplicates in each file, $\texttt{rp}=\sum_{k<k'}r_k r_{k'}$. For in between situations where we assume there are no duplicates in some files and duplicates in the remaining files, it can be shown that $\sum_{k<k'}r_k r_{k'}<\texttt{rp}<r(r-1)/2$. Thus in the most general case, pre-computing comparison vectors scales quadratically in the number of records. We discuss in the following section how the cost of this step can be reduced through the usage of blocking.
    \item The computational complexity of step 1 of the Gibbs sampler presented in Appendix \ref{sec:gibbs}, i.e. sampling the $\bem$ and $\bu$ parameters, is $\mathcal{O}(\texttt{rp}*\texttt{fl} + \texttt{fp}*\texttt{fl})$, where $\texttt{fp}$ is the number of valid file pairs, and $\texttt{fl}=\sum_{f=1}^F(L_f + 1)$ is the total number of agreement levels across all fields. We have that $\texttt{fp}=\binom{K}{2}+K_d$, where $K_d$ is the number of files that are assumed to have duplicates. This follows as the $a$ and $b$ summaries of the partition can be calculated from a matrix multiplication of a $\texttt{fl}\times\texttt{rp}$ matrix and a $\texttt{rp}\times1$ matrix, and given these $a$ and $b$ summaries the complexity of sampling the $\bem$ and $\bu$ parameters from their full conditionals is $\mathcal{O}( \texttt{fp}*\texttt{fl})$.
    \item The computational complexity of step 2 of the Gibbs sampler presented in Appendix \ref{sec:gibbs}, i.e. sampling the partition $\C$, is difficult to analyze in general. In the most general case, where we assume there are duplicates in each file, in the worst case scenario, each record could be placed in its own cluster. The complexity of sampling the cluster assignment for a single record would then be $\mathcal{O}(r)$, and the complexity of sampling the cluster assignment for all records would then be $\mathcal{O}(r^2)$. However, the number of clusters potentially changes whenever a new cluster assignment is sampled, which complicates this analysis. Further, once introduces constraints on the partition space, either through assuming there are no duplicates in some files, or using indexing as described in the next section, the number of clusters available for a specific record's cluster assignment step will depend on these constraints. In general the best we can say is that this step will be faster when each record has on average (with respect to the posterior) a small number of clusters to which it can be assigned, and slower when each record has on average a large number of clusters to which it can be assigned.
\end{itemize}

In our current implementation of the proposed approach, we have found that even though both steps of the Gibbs sampler scale quadratically in the number of records in the worst case, the cost of sampling the partition generally dominates the cost of sampling the $\bem$ and $\bu$ parameters, and is the main bottleneck of our approach. We note that the sampling of the partition will essentially have the same computational complexity regardless of whether one uses a comparison-based model for records, as we have proposed in the main text, or one uses a direct-modeling approach, as in \cite{Steorts_2016}.

\subsection{Blocking, Indexing, and Scalability}
\label{sec:indexing}
As described in the previous section, there are two main bottlenecks to scalability in our proposed approach: pre-computing the comparison vectors and sampling the partition from its full conditional in the Gibbs sampler presented in Appendix \ref{sec:gibbs}. Both of these bottlenecks can be sped up through the use of indexing techniques, which declare certain pairs of records non-coreferent a priori based on comparisons of a small number of fields \citep{Christen_2012, Steorts_2014, Murray_2015}. This both reduces the number of record pairs under consideration, and reduces on average the number of clusters to which each record can be assigned in the Gibbs sampler presented in Appendix \ref{sec:gibbs}.

In particular, if $\pairP=\cup_{k\leq k'}\pairP_{kk'}$ is the set of all possible record pairs, indexing techniques generate a set $\pairP^*\subset \pairP$, such that $|\pairP^*|\ll |\pairP|$, where record pairs in $\pairP^*$ are candidate coreferent pairs, and record pairs in $\pairP\setminus \pairP^*$ are fixed as non-coreferent. Thus when performing posterior inference, this truncates our prior on multifile partitions to the set $\{\C : \C(i)\neq\C(j), \forall (i,j)\in\pairP\setminus \pairP^*\}$. 

We briefly review two common indexing techniques from \cite{Murray_2015}, blocking and indexing by disjunction. Blocking declares pairs of records to be non-coreferent when they disagree on a set of error-free fields.
The use of error-free fields guarantees that the candidate coreferent pairs output from blocking are transitive, so that $\pairP^*$ forms a partition of the records. Indexing by disjunction declares pairs of records to be non-coreferent when they disagree at a certain threshold for each field in a given set of reliable fields. Candidate coreferent pairs output from indexing by disjunction are not guaranteed to be transitive.

When a set of error-free fields are available, we recommend blocking.  Blocking schemes can be implemented without constructing comparison vectors for each record pair, thus reducing the cost of pre-computing comparison vectors for all record pairs to just the cost of pre-computing comparison vectors for record pairs within each block. Our proposed approach can then be run independently in each block, drastically reducing on average the number of clusters to which each record can be assigned in the Gibbs sampler presented in Appendix \ref{sec:gibbs}.

Within blocks, there is no further way to reduce cost of the pre-computing the comparison vectors. However, it is still possible to reduce the cost of sampling the partition from its full conditional in the Gibbs sampler presented in Appendix \ref{sec:gibbs} through the use of indexing by disjunction. By fixing certain pairs of records to be non-coreferent, one reduces on average the number of clusters to which each record can be assigned in the Gibbs sampler presented in Appendix \ref{sec:gibbs}. Note that the comparisons for record pairs fixed as non-coreferent in $\pairP\setminus \pairP^*$ still contribute to the model through the $b_{kk'}^{fl}$ term in the likelihood in 
Equation (2) of the main text, avoiding many of the issues presented in \cite{Murray_2015}.

However, the non-transitivity of $\mathcal{P}^*$ output from indexing by disjunction can be problematic, as it suggests that the thresholds used in indexing by disjunction are too stringent, and that they may be excluding true coreferent pairs. 
Non-transitivity can also cause problems for Markov chain Monte Carlo samplers (like our Gibbs sampler in Appendix \ref{sec:gibbs}), as it can make traversing the constrained space of multifile partitions difficult. 
To avoid the issue of non-transitivity,  we propose to use the transitive closure of the candidate coreferent pairs, $\mathcal{P}^*$, generated by indexing by disjunction, which we refer to as \textit{transitive indexing}. Transitive indexing has been used before in the post-hoc blocking methodology of \cite{McVeigh_2019} for two-file record linkage. 

\subsection{Initialization}
Due to the nature of the Gibbs sampler in Appendix \ref{sec:gibbs}, we can initialize the multifile partition $\C$ without needing to initialize $\Phi$. A simple initialization for $\C$ is to let each record belong to its own cluster, which works well when indexing is used. However, we observed during some preliminary simulations that when sampling $K$-partite matchings without using indexing, the sampler can take a large number of iterations to mix if we initialize $\C$ in this way, where the number of iterations depends on the partition the data was simulated from. This problem can not be avoided as in Appendix \ref{sec:indexing}, as the constraints on the space of $K$-partite matchings cannot be relaxed.
In this case, we constructed a simple alternative initialization.  The idea is to use an indexing scheme for initialization, even if indexing is not being used to reduce the number of candidate coreferent pairs. In particular, we first generate a set of record pairs $\mathcal{P}^*\subset\mathcal{P}$ through transitive indexing as described in Appendix \ref{sec:indexing}. For each block of records in $\mathcal{P}^*$, we sample a random $K$-way matching of records in that block. We then initialize $\C$ such that each record belongs to its own cluster, except for the sampled $K$-way matchings.


\section{Point Estimation Appendix}
In this appendix we discuss how our proposed loss function differs from the loss function of \cite{Sadinle_2017}, and propose a strategy for approximating the Bayes estimate under our proposed loss function.

\subsection{Comparison to \texorpdfstring{\cite{Sadinle_2017}}{Sadinle (2017}}
Unlike our loss function construction, in the two-file set-up \cite{Sadinle_2017} constructed the loss function from individual losses for the records in the smaller datafile only.
Such construction however leads to an asymmetry in the loss function that is arbitrary.  Consider an example of two datafiles, where the first file has records $a$ and $b$, and the second file has records $c$ and $d$.  In that case the role of the datafiles can be arbitrarily interchanged.  If the true matching has a link between $a$ and $c$ but the matching estimate has a link between $b$ and $c$, the loss will be $\lfnm+\lfma$ if file two is chosen not to contribute to the loss, but it will be $\lfmb$ if file one is chosen not to contribute to the loss.  Our new construction presented in the main text does not lead to such issues.  

\subsection{Approximating the Bayes Estimate}
\label{sec:derive}
Finding a partition $\hat\bZ$ such that $\mathbb{R}(\hat\bZ)$ is minimized corresponds to an optimization problem closely related to graph partitioning problems \citep[e.g.,][]{Brandes_2007, Lancichinetti_2009, Newman_2013} or correlation clustering \citep[e.g.,][]{Bansal_2004,Demaine_2006}, both of which are known to be NP-complete. 
Thus, unlike in \cite{Sadinle_2017}, we cannot minimize $\mathbb{R}(\hat\bZ)$ exactly in general. All approaches for graph partitioning problems or for correlation clustering instead rely on heuristic algorithms whose performance is evaluated empirically via simulation studies and benchmark datasets. We will follow a similar approach.

We take advantage of the fact that in practice a large number of record pairs will have zero or close to zero posterior probability of matching $\PP(\Delta_{ij}=1\mid \bgamma^{obs})$. 
Based on this, we propose to threshold $\PP(\Delta_{ij}=1\mid \bgamma^{obs})$ at a small value $\delta$ to create a graph where an edge represents a non-negligible probability of matching between two records.  We then break the records up into connected components of this graph, each component representing groups of records that are more likely to be coreferent.  We then find the Bayes estimate by minimizing $\mathbb{R}(\hat{\bZ})$ separately within each of these connected components.  We can think of $\delta$ as a way of trading-off between accuracy of the Bayes estimate and computational tractability: larger values of $\delta$ decrease the size of the resulting connected components, making the minimization more tractable within each component, while smaller $\delta$ make the resulting approximation more accurate as using the threshold $\delta=0$ is no longer an approximation.  We recommend setting $\delta$ as the smallest probability such that the largest connected component is smaller than some pre-specified upper bound that captures a computational budget. To minimize $\mathbb{R}(\hat{\bZ})$ within the connected components, we propose  
to do so over posterior samples, $\{\bZ^{[t]}\}_{t=1}^T$, and find the sample which minimizes $\mathbb{R}(\bZ^{[t]})$. As this minimization is happening separately within each connected component, the final Bayes estimate of the partition of all $r$ records does not itself have to be a posterior sample. 

To minimize $\mathbb{R}(\hat{\bZ})$ when searching for partial estimates, let $\Omega$ denote the power set of $[r]$, and let $\bZ^{[t]}_{\omega}$ denote the posterior draw $\bZ^{[t]}$ where the records in $\omega\in\Omega$ are set to abstain, $A$. Then in order to accommodate partial estimates, we can minimize $\mathbb{R}(\hat{\bZ})$ over $\{\bZ^{[t]}_{\omega}\mid t\in[T], \omega\in\Omega\}$. 

Unless stated otherwise, in all simulations and in the application, for full (partial) estimates, we find the Bayes estimate separately within connected components of records with posterior probability larger than $\delta$ of matching, where $\delta$ is the smallest probability such that the largest connected component is smaller than $50$ ($12$).


\section{Simulation Appendix}

\subsection{Tables 3 and 4 from \texorpdfstring{\cite{Sadinle_2014}}{Sadinle (2014}}

Table 3 of \cite{Sadinle_2014} is reproduced in Table \ref{tab:error_types}. In this table, edit errors are insertions, deletions, or substitutions of characters in a string, OCR errors are optical character recognition errors, keyboard errors are  typing errors that rely on a certain keyboard layout, and phonetic errors  are errors using a list of predefined phonetic rules. Table 4 of \cite{Sadinle_2014} is reproduced in Table \ref{tab:compdata}.

\begin{table}[h]
\caption{Types of errors per field in the simulation studies.}
\label{tab:error_types}
\begin{tabular*}{\textwidth}{@{\extracolsep{\fill}}lcccccc@{}}
\cline{1-7}
& \multicolumn{6}{c@{}}{\textbf{Type of error}}\\
\cline{2-7}
\textbf{Field} & \textbf{Missing values} & \textbf{Edits} & \textbf{OCR} &
\textbf{Keyboard} & \textbf{Phonetic} & \textbf{Misspelling} \\
\cline{1-7}
Given name & & $\checkmark$ & $\checkmark$ & $\checkmark$ & $\checkmark $ & \\
Family name & & $\checkmark$ & $\checkmark$ & $\checkmark$ &
$\checkmark$ & $\checkmark$ \\
Age interval & $\checkmark$ & &&&& \\
Sex & $\checkmark$ & &&&& \\
Occupation & $\checkmark$ & &&&& \\
Phone number & $\checkmark$ & $\checkmark$ & $\checkmark$ & $\checkmark $ && \\
Postal code & $\checkmark$ & $\checkmark$ & $\checkmark$ & $\checkmark $ && \\
\cline{1-7}
\end{tabular*}
\end{table}

\begin{table}[!h]
\caption{Construction of levels of disagreement for the simulation studies.} \label{tab:compdata}
\begin{tabular*}{\textwidth}{@{\extracolsep{\fill}}lccccc@{}}
\hline
& & \multicolumn{4}{c@{}}{\textbf{Levels of disagreement}}\\
\cline{3-6}
\textbf{Field} & \textbf{Similarity measure} & $\bm{0}$ & $\bm{1}$ &
$\bm{2}$ & \multicolumn{1}{c@{}}{$\bm{3}$} \\
\hline
Given name & Levenshtein & 0 & $(0,0.25]$ & $(0.25,0.5]$ & $(0.5,1]$ \\
Family name & Levenshtein & 0 & $(0,0.25]$ & $(0.25,0.5]$ & $(0.5,1]$
\\
Age interval & Binary comparison & Agree & Disagree &&\\
Sex & Binary comparison & Agree & Disagree &&\\
Occupation & Binary comparison & Agree & Disagree &&\\
Phone number & Levenshtein & 0 & $(0,0.25]$ & $(0.25,0.5]$ & $(0.5,1]$
\\
Postal code & Levenshtein & 0 & $(0,0.25]$ & $(0.25,0.5]$ & $(0.5,1]$
\\
\hline
\end{tabular*}
\end{table}

\subsection{Prior Sensitivity Analysis for Simulation with Duplicate-Free Files, Equal Errors Across Files}
\label{sec:sim_1_sens}
In Section 6.2 of the main text,
we saw that our proposed approach struggled in the no-three-file overlap setting when there was high measurement error. In practice, if we knew that no entity is represented in all three datafiles we could enforce that restriction just like we enforce that there are no duplicates in given files, which would likely lead to better performance. While it is reasonable to assume in some applications that there are no duplicates in a given file (for example the application considered in Section 7 of the main text),
it is less reasonable to assume with absolute certainty that there is no entity represented in all three datafiles. Thus we want to instead incorporate the weaker information that there is a low amount of three way overlap. We can achieve this through an informative specification of $\balpha$.

In the no-three-file overlap setting, the overlap table was generated from a multinomial distribution with probability vector $\bp=(p_{001}, p_{010}, p_{011}, p_{100}, p_{101}, p_{110}, p_{111})$ where $p_{001}=p_{010}=p_{100}=0.55/3, p_{011}=p_{101}=p_{110}=0.15, p_{111}=0$. Note that our Dirichlet-multinomial prior can be motivated as the result of first drawing $\{q_{\bh}\}_{{\bh}\in\HH }$ from a Dirichlet distribution with hyperparameters $\balpha$, then drawing $\bn$ from a multinomial distribution of size $n$ with probabilities $\{q_{\bh}\}_{{\bh}\in\HH }$. As discussed in Appendix \ref{ref:prior_guide}, $\balpha$ can be interpreted as prior cell counts, which can be used to incorporate prior information about the amount of overlap between datafiles. Thus when specifying an informative $\balpha$, we want the Dirichlet prior for $\{q_{\bh}\}_{{\bh}\in\HH }$ to be centered roughly around $\bp$. We can accomplish this by setting $\balpha=\kappa\times(p_{001}, p_{010}, p_{011}, p_{100}, p_{101}, p_{110}, 1/\kappa)$. $\kappa+1$ represents the sum of prior cell counts. As $\kappa$ increases, the Dirichlet prior for $\{q_{\bh}\}_{{\bh}\in\HH }$ becomes more concentrated near $\bp$.

\begin{figure}[!h]
\centering
\includegraphics[width=0.65\linewidth]{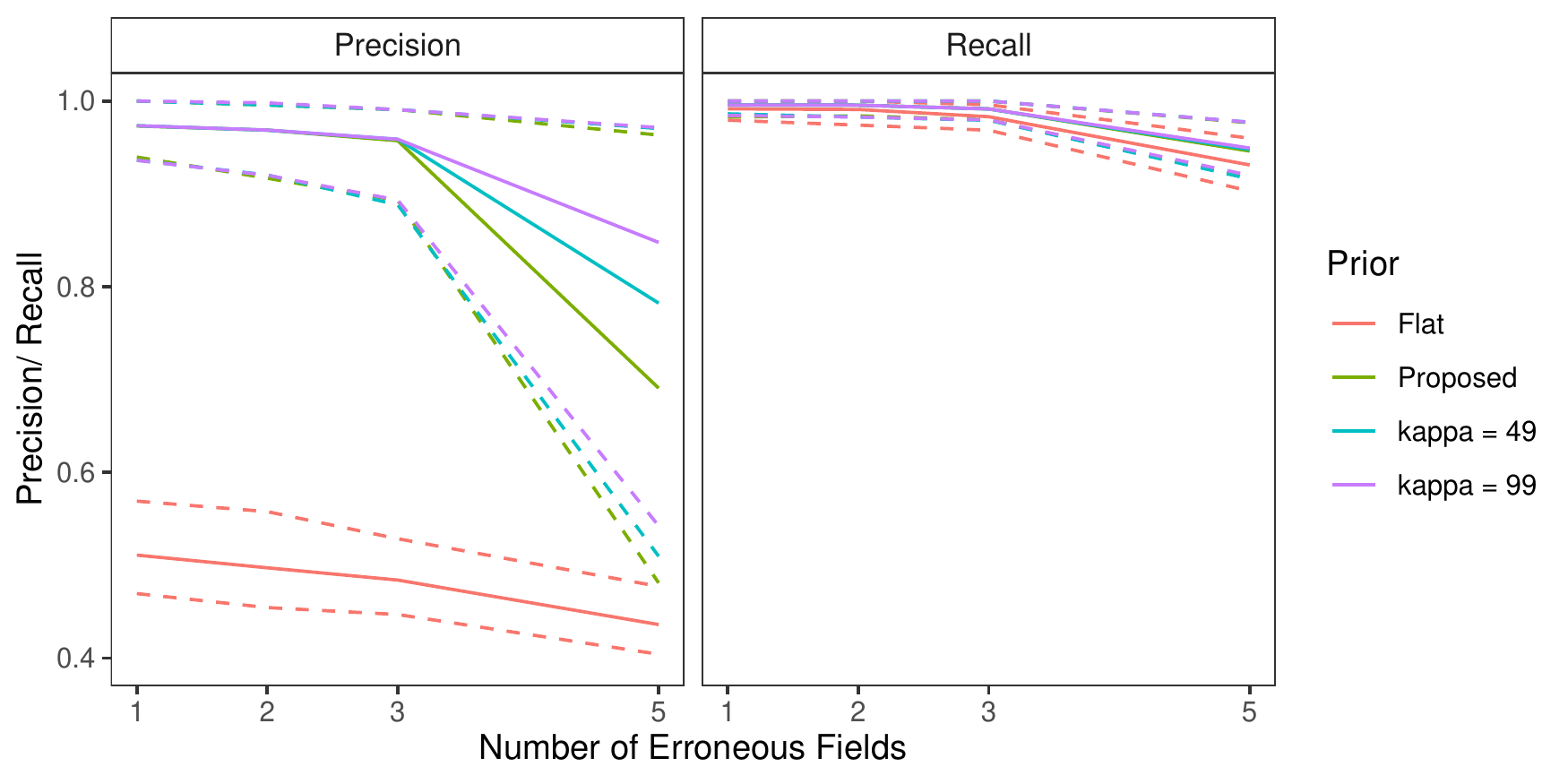}
\begin{minipage}[b]{1\textwidth}
    \caption{Performance comparison for no-three-file overlap simulation with more informative settings of $\balpha$. Solid lines show medians, and dashed lines show 2nd and 98th percentiles. ``Flat" refers to a flat prior on tripartite matchings, ``Proposed" refers to our structured prior for partitions when $\balpha=(1,\cdots, 1)$, and ``kappa = 49" and ``kappa = 99" refer to the more informative specifications of $\balpha$ with $\kappa \in\{49, 99\}$.}
    \label{fig:no_dup_sim_sens}
\end{minipage} 
\end{figure}

We repeated the no-three-file overlap setting simulation from Section 6.2 of the main text
using this informative specification with $\kappa \in\{49, 99\}$. The results are presented in Figure \ref{fig:no_dup_sim_sens}. We see that when there is high measurement error, these more informative specifications improve upon the performance of the default specification of $\balpha=(1,\cdots, 1)$.

\subsection{Files with Duplicates, Full Estimates}
\label{sec:sim2}
This simulation study consists of linkage and duplicate detection for three datafiles, so that the target of inference is a general multifile partition.  We conduct this study with probabilities fixed at $p_{001}=p_{010}=p_{100}=0.3, p_{011}=p_{101}=p_{110}=0.025, p_{111}=0.025$, representing a very low overlap setting, which can be challenging as seen in Section 6.2 of the main text.  For each entity represented in the datafiles we generated a within-file cluster size from a Poisson distribution with mean $\lambda$ truncated to $\{1,\cdots, 5\}$.  In this study, in addition to varying the number of erroneous fields per record over $\{1,2,3,5\}$ to explore different amounts of measurement error, and we vary $\lambda$ over $\{0.1, 1, 2\}$ to explore low, medium, and high amounts of duplication.

To implement our methodology, in addition to the general set-up described in Section 6.1 of the main text,
we use a Poisson prior with mean $1$ truncated to $\{1,\cdots, 10\}$ on the within-file cluster sizes. For comparison we use the comparison-based model of \cite{Sadinle_2014}, which treats all of the records as coming from one file and uses a flat prior on partitions. For both models we use transitive indexing as in described Appendix \ref{sec:indexing} to reduce the number of comparisons, where the initial indexing scheme declares record pairs as non-coreferent if they disagree in either given or family name at the highest level (according to Table 4 of \cite{Sadinle_2014}).

\begin{figure}[!t]
\centering
\includegraphics[width=0.9\linewidth]{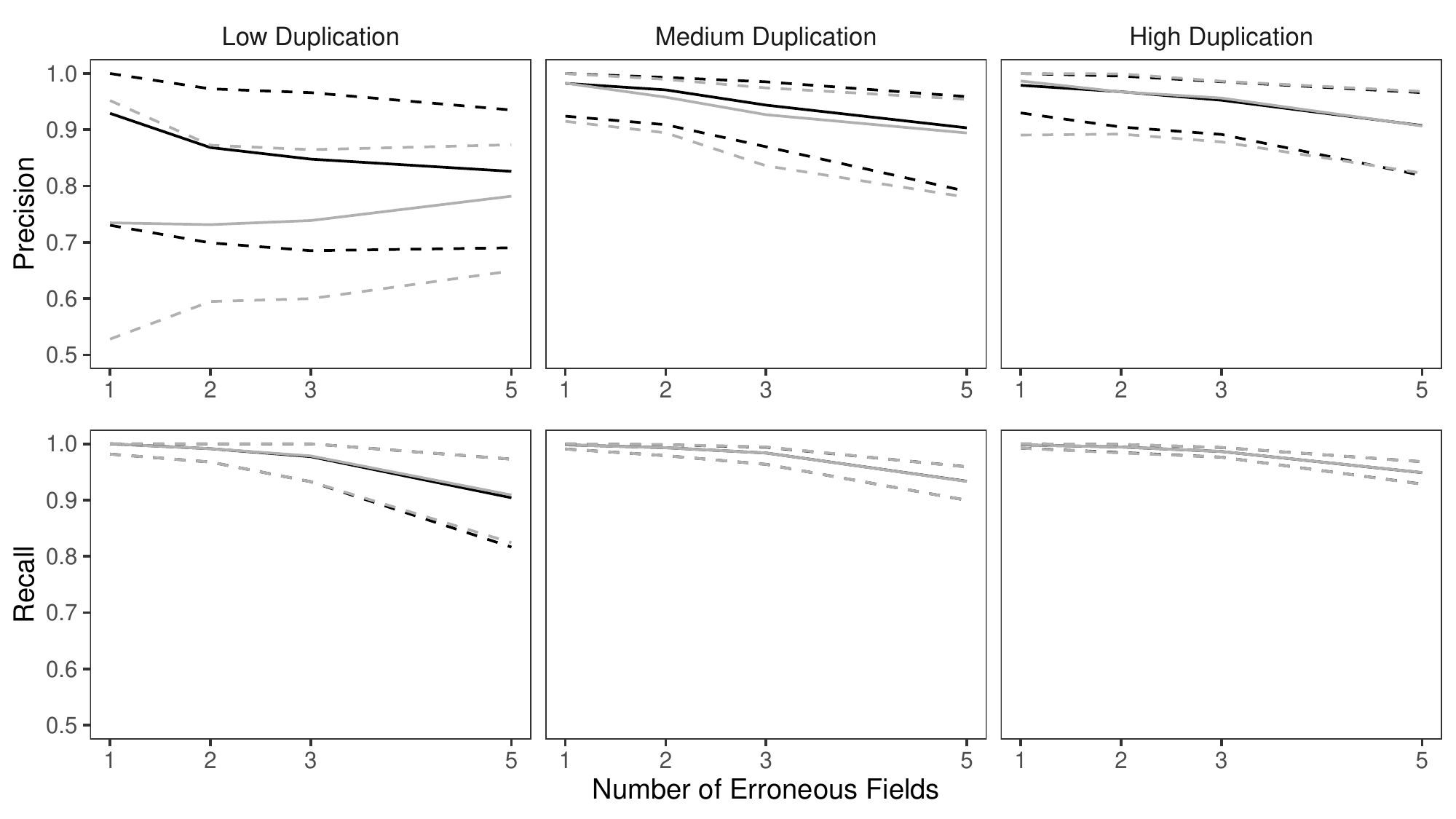}
\begin{minipage}[b]{1\textwidth}
    \caption{Performance comparison for simulation with datafiles with duplicates and full estimates. Black lines refer to results under our structured prior, grey lines refer to the approach of \cite{Sadinle_2014}, solid lines show medians, and dashed lines show 2nd and 98th percentiles. }
    \label{fig:dup_sim}
\end{minipage} 
\end{figure}

The results of the simulation are seen in Figure \ref{fig:dup_sim}. In the medium and high duplication settings, the models have similarly good performance. We believe that the similar performance between models in these settings is due to the use of the indexing, which significantly reduces the size of the space of possible multifile partitions, so that the influence of the structured prior is minimized. However, in the low duplication setting we see that the precision of the proposed model is better across the varying measurement error settings than the model of \cite{Sadinle_2014}. This suggests that in low duplication settings, our approach once again improves upon an approach that uses flat priors for partitions by protecting against over-matching. 

\begin{figure}[!h]
\centering
\includegraphics[width=1\linewidth]{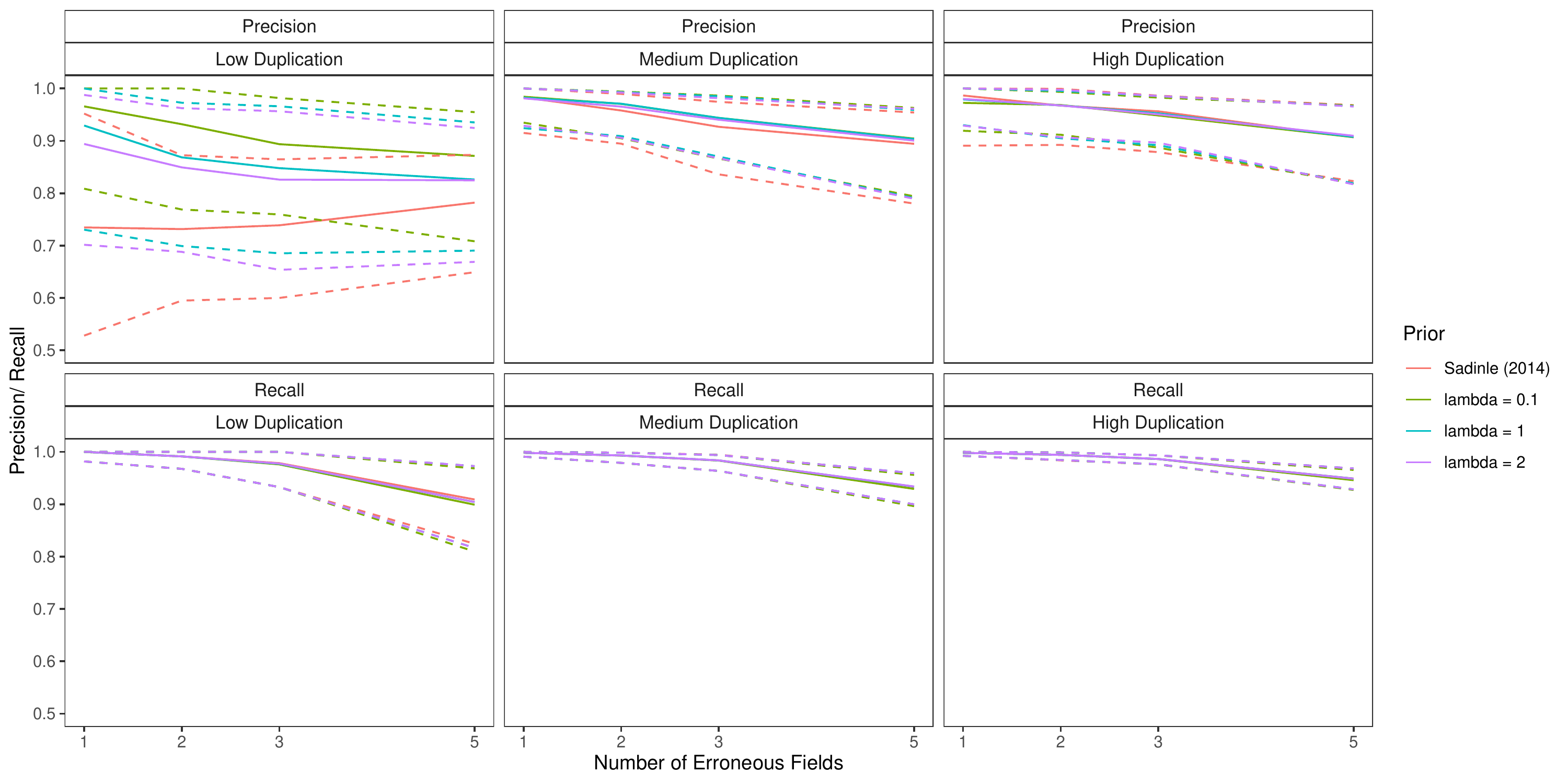}
\begin{minipage}[b]{1\textwidth}
    \caption{Performance comparison for simulation with datafiles with duplicates and full estimates, with varying priors for the within-file cluster sizes. Solid lines show medians, and dashed lines show 2nd and 98th percentiles. }
    \label{fig:dup_sim_sens}
\end{minipage} 
\end{figure}

We now explore the sensitivity of our approach to changes in the prior for the number within-file cluster sizes, and demonstrate how the performance in the low duplication setting can be further improved through the incorporation of an informative prior for the within-file cluster sizes. In the simulation that was just described, we used a Poisson prior with mean $\lambda=1$ truncated to $\{1,\cdots, 10\}$ for the within-file cluster sizes. In the simulation, the within-file cluster sizes were generated from Poisson distributions with mean $\lambda$ truncated to $\{1,\cdots, 5\}$, where $\lambda$ varied over $\{0.1, 1, 2\}$. We now repeat the same simulation using Poisson priors with mean $\lambda$ truncated to $\{1,\cdots, 10\}$ for the within-file cluster sizes, where $\lambda \in \{0.1, 1, 2\}$. The results are presented in Figure \ref{fig:dup_sim_sens}. The results for the medium and high duplication settings are very robust to the within-file cluster size prior specification. In the low duplication setting we see that the performance among the different within-file cluster size prior specifications is best when $\lambda=0.1$, and worst when $\lambda =2$ (but still better than the approach of \cite{Sadinle_2014}). This behavior is expected as the within-file cluster size prior with $\lambda=0.1$ is informative in the low duplication setting.

\subsection{Files with Duplicates, Partial Estimates}
\label{sec:sim3}

We now examine the performance of partial estimates in the low duplication setting of the simulation presented in the previous section, where both the proposed approach and the approach of \cite{Sadinle_2014} struggled the most. For partial estimates, we use $\lfnm=\lfma=1$, $\lfmb=2$, and $\lambda_A=0.1$, so that abstaining from making a linkage decision is $10\%$ as costly as making a false non-match. We will assess the performance of the Bayes estimate using precision and the \textit{abstention rate}, $\sum_{i=1}^rI(\hat{Z}_i=A)/r$, the proportion of records which the Bayes estimate abstained from making a linkage decision. Recall is no longer useful when using partial estimates, as we are not trying to find all true matches.

\begin{figure}[!h]
\centering
\includegraphics[width=0.75\linewidth]{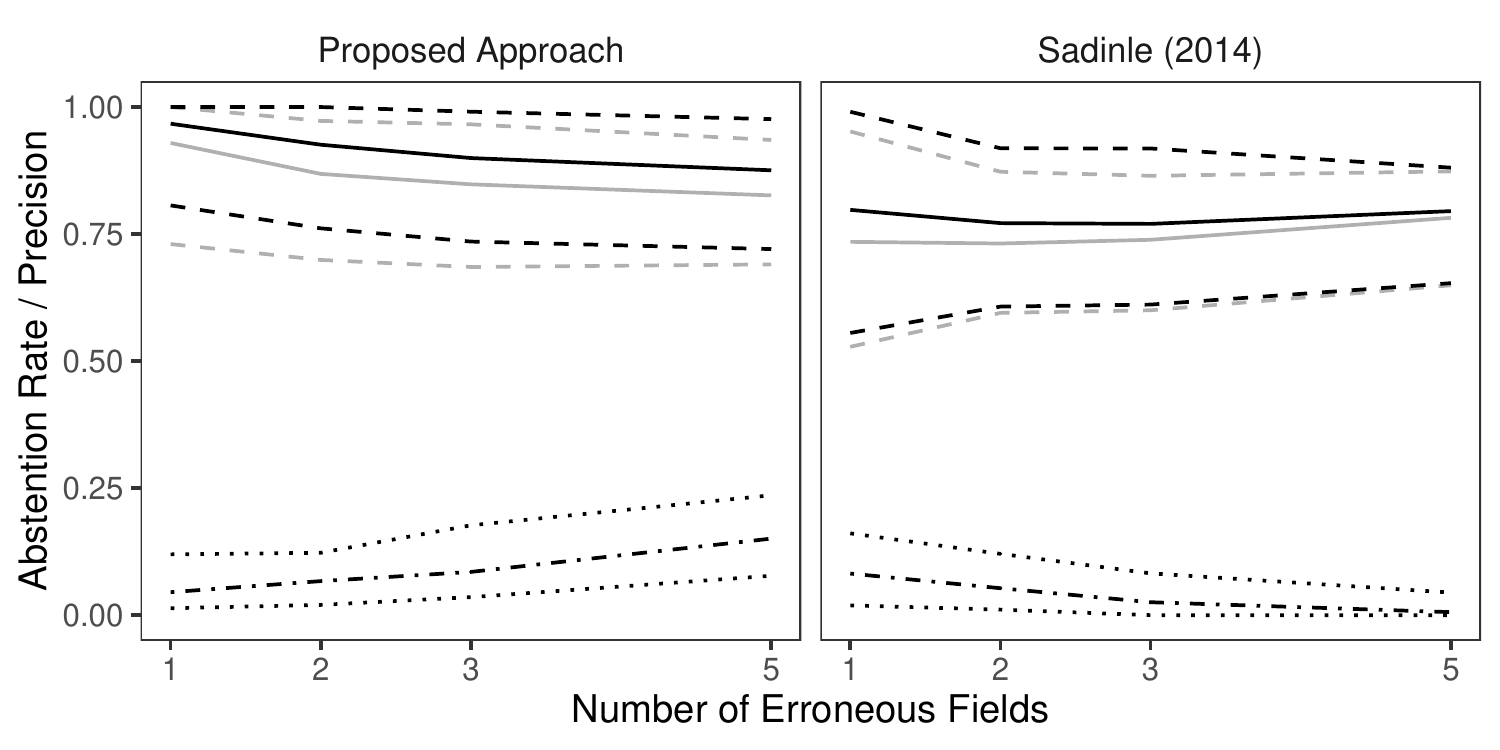}
\begin{minipage}[b]{1\textwidth}
    \caption{Performance comparison for simulation with datafiles with duplicates and
partial estimates.  Black solid and dashed lines refer to precision for partial estimates, grey solid and dashed lines refer to precision for full estimates, and dot-dashed and dotted lines refer to the abstention rate for partial estimates. Solid and dot-dashed lines show medians, and dashed and dotted lines show 2nd and 98th percentiles. }
    \label{fig:dup_sim_partial}
\end{minipage} 
\end{figure}

In Figure \ref{fig:dup_sim_partial} we see that, for both approaches, using partial estimates leads to improved precision in comparison with full estimates, while maintaining a relatively low abstention rate.  This result is expected, as the records to which our partial estimate assigns the abstain option are the most ambiguous in terms of which records they should be linked to, and therefore they are the most likely to lead to false matches which decrease the precision.  Using partial estimates with an abstain option is therefore a good way of compromising between automated and manual linkage: records for which linkage decisions are difficult are left to be handled via clerical review.  

\subsection{Results for an Alternative Metric}
When evaluating the performance of the full estimates in the simulations thus far, we have focused on the metrics of precision and recall, which are global measures of how well the true partition is being estimated. One could also be interested in how well other summaries of the partition are being estimated, e.g. the number of entities (i.e. the number of clusters), the sizes of the clusters, the overlap table, etc.
In this section we report the performance of the full estimates in the previous simulations when estimating the number of entities. 

For each replicate data set in each simulation scenario considered thus far, we obtained a full estimate of the partition, which can be used to derive an estimate of the number of entities. For a given simulation scenario, let $n_0$ denote the true number of entities (in all the scenarios considered thus far, $n_0=500$), and let $\hat{n}_s$ denote the estimate of the number of entities based on the full estimate of the partition for replicate data set $s\in\{1,\cdots, 100\}$. For each simulation scenario, we can thus estimate the bias of these estimates, $\sum_{s=1}^{100}\hat{n}_s-n_0$, and the mean-squared error of these estimates, $\sum_{s=1}^{100}(\hat{n}_s-n_0)^2$.

\subsubsection{Duplicate-Free Files, Equal Errors Across Files}
\begin{figure}[!h]
\centering
\includegraphics[width=0.95\linewidth]{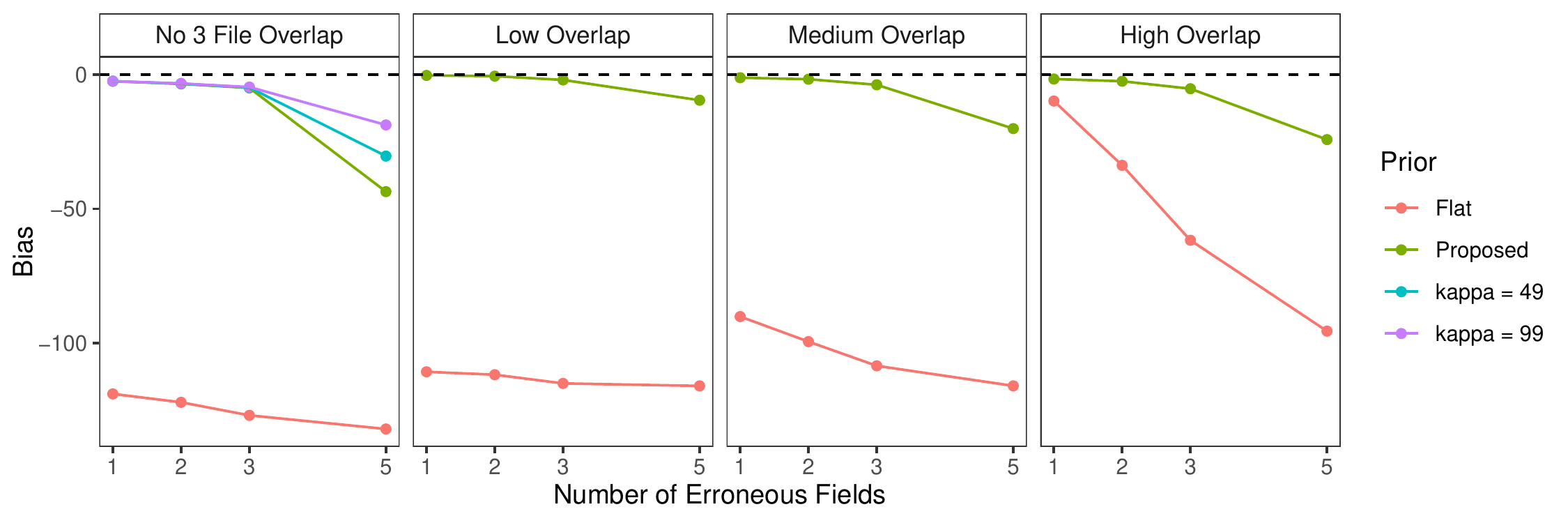}
\begin{minipage}[b]{1\textwidth}
    \caption{Bias estimates for simulation with duplicate-free files and equal errors across files. ``Flat" refers to a flat prior on tripartite matchings, ``Proposed" refers to our structured prior for partitions when $\balpha=(1,\cdots, 1)$, and ``kappa = 49" and ``kappa = 99" refer to the more informative specifications of $\balpha$ discussed in Appendix \ref{sec:sim_1_sens}.}
    \label{fig:no_dup_sim_extra_sum_bias}
\end{minipage} 
\end{figure}

\begin{figure}[!h]
\centering
\includegraphics[width=0.95\linewidth]{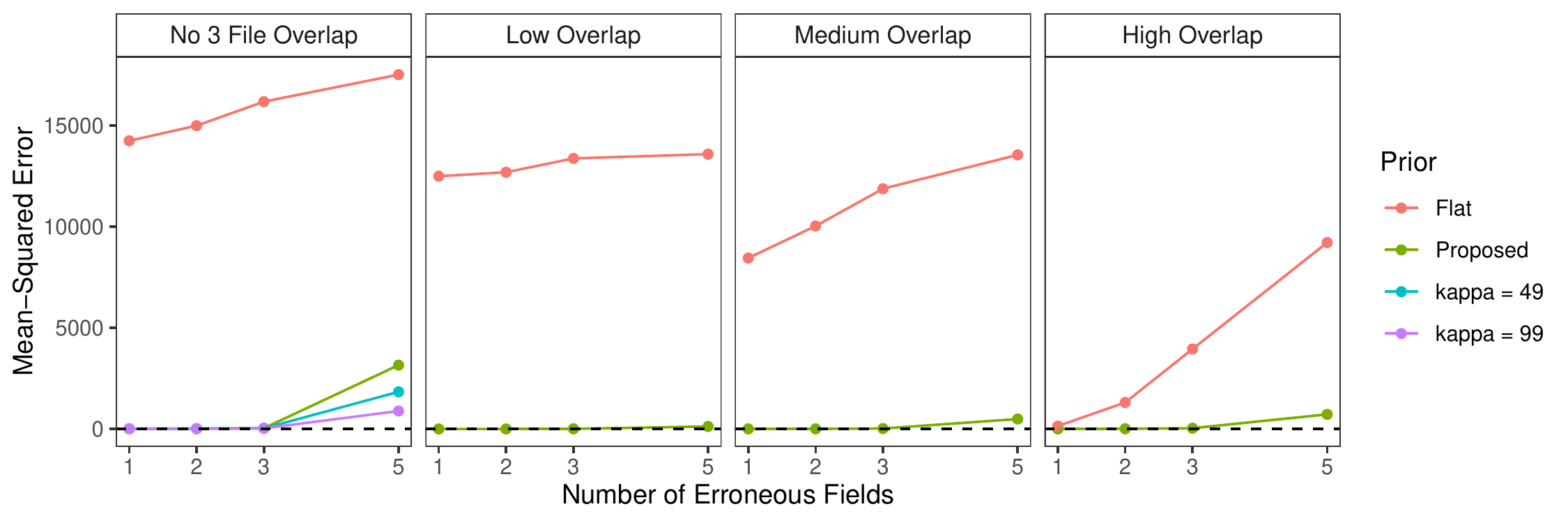}
\begin{minipage}[b]{1\textwidth}
    \caption{Mean-squared error estimates for simulation with duplicate-free files and equal errors across files. ``Flat" refers to a flat prior on tripartite matchings, ``Proposed" refers to our structured prior for partitions when $\balpha=(1,\cdots, 1)$, and ``kappa = 49" and ``kappa = 99" refer to the more informative specifications of $\balpha$ discussed in Appendix \ref{sec:sim_1_sens}.}
    \label{fig:no_dup_sim_extra_sum_mse}
\end{minipage} 
\end{figure}

The results for estimating the number of latent entities in the simulations conducted in Section 6.2 of the main text and Appendix \ref{sec:sim_1_sens} are seen in Figures \ref{fig:no_dup_sim_extra_sum_bias} and \ref{fig:no_dup_sim_extra_sum_mse}. We see that that across the different simulation settings the proposed approach has a slight negative bias, and the approach using a flat prior has a very large negative bias. In the no-three-file overlap settings, we see that the more informative prior specifications are less biased than the proposed approach when there are a larger number of erroneous fields, which mirrors the results from Appendix \ref{sec:sim_1_sens}. Across all approaches, the bias increases as the number of erroneous fields increases. The results for the mean-squared error estimates are very similar to the results for the bias estimates.

\subsubsection{Duplicate-Free Files, Unequal Errors Across Files}
The results for estimating the number of latent entities in the simulation conducted in Section 6.3 of the main text are seen in Figures \ref{fig:no_dup_sim_extra_sum_bias} and \ref{fig:no_dup_sim_extra_sum_mse}. We see that that across the two simulation settings all approaches have a negative bias, with the proposed approach having the smallest bias and the approach using a flat prior having the largest bias in both scenarios. The results for the mean-squared error estimates are very similar to the results for the bias estimates.

\begin{figure}[!t]
\centering
\includegraphics[width=0.7\linewidth]{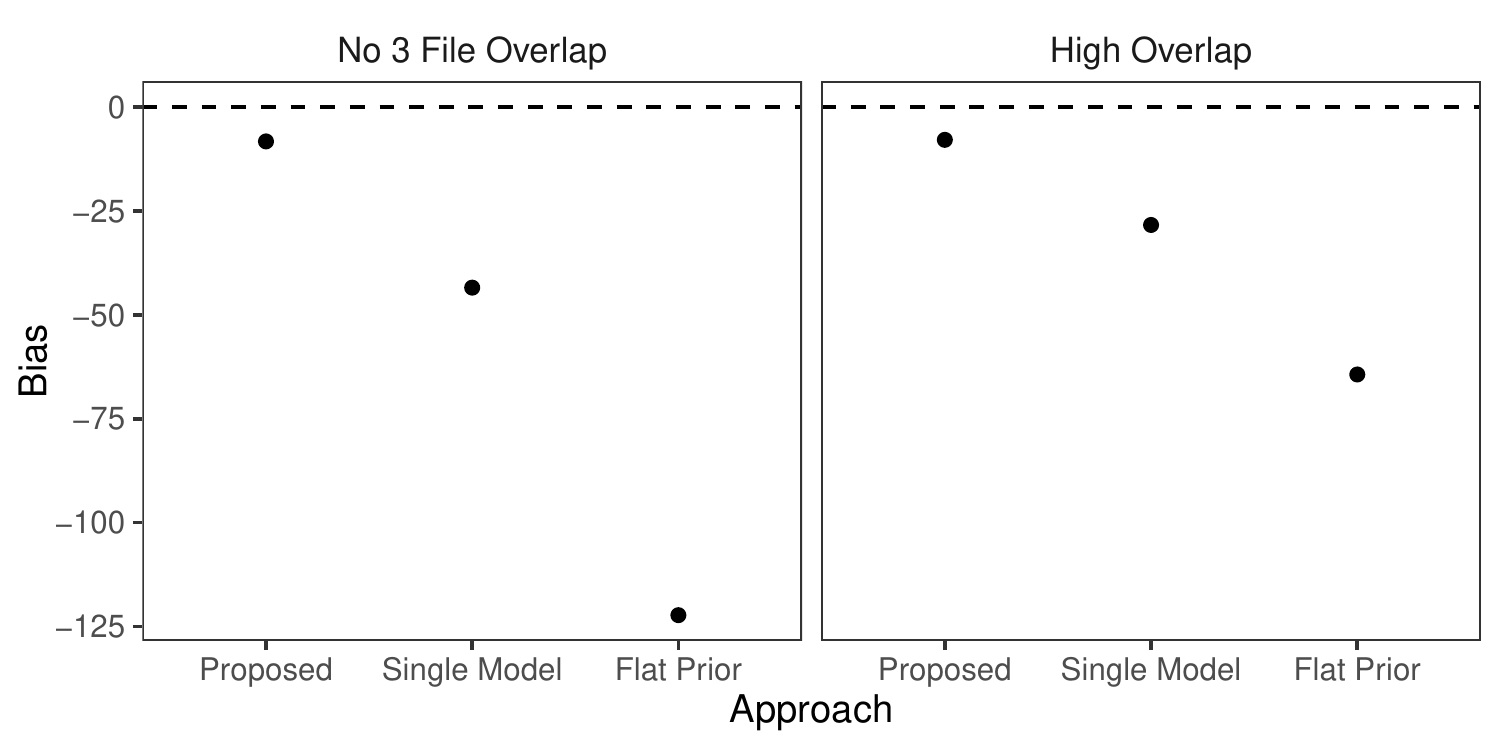}
\begin{minipage}[b]{1\textwidth}
    \caption{Bias estimates for simulation with duplicate-free files and unequal errors across files. ``Proposed" refers to our proposed approach, ``Single Model" refers to the approach using a single model for all file-pairs and our structured prior for partitions, and ``Flat Prior" refers to the approach using our model for comparison data with a flat prior on tripartite matchings.}
    \label{fig:no_dup_error_sim_extra_sum_bias}
\end{minipage} 
\end{figure}

\begin{figure}[!h]
\centering
\includegraphics[width=0.7\linewidth]{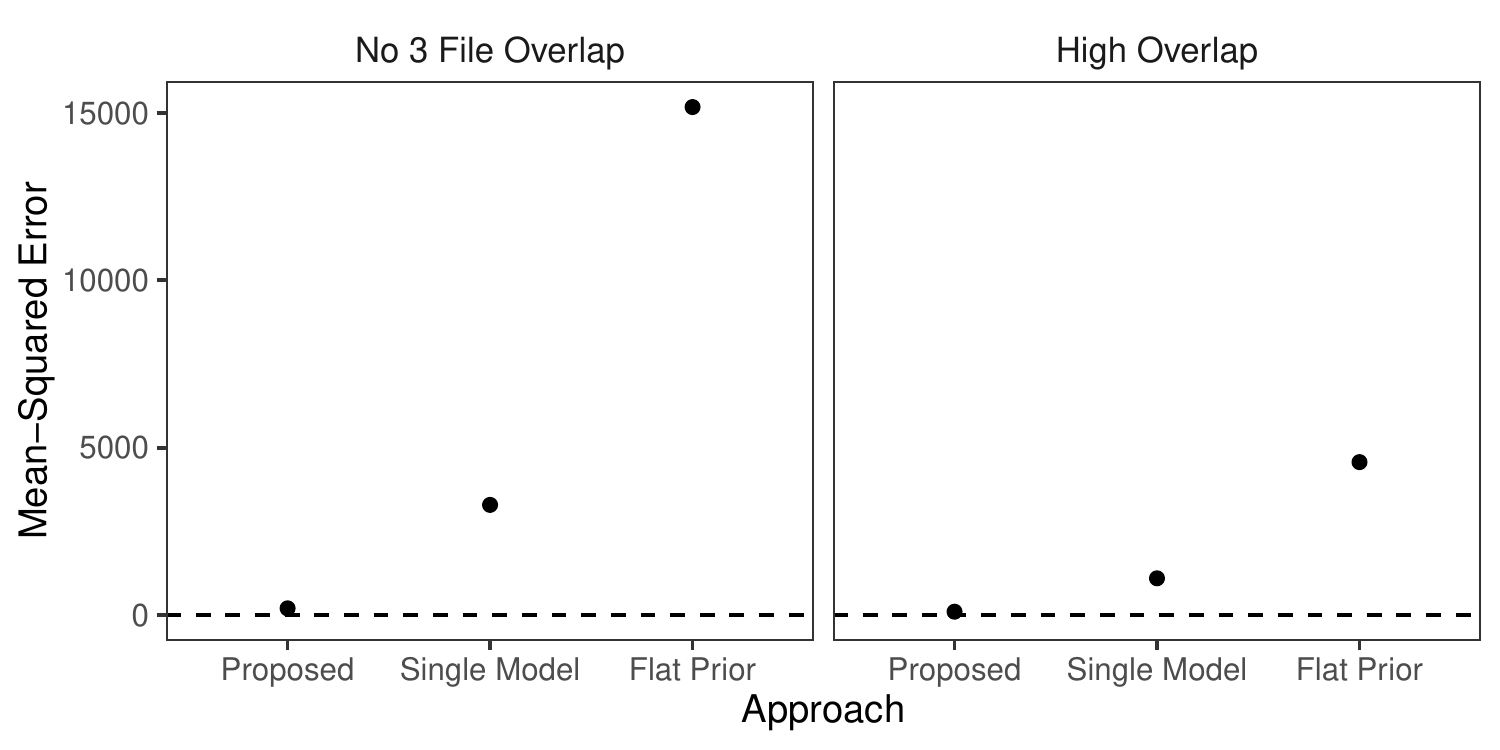}
\begin{minipage}[b]{1\textwidth}
    \caption{Mean-squared error estimates for simulation with duplicate-free files and unequal errors across files. ``Proposed" refers to our proposed approach, ``Single Model" refers to the approach using a single model for all file-pairs and our structured prior for partitions, and ``Flat Prior" refers to the approach using our model for comparison data with a flat prior on tripartite matchings.}
    \label{fig:no_dup_error_sim_extra_sum_mse}
\end{minipage} 
\end{figure}

\subsubsection{Files with Duplicates, Full Estimates}
\begin{figure}[!h]
\centering
\includegraphics[width=0.75\linewidth]{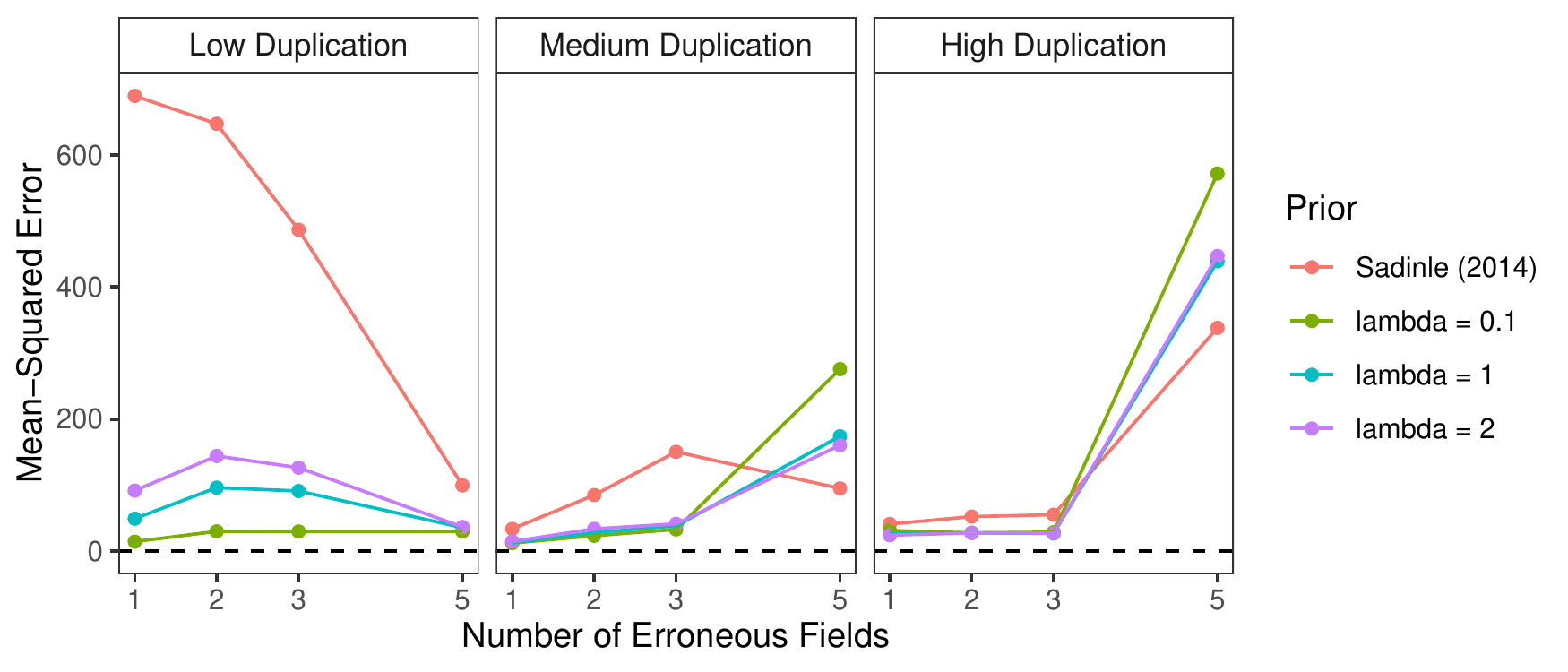}
\begin{minipage}[b]{1\textwidth}
    \caption{Mean-squared error estimates for simulation with files with duplicates and equal errors across files. ``Sadinle (2014)" refers to the approach of \cite{Sadinle_2014} and ``lambda=..." refers to the proposed approach, varying the prior over within-file cluster sizes.}
    \label{fig:dup_sim_extra_sum_mse}
\end{minipage} 
\end{figure}

The results for estimating the number of latent entities in the simulation conducted in Appendix \ref{sec:sim2} are seen in Figures \ref{fig:dup_sim_extra_sum_bias} and \ref{fig:dup_sim_extra_sum_mse}. In the low duplication setting, we see that the proposed approach with $\lambda=0.1$ has the smallest bias across error settings, followed by the proposed approach with $\lambda=1$, then the proposed approach with $\lambda=2$, and then the approach of \cite{Sadinle_2014} with the largest bias across error settings. This mirrors the results from Appendix \ref{sec:sim2}. In the medium and high duplication settings we see that all approaches have a slight negative bias when there are a small number of erroneous fields, and a slight positive bias when there are a large number of erroneous fields. The different variants of the proposed approach all have similar performance in these settings. The proposed approach performs best when there are a small number of erroneous fields, and the approach of \cite{Sadinle_2014} performs best when there are a large number of erroneous fields. The results for the mean-squared error estimates are very similar to the results for the bias estimates.

\begin{figure}[!h]
\centering
\includegraphics[width=0.75\linewidth]{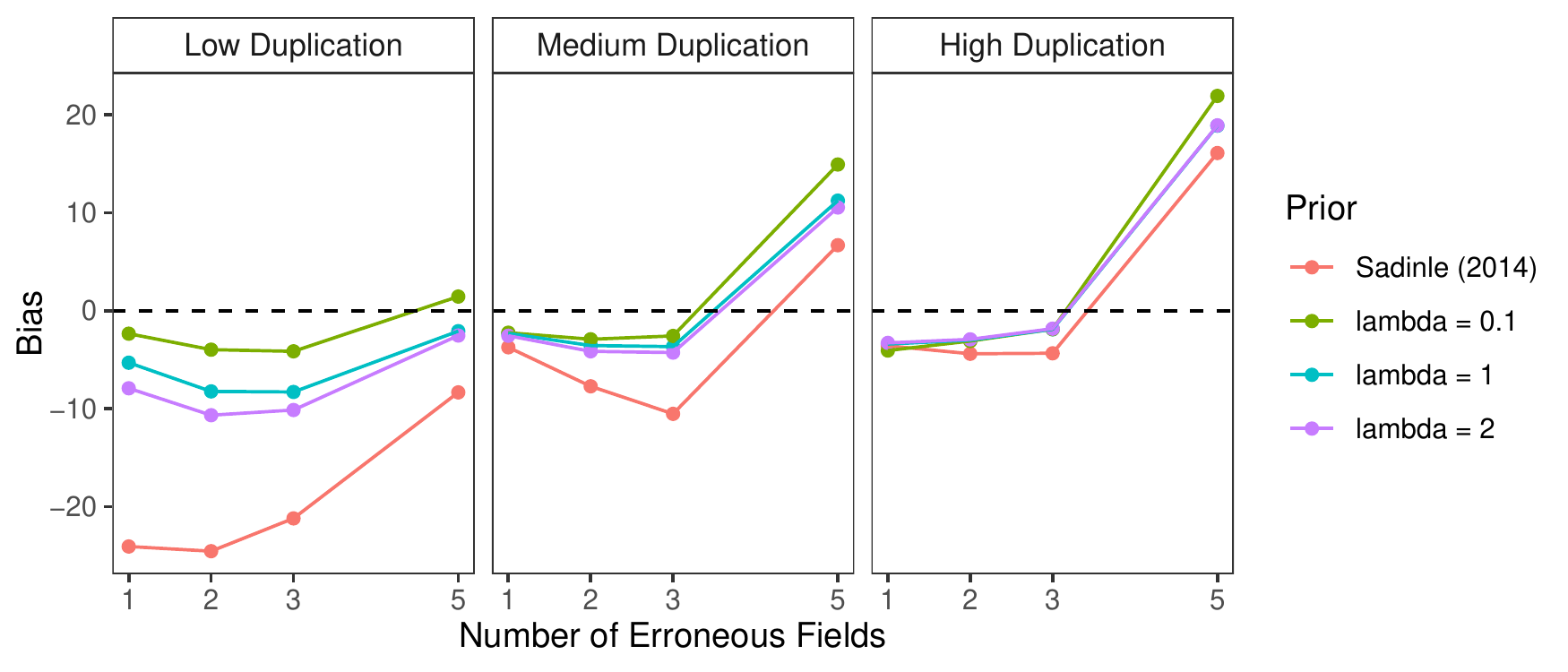}
\begin{minipage}[b]{1\textwidth}
    \caption{Bias estimates for simulation with files with duplicates and equal errors across files. ``Sadinle (2014)" refers to the approach of \cite{Sadinle_2014} and ``lambda=..." refers to the proposed approach, varying the prior over within-file cluster sizes.}
    \label{fig:dup_sim_extra_sum_bias}
\end{minipage} 
\end{figure}

\subsection{Simulation Running Times}
In this section we present the average running time of our proposed Gibbs sampler across the various simulations settings (i.e. the average time it takes to draw 1000 samples for each simulation setting). 
The running times are based on the implementation in the R package multilink which can be downloaded on GitHub at \url{https://github.com/aleshing/multilink}, 
with the Gibbs sampler described in Appendix \ref{sec:gibbs} written in C++, running on a laptop with a 3.1 GHz processor. 

\subsubsection{Duplicate-Free Files, Equal Errors Across Files}
The average running times of our approach in the simulations conducted in Section 6.2 of the main text are presented in Table \ref{tab:no_dup_sim_times}. The average number of records was 725 in the settings with no-three-file overlap, 676 in the settings with low overlap, 725 in the settings with medium overlap, and 875 in the settings with high overlap.

\begin{table}[ht]
\caption{Average running time in seconds for proposed approach in simulations with duplicate-free files and equal errors across files.} \label{tab:no_dup_sim_times}
\centering
\resizebox{\columnwidth}{!}{
\begin{tabular}{lcccc}
  \hline
Number of Erroneous Fields & No 3 File Overlap & Low Overlap & Medium Overlap & High Overlap \\ 
  \hline
1 & 111.0 & 77.6 & 83.7 & 108.6 \\ 
  2 & 112.9 & 77.7 & 83.7 & 109.2 \\ 
  3 & 111.8 & 77.4 & 83.0 & 109.4 \\ 
  5 & 95.7 & 73.8 & 77.3 & 101.1 \\ 
   \hline
\end{tabular}}
\end{table}

\subsubsection{Duplicate-Free Files, Unequal Errors Across Files}
The average running time of our proposed approach in the simulations conducted in Section 6.3 of the main text was 121.8 seconds in the no-three-file overlap setting and 104.1 seconds in the high overlap setting. The average number of records was 725 in the settings with no-three-file overlap and 875 in the settings with high overlap.

\subsubsection{Files with Duplicates, Full Estimates}
The average running times of our approach in the simulations conducted in Appendix \ref{sec:sim2} are presented in Table \ref{tab:dup_sim_times}. The average number of records was 590 in the settings with low duplication, and 889 in the settings with medium duplication, and 1260 in the settings with high duplication. We note here that in this simulation, compared to the simulations with duplicate-free files, we used indexing, which sped up the running time.

\begin{table}[ht]
\caption{Average running time in seconds for proposed approach in with files with duplicates and equal errors across files.} \label{tab:dup_sim_times}
\centering
\resizebox{\columnwidth}{!}{
\begin{tabular}{lccc}
  \hline
Number of Erroneous Fields & Low Duplication & Medium Duplication & High Duplication \\ 
  \hline
1 & 1.5 & 8.0 & 20.9 \\ 
  2 & 1.1 & 7.7 & 20.8 \\ 
  3 & 1.0 & 7.4 & 21.0 \\ 
  5 & 0.9 & 6.6 & 19.2 \\ 
   \hline
\end{tabular}}
\end{table}

\subsection{Larger Sample Size Simulation}
All of the simulations thus far had fixed the true number of latent entities, $n$, to $500$. To explore the running time of our proposed approach further, we now present an additional set of simulations where the number of latent entities varies over $\{100, 500, 1000, 2500\}$. For concreteness we will focus on the simulation setting with duplicate-free files, equal errors across files, medium overlap, and $1$ erroneous field per record (i.e. one of the settings from the simulation conducted in Section 6.2 of the main text). For this chosen simulation setting, we repeat the simulation as conducted in Section 6.2 of the main text, varying the number of latent entities over $\{100, 500, 1000, 2500\}$. For the $n=2500$ setting we conduct $25$, rather than $100$, replications (due to how long each replication takes). The average number of records was 146 when $n=100$, 725 when $n=500$, 1452 when $n=1000$, and 3629 when $n=2500$. In addition to fitting the proposed approach without indexing as in Section 6.2 of the main text, we additionally fit the proposed approach using the indexing scheme described in Appendix \ref{sec:sim2} to demonstrate the utility of indexing for improving the running time or our proposed approach, as discussed in Appendix \ref{sec:indexing}.

\begin{figure}[!t]
\centering
\includegraphics[width=0.615\linewidth]{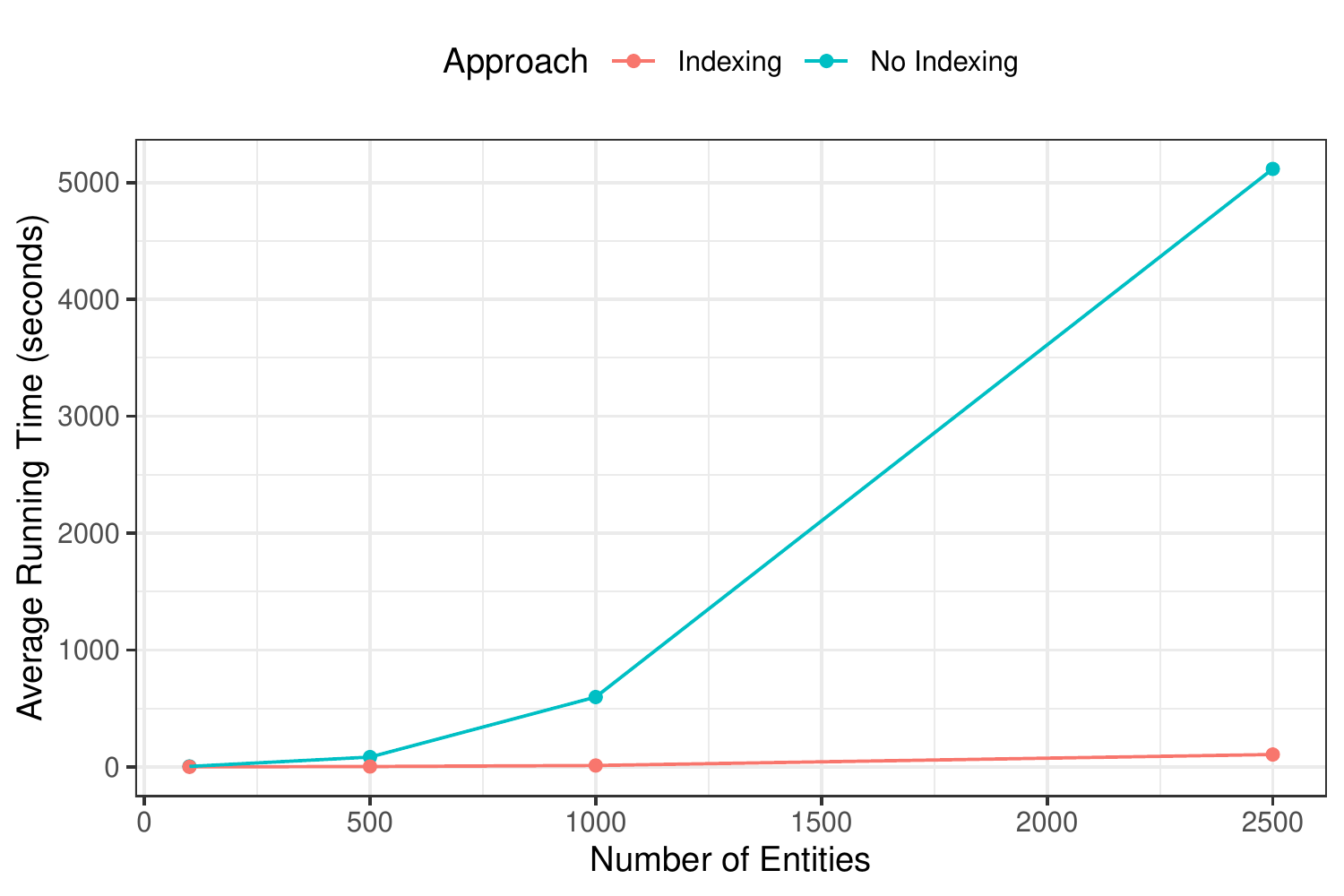}
\begin{minipage}[b]{0.95\textwidth}
    \caption{Average running time for simulation varying the number of latent entities.}
    \label{fig:big_sim}
\end{minipage} 
\end{figure}

The average running time for each setting of the number of entities, $n$, is presented in Figure \ref{fig:big_sim}. Our proposed approach without indexing runs relatively quickly in the settings with $n\in\{100, 500, 1000\}$, for example it only takes around $10$ minutes on average to draw $1000$ samples when $n=1000$. However, when $n=2500$ our proposed approach without indexing runs takes roughly an hour and a half on average to draw $1000$ samples. While this running time is manageable, it indicates that the running time in settings with more records than this would prohibitively slow. When looking at the running time of our proposed approach using indexing, we see that the average running is reduced drastically. For example, when $n=2500$, the average running time is only around 100 seconds.

These results suggest that our approach without indexing can be run in a manageable amount of time for thousands of records, but would be prohibitively slow when the number of records is much larger than a few thousand. Our proposed approach with indexing however can scale to larger file sizes, potentially in the tens of thousands. Using our approach, with or without indexing, in conjunction with blocking, as suggested in Appendix \ref{sec:indexing}, can help to further scale our approach to large file sizes.

We note that the precision and recall in these simulations, across the different settings of the number of entities, were comparable to the results for the medium overlap, $1$ erroneous field per record setting from the simulation results in Section 6.2 of the main text.

\subsection{Alternative Loss Function Specifications}
In this section explore the impact of varying the specification of the losses $\lfnm, \lfma, \lfmb,$ and $\lambda_A$ on the performance of our proposed approach across the different simulation settings. For full estimates, we follow \cite{Sadinle_2017} and consider the following  specifications of the losses $\lfnm, \lfma,$ and  $\lfmb$ (with $\lambda_A=\infty$).
\begin{enumerate}[label=\Alph*)]
    \item $\lfnm=1, \lfma=1, \lfmb=2$. This is the specification used in all of the simulations thus far.
    \item $\lfnm=\lfma=\lfmb=1$. Compared to specification A, this specification does not penalize type 2 false matches as heavily. 
    \item $\lfnm=4, \lfma=\lfmb=1$. This specification penalizes false non-matches more heavily than false matches.
    \item $\lfnm=1, \lfma=3, \lfmb=5$. This specification penalizes false matches more heavily than false non-matches, and type 2 false matches more heavily than type 1 false matches.
    \item $\lfnm=1, \lfma=2, \lfmb=3$. Compared to specification E, this specification does not penalizes false matches as heavily.
    \item $\lfnm=\lfma=1, \lfmb=4$.  Compared to specification A, this specification penalizes type 2 false matches more heavily. 
\end{enumerate}
For partial estimates, we consider combining $\lambda_A\in\{0.05, 0.1,0.25\}$ with the six specifications of $\lfnm, \lfma,$ and  $\lfmb$ that we have just introduced.

\subsubsection{Duplicate-Free Files, Equal Errors Across Files}

\begin{figure}[h]
\centering
\includegraphics[width=1\linewidth]{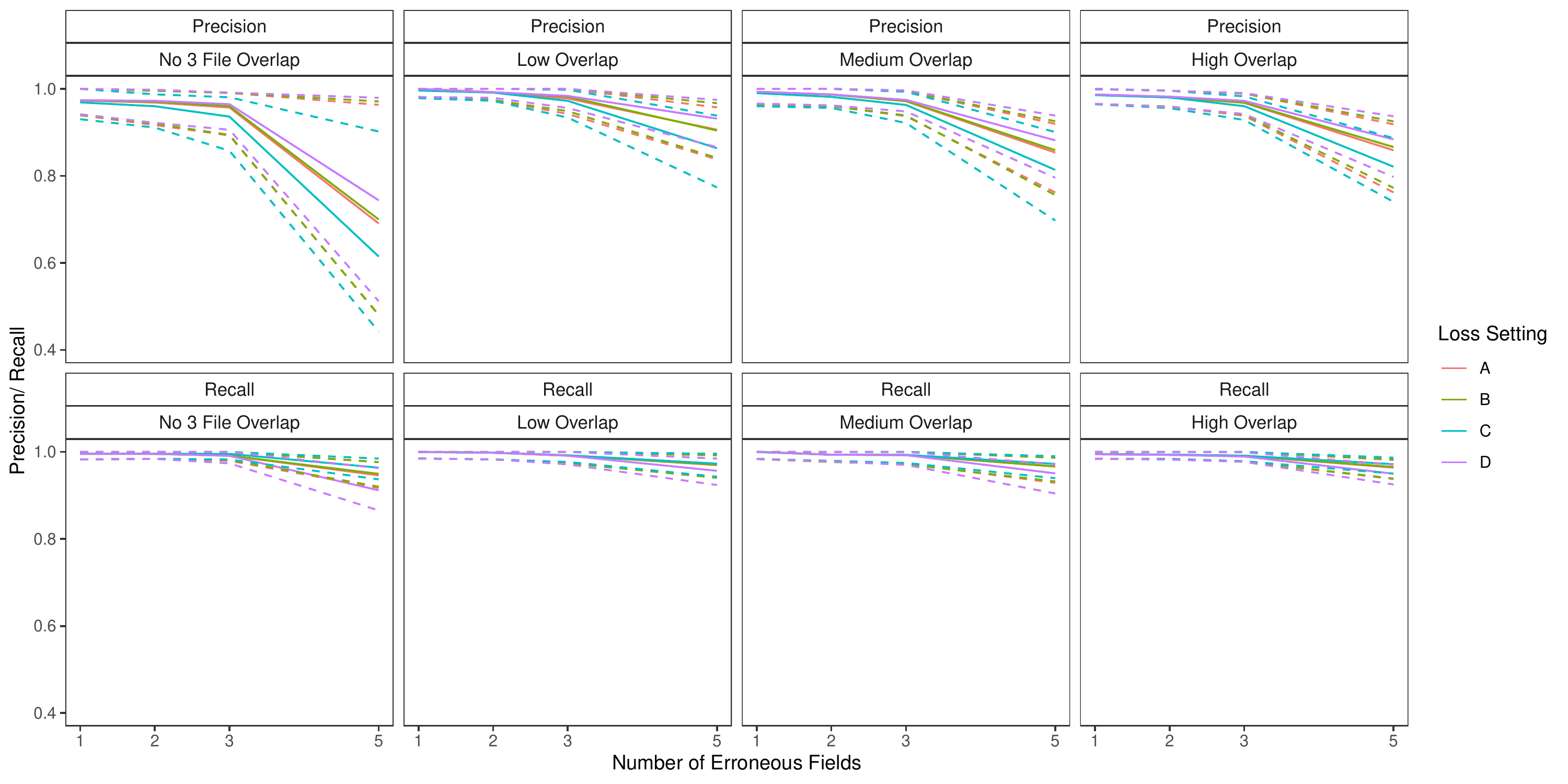}
\begin{minipage}[b]{1\textwidth}
    \caption{Performance comparison across different loss function specifications for simulation with duplicate-free files and equal measurement error across files. Solid lines show medians, and dashed lines show 2nd and 98th percentiles.}
    \label{fig:no_dup_sim_extra_losses}
\end{minipage} 
\end{figure}

The performance of our proposed approach in the simulations conducted in Section 6.2 of the main text, using the different loss function specifications, are seen in Figure \ref{fig:no_dup_sim_extra_losses}. So that the figure is easier to scrutinize, we do not plot the results under specifications E and F, as the results under specifications E are very similar to the results under specification D, and the results under specifications F are very similar to the results under specifications A and B.

We see that when there are a low number of erroneous fields, the results are fairly robust to the loss function specification. When there are a high number of erroneous fields, we see that specification D leads to the highest precision and the lowest recall, specification C leads to the lowest precision and the highest recall, and specifications A and B are somewhere in between. These results are expected, as specification C penalizes false non-matches much more than false matches, and will thus decide to match records more often than the other specifications, and specification D penalizes false matches much more than false non-matches, and thus will decide to match records less often than the other specifications.

\subsubsection{Duplicate-Free Files, Unequal Errors Across Files}
\begin{figure}[h]
\centering
\includegraphics[width=1\linewidth]{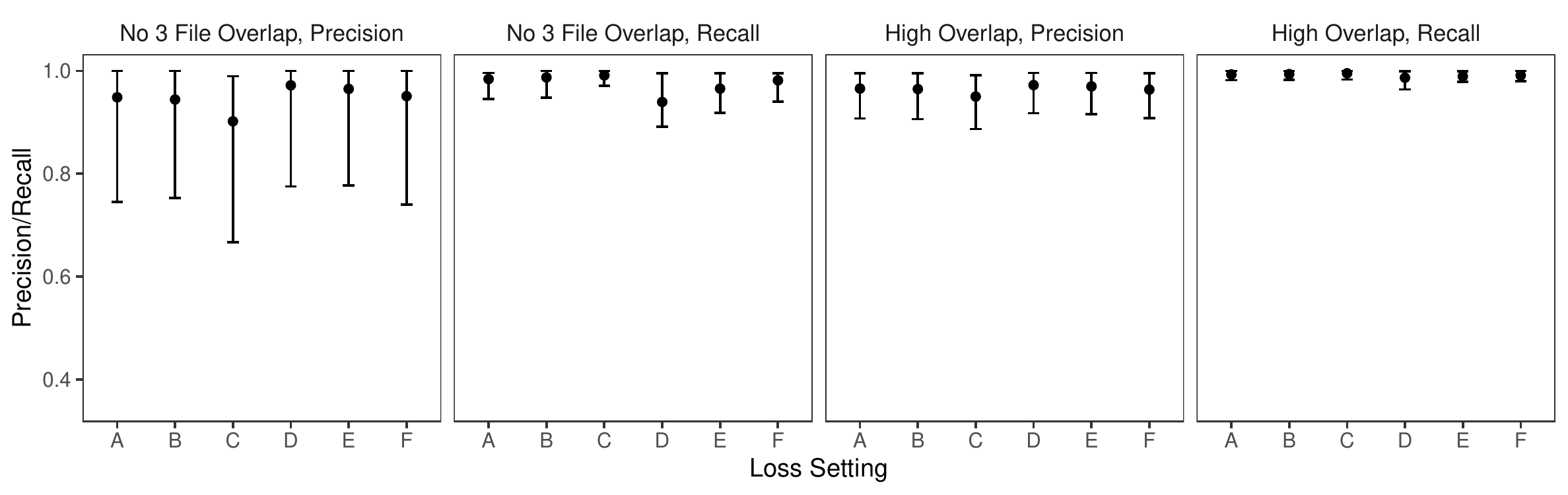}
\begin{minipage}[b]{1\textwidth}
    \caption{Performance comparison across different loss function specifications for simulation with duplicate-free files and unequal measurement error across files. Solid lines show medians, and dashed lines show 2nd and 98th percentiles.}
    \label{fig:no_dup_error_sim_extra_losses}
\end{minipage} 
\end{figure}

The performance of our proposed approach in the simulations conducted in Section 6.3 of the main text, using the different loss function specifications, are seen in Figure \ref{fig:no_dup_error_sim_extra_losses}. Overall the results are fairly robust to the loss function specification. The results under specifications A, B and F are very similar and the results under specifications D and E are very similar. Similar to the last section, specifications D and E lead to the highest precision and the lowest recall, specification C leads to the lowest precision and the highest recall, and specifications A, B, and F are somewhere in between.

\subsubsection{Files with Duplicates, Full Estimates}
The performance of our proposed approach in the simulations conducted in Appendix \ref{sec:sim2}, using the different loss function specifications, are seen in Figure \ref{fig:no_dup_sim_extra_losses}. So that the figure is easier to scrutinize, we do not plot the results under specifications E and F, as the results under specifications E are very similar to the results under specification D, and the results under specifications F are very similar to the results under specifications A and B.

We see that when there is medium and high duplication, the results are fairly robust to the loss function specification. When there is low duplication, we see that specification D leads to the highest precision and the lowest recall, specification C leads to the lowest precision and the highest recall, and specifications A and B are somewhere in between. These results are expected, as described in the previous sections.

\begin{figure}[h]
\centering
\includegraphics[width=0.95\linewidth]{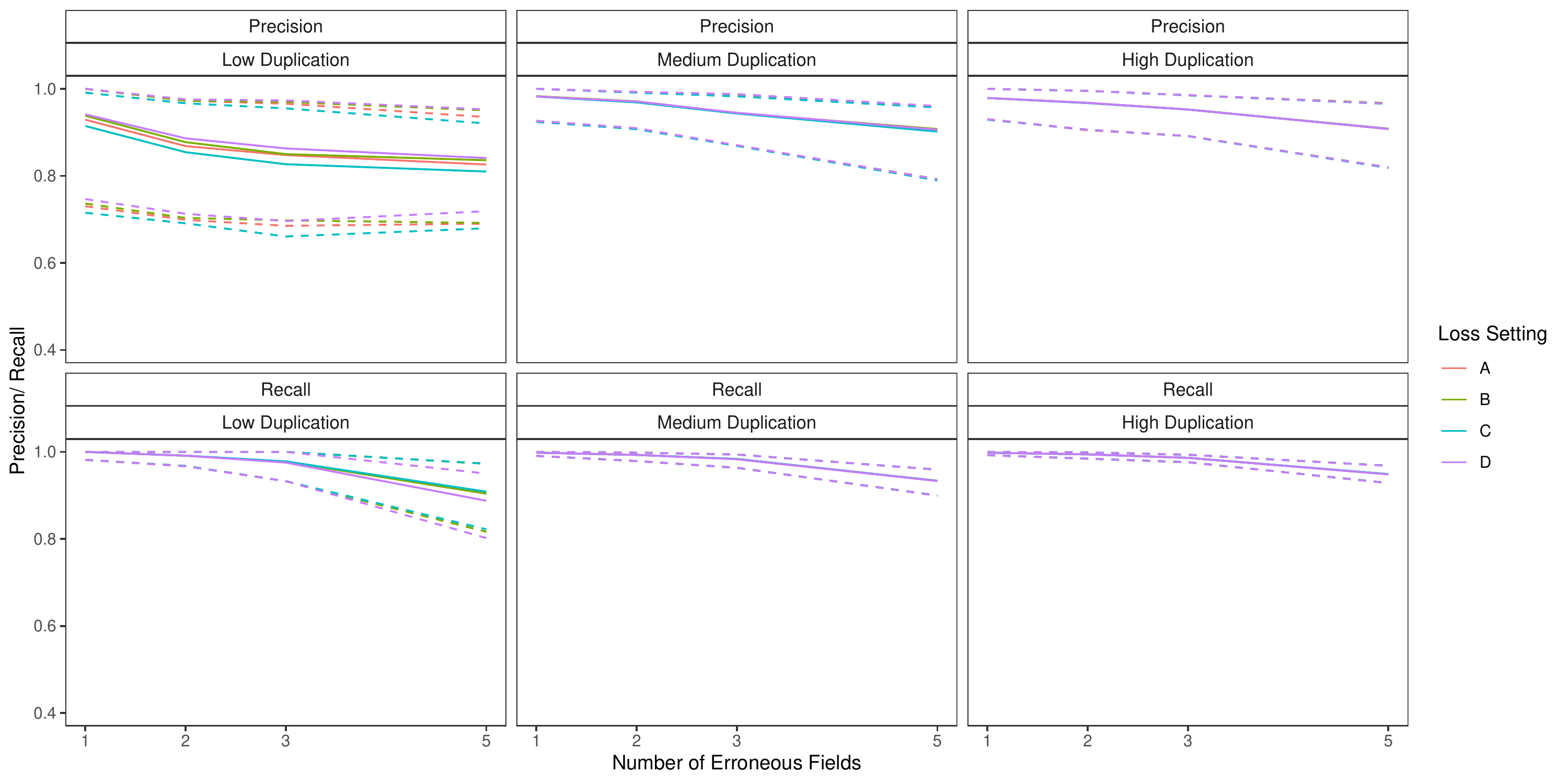}
\begin{minipage}[b]{1\textwidth}
    \caption{Performance comparison across different loss function specifications for simulation with files with duplicates and equal measurement error across files, when using full estimates. Solid lines show medians, and dashed lines show 2nd and 98th percentiles.}
    \label{fig:dup_sim_extra_losses}
\end{minipage} 
\end{figure}

\subsubsection{Files with Duplicates, Partial Estimates}

\begin{figure}[h]
\centering
\includegraphics[width=1\linewidth]{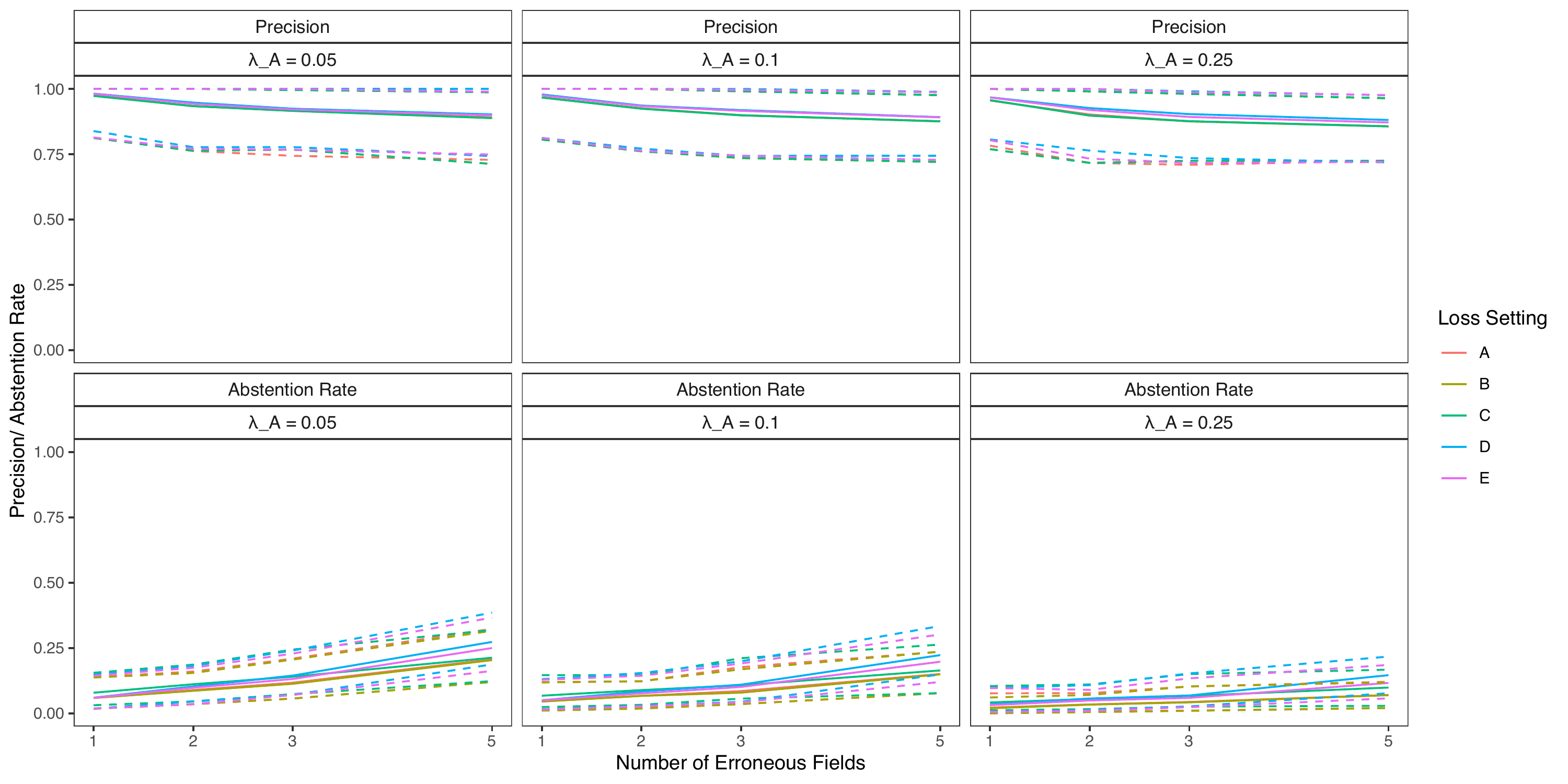}
\begin{minipage}[b]{1\textwidth}
    \caption{Performance comparison across different loss function specifications for simulation with files with duplicates and equal measurement error across files, when using partial estimates. Solid lines show medians, and dashed lines show 2nd and 98th percentiles.}
    \label{fig:dup_sim_partial_extra_losses}
\end{minipage} 
\end{figure}

The performance of our proposed approach in the simulations conducted in Appendix \ref{sec:sim3}, using the different loss function specifications, are seen in Figure \ref{fig:dup_sim_partial_extra_losses}. So that the figure is easier to scrutinize, we do not plot the results under specifications F, as the results under specifications F are very similar to the results under specifications A and B.

We see that as we increase $\lambda_A$, the abstention rate decreases and the precision decreases across the different specifications of $\lfnm, \lfma,$ and  $\lfmb$. Further, for each loss specification the abstention rate increases as the number of erroneous fields increases, as there is more uncertainty in the linkage. For a given setting of $\lambda_A$, the precision is fairly robust to the different specifications of $\lfnm, \lfma,$ and  $\lfmb$, with specifications D and E having slightly higher precision that specifications A, B and C. For a given setting of $\lambda_A$, the abstention rate is highest under specifications D and E, lowest under specifications A and B, with specification C somewhere in between.

\subsection{Convergence Diagnostics}
For all simulations we ran the Gibbs sampler described in Appendix \ref{sec:gibbs} for $1,000$ iterations, discarding the first $100$ samples as burn-in. We initially came up with these sampling and burn-in lengths based on a small number of test runs for each simulation scenario. In particular, for each simulation scenario we ran a small number of test runs and examined the trace plots for the number of entities, $n$. As the chains for $n$ appeared to converge quickly based on these trace plots, we determined that a burn-in period of $100$ samples was appropriate. To illustrate this procedure, for each simulation scenario we now present trace plots for the number of entities, $n$, for a small number of runs.

\subsubsection{Duplicate-Free Files, Equal Errors Across Files}
\begin{figure}[!h]
\centering
\includegraphics[width=0.9\linewidth]{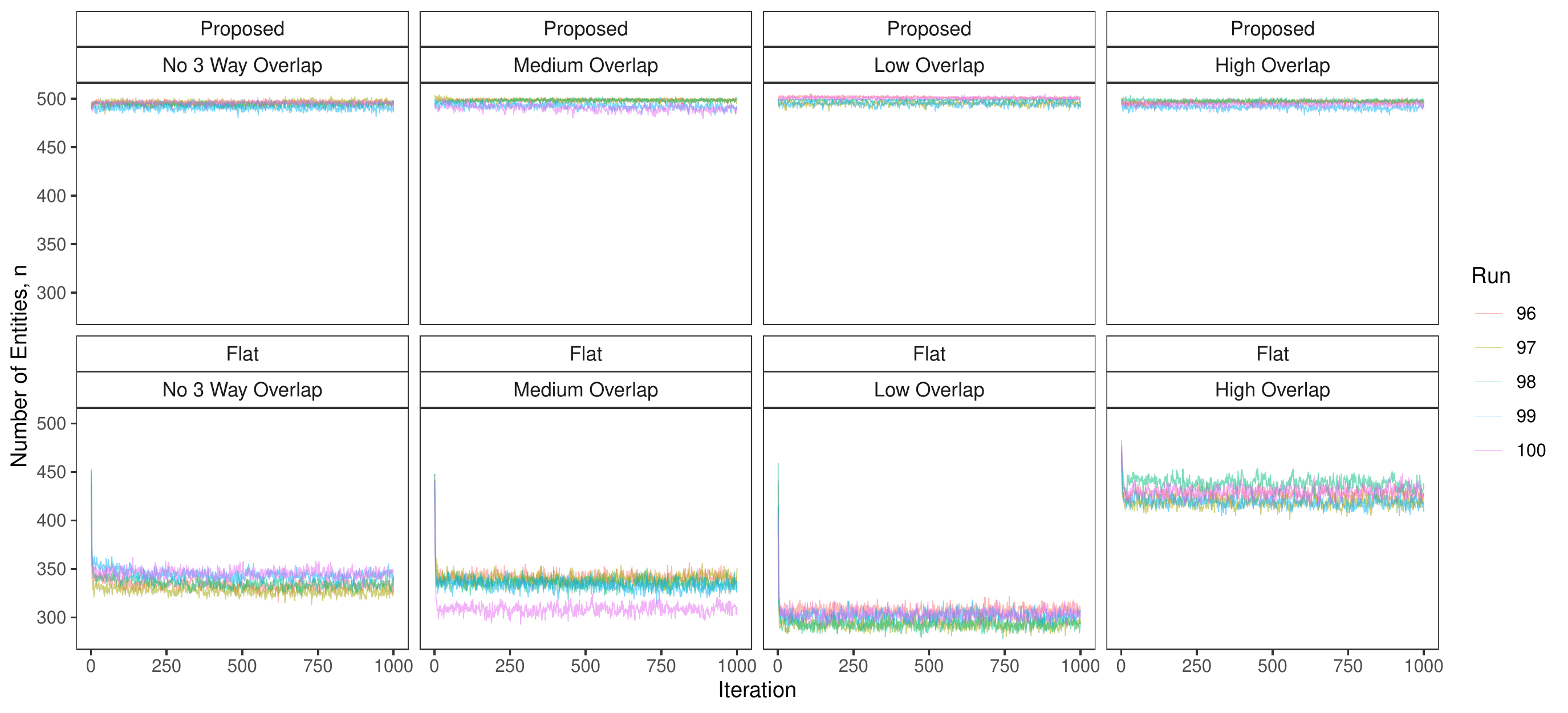}
\begin{minipage}[b]{1\textwidth}
    \caption{Trace plots for $n$ for the last $5$ of $100$ runs for the simulation with duplicate-free files and equal measurement error across files. ``Proposed" refers to the proposed approach and ``Flat" refers to the approach using a flat prior for tripartite matchings.}
    \label{fig:no_dup_sim_trace}
\end{minipage} 
\end{figure}

For the simulations conducted in Section 6.2 of the main text with $3$ erroneous fields per record, for each overlap setting we present the trace plots for $n$ for the last $5$ of the $100$ simulation runs in Figure \ref{fig:no_dup_sim_trace}. The chains for the other scenarios with $1,2,$ and $5$ erroneous fields per record converged similarly quickly.

\subsubsection{Duplicate-Free Files, Unequal Errors Across Files}

\begin{figure}[!h]
\centering
\includegraphics[width=0.85\linewidth]{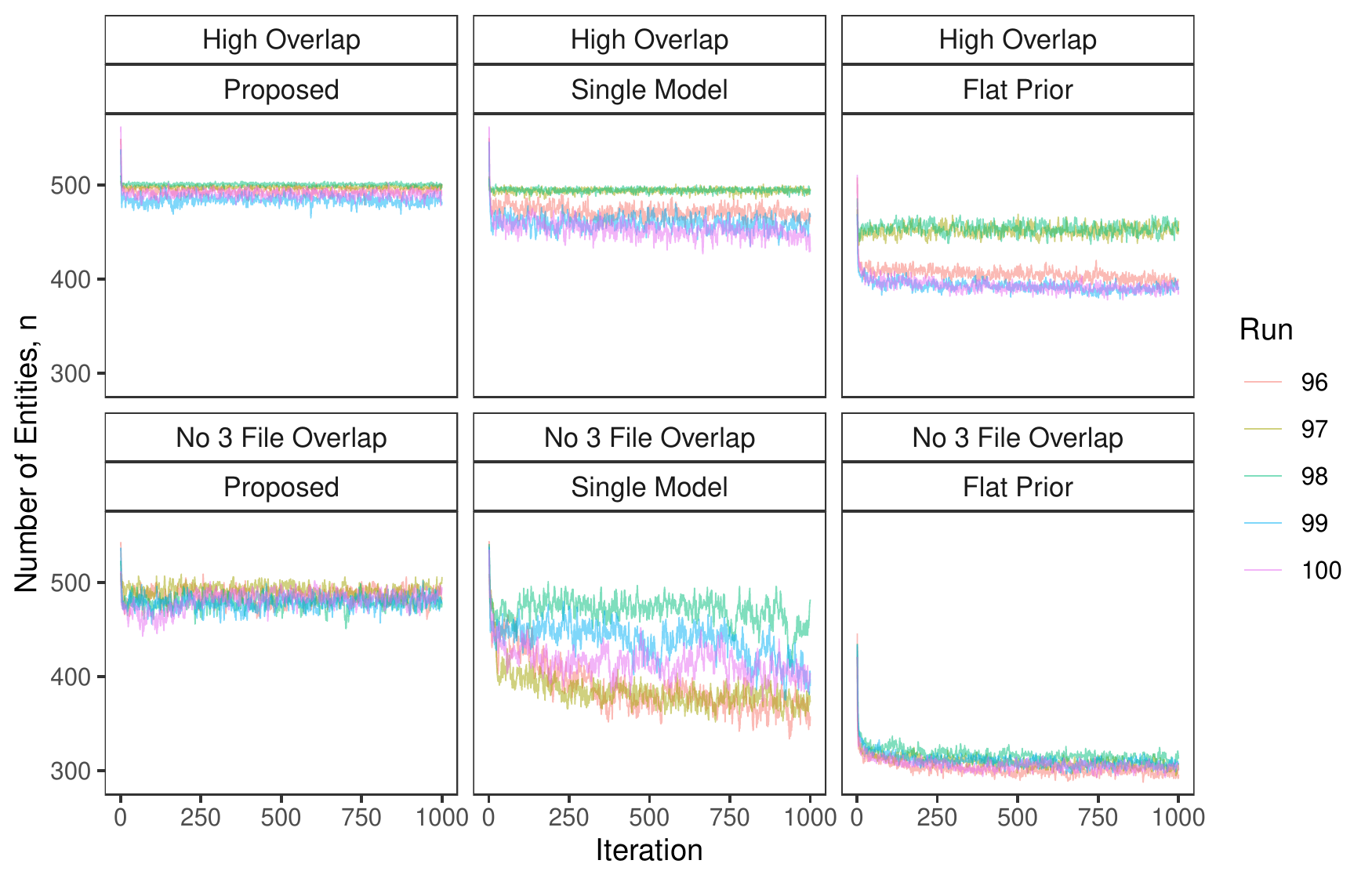}
\begin{minipage}[b]{1\textwidth}
    \caption{Trace plots for $n$ for the last $5$ of $100$ runs for the simulation with duplicate-free files and unequal measurement error across files. ``Proposed" refers to our proposed approach, ``Single Model" refers to the approach using a single model for all file-pairs and our structured prior for partitions, and ``Flat Prior" refers to the approach using our model for comparison data with a flat prior on tripartite matchings.}
    \label{fig:no_dup_error_sim_trace}
\end{minipage} 
\end{figure}

For the simulations conducted in Section 6.3 of the main text, for each overlap setting we present the trace plots for $n$ for the last $5$ of the $100$ simulation runs in Figure \ref{fig:no_dup_error_sim_trace}. 

\subsubsection{Files with Duplicates, Full Estimates}
\begin{figure}[!h]
\centering
\includegraphics[width=1\linewidth]{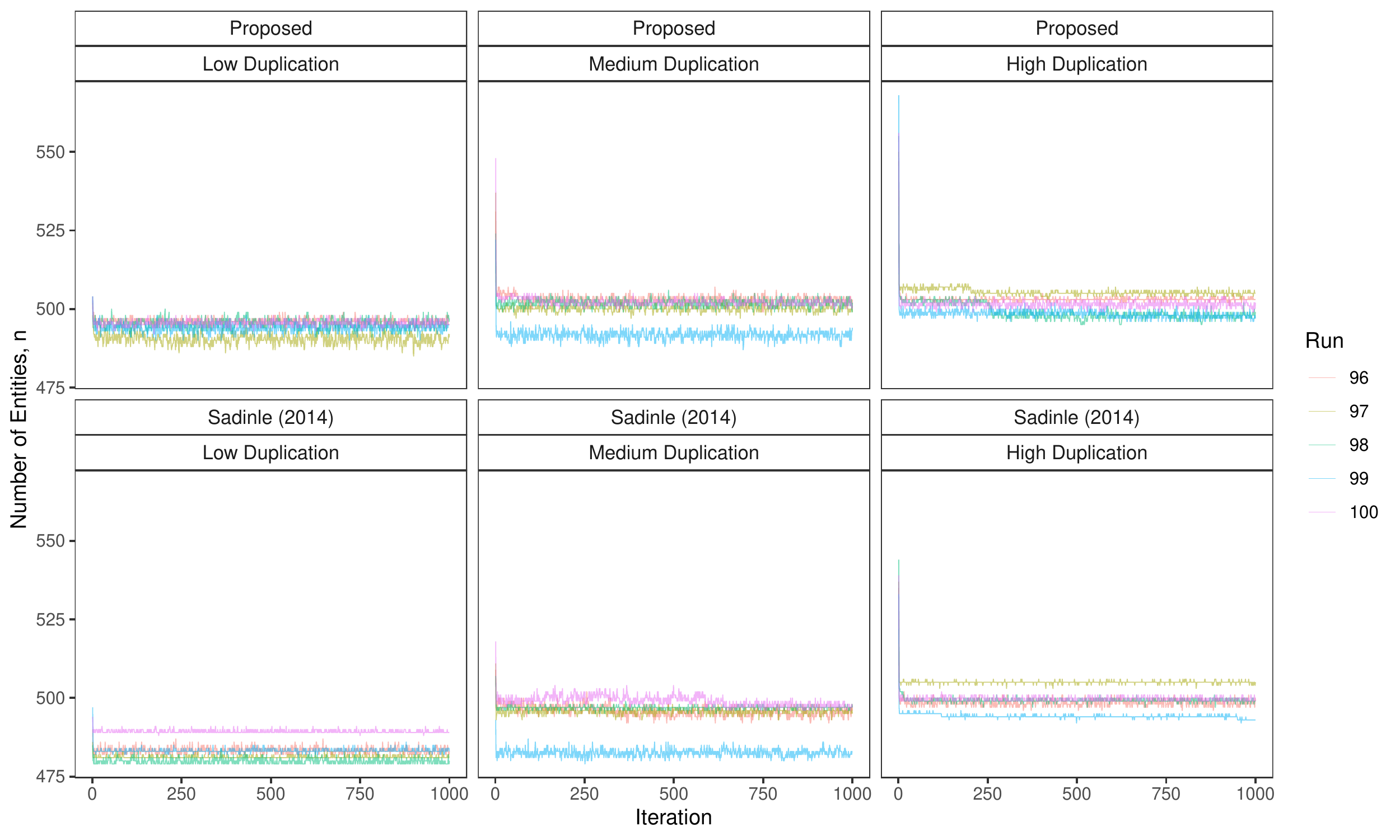}
\begin{minipage}[b]{1\textwidth}
    \caption{Trace plots for $n$ for the last $5$ of $100$ runs for the simulation with files with duplicates and equal measurement error across files, when using full estimates. ``Proposed" refers to the proposed approach and ``Sadinle (2014)" refers to the approach of \protect\cite{Sadinle_2014}.}
    \label{fig:dup_sim_trace}
\end{minipage} 
\end{figure}

For the simulations conducted in Appendix \ref{sec:sim2} with $3$ erroneous fields per record, for each overlap setting we present the trace plots for $n$ for the last $5$ of the $100$ simulation runs in Figure \ref{fig:dup_sim_trace}. The chains for the other scenarios with $1,2,$ and $5$ erroneous fields per record converged similarly quickly.

\clearpage

\section{Colombia Application}
\label{sec:apps}
In this appendix we re-examine the three record systems containing information on homicides from 2004 in the Quindio province of Colombia, previously studied in \cite{Sadinle_2013}. These record systems, provided by the Conflict Analysis Resource Center (CERAC), were maintained by the Colombian National Statistics Office (Departamento Administrativo Nacional de Estadistica, DANE), the Colombian National Police (Policia Nacional de Colombia, PN), and the Colombian Forensics Institute (Instituto Nacional de Medicina Legal y Ciencias Forenses, ML). While the purpose of DANE is to record all homicides occurring in Colombia, PN and ML only record homicides obtained from their daily activities \citep{DANEvitales, Restrepo_2007}.  Linking DANE to PN and ML is thus an important step in assessing the coverage of DANE and arriving at better estimates of the number of homicides in Colombia. Previously the records were linked by hand, which gives us a ground truth to assess the performance of our proposed approach. The linkage methodology of \cite{Sadinle_2013} did not scale to a large number of records, so the authors restricted their analysis to the $162$ records from the last three months of 2004.  We will now use our proposed approach to link all the $769$ records from 2004.

\subsection{Implementation Details}
The three record systems are believed to be free of duplicates, so the target of inference is a tripartite matching. The fields available from all three record systems are municipality and date of the homicide, whether the location of the homicide was urban or rural, and the age, sex, and marital status of the victim. Additionally, educational status of the victim is available in DANE and ML, which we are able to use despite it being missing in PN, as explained in Section 4. Although we have seven fields available for the linkage, none of them provide a large amount of discriminative information, which comparison-based approaches rely on. Thus we expect there to be significant uncertainty in the linkage, making the proposed approach particularly relevant. 

There are $r_1=323$ records in DANE, $r_2=157$ records in ML, and $r_3=289$ records in PN, so there are $189,431$ record pairs for which we construct comparison data, according to Table \ref{tab:colombia_comp}. 
We use transitive indexing as described in Appendix B, where the initial indexing scheme declares record pairs as non-coreferent if they disagree in either municipality, sex, date by more than 60 days, or age by more than 9 years. This reduces the number of candidate coreferent record pairs down to $60,324$. 

\begin{table}
\caption{Construction of levels of disagreement for the Colombian homocide record systems.} \label{tab:colombia_comp}
\begin{tabular*}{\textwidth}{@{\extracolsep{\fill}}lccccc@{}}
\hline
& & \multicolumn{4}{c@{}}{\textbf{Levels of disagreement}}\\
\cline{3-6}
\textbf{Field} & \textbf{Similarity measure} & $\bm{0}$ & $\bm{1}$ &
$\bm{2}$ & \multicolumn{1}{c@{}}{$\bm{3}$} \\
\hline
Date & Absolute Difference & 0 & $1-2$ & $3-7$ & $8+$ \\
Age & Absolute Difference & 0& $1-2$ &$3-9$&$10+$\\
Other Fields & Binary comparison & Agree & Disagree &&\\
\hline
\end{tabular*}
\end{table}

We present results from our approach under two prior specifications. The first is the default specification used in the simulations in Sections 6.2 and 6.3. The second specification differs from the default by placing a more informative prior on the overlap table through $\balpha$. In particular, based on characteristics of the record systems described in \cite{Restrepo_2007}, we specify a prior that captures the beliefs that: 1) if a homicide is recorded in PN or ML, it is highly likely to also be recorded in DANE, and 2) DANE and PN are expected to have a high coverage of homicides. This prior is described in more detail in the following section. We used the same loss function specification as outlined in Section 6.1 and Appendix D. We ran $3,000$ iterations of the Gibbs sampler presented in Appendix B, discarding the first $1,000$ as burn-in. In Section \ref{sec:trace} we discuss convergence of the Gibbs sampler for this application.

\subsection{An Informative Prior Specification}
We will guide our prior specification using the following two passages from \cite{Restrepo_2007} (translated to English):
\begin{itemize}
    \item ``According to DANE, `The differences in the Legal Medicine and DANE data are mainly due to the fact that the latter organization receives, in addition to the death certificates sent by Legal Medicine (which are sent to DANE after a technical examination), the homicide reports made by police inspectors, nurses or health promoters - in places where there are no legal doctors - who arrive at the site where the body is found and register the cases as homicide without a technical examination and according to the picture that presents the corpse'. So we find here a reason for the increased coverage of vital statistics"  \cite[][p. 331]{Restrepo_2007}.
    \item ``The National Police assures to have coverage at the national level, making an institutional presence in all the municipalities of the country: `We have a presence throughout the country and we know all the cases; Legal Medicine does not have the same coverage and thus their data is not exact' " \cite[][p. 330]{Restrepo_2007}.
\end{itemize}

Note that our Dirichlet-Multinomial prior for the overlap table can be motivated as the result of first drawing  $\{q_{\bh}\}_{{\bh}\in\HH }$ from a Dirichlet distribution with hyperparameters $\balpha$, then drawing $\bn$ from a multinomial distribution of size $n$ with probabilities $\{q_{\bh}\}_{{\bh}\in\HH }$. Based on these passages we specify $\balpha$ as follows, referring in our notation to DANE as list $1$, ML as list $2$, and PN as list $3$:
\begin{itemize}
    \item If a homicide is going to be recorded by one of the three record systems, it is very likely that it will be known by DANE. Therefore, we choose $\alpha_{1++}=\alpha_{100} + \alpha_{101} + \alpha_{110} + \alpha_{111}$ and $\alpha_{0++} = \alpha_{001} + \alpha_{010} + \alpha_{011}$ such that $\text{mode}(q_{1++})=0.95$ and $\PP(q_{1++}>0.9) = 0.95$, where $q_{1++} =q_{100} + q_{101} + q_{110} + q_{111}\sim\textsf{Beta}(\alpha_{1++}, \alpha_{0++})$ is the prior probability of a homicide being recorded in DANE given it is recorded in one of the three systems.

    \item If a homicide is recorded by PN, then it is very likely that it will be recorded by DANE.  Therefore, we choose $\alpha_{1+1}=\alpha_{101}+\alpha_{111}$ and $\alpha_{0+1}=\alpha_{001}+\alpha_{011}$ such that $\text{mode}(q_{1+1})=0.95$ and $P(q_{1+1}>0.9) = 0.9$, where 
    $q_{1+1}=q_{101}+q_{111}\sim\textsf{Beta}(\alpha_{1+1}, \alpha_{0+1})$ is the prior probability of a homicide being recorded in DANE given it is recorded in PN.

    \item If a homicide is recorded by ML, then it is very likely that it will be known by DANE.  Therefore, we choose $\alpha_{11+}=\alpha_{110}+\alpha_{111}$ and $\alpha_{01+}=\alpha_{010}+\alpha_{011}$ such that $\text{mode}(q_{11+})=0.95$ and $P(q_{11+}>0.9) = 0.9$, where 
    $q_{11+}=q_{110}+q_{111}\sim\textsf{Beta}(\alpha_{11+}, \alpha_{01+})$ is the prior probability of a homicide being recorded in DANE given it is recorded in ML.

    \item The above induce six constraints, which determine $\alpha_{011}$, $\alpha_{001}$, and $\alpha_{010}$,  but we need one extra constraint to determine the remaining $\alpha$. We thus choose the configuration of $\alpha_{100}$, $\alpha_{101}$, $\alpha_{110}$, and $\alpha_{111}$ that maximizes $\alpha_{101}$, which controls the prior probability of a homicide being jointly recorded by DANE and PN, given that these systems that are supposed to have the largest coverage of homicides.
\end{itemize}
Under this specification we arrive at the setting $\alpha_{001}=1.63, \alpha_{010}=1.63, \alpha_{011}=2.94, \alpha_{100}=0, \alpha_{101}=30.96, \alpha_{110}=30.96, \alpha_{111}=37.78$. For the sake of propriety, we set
$\alpha_{100}=0.1$.

\subsection{Results}
Under the default prior specification the precision and recall of the full estimate are $0.90$ and $0.93$ respectively. Recall is no longer useful when using partial estimates, as we are not trying to find all true matches. Thus we will assess the performance of the partial estimates using precision and the \textit{abstention rate}, the proportion of records which the Bayes estimate abstained from making a linkage decision. The partial estimate has an abstention rate of $11\%$, and improves the precision of the estimate to $0.93$. Under the informative prior specification the precision and recall of the full estimate are $0.93$ and $0.96$ respectively. The partial estimate has an abstention rate of $11\%$, and improves the precision of the estimate to $0.95$. Due to the performance difference, we will focus on the results under the informative prior specification for the rest of this section. Analogous results under the default specification are provided in the next section. 

The total number of homicides based on the hand labelling is $383$. Under the informative prior specification a $95\% $ credible interval for the number of unique homicides $n$ is $[372, 383]$, with an estimate based on the full estimate of the tripartite matching of $376$.  In Table \ref{tab:post_over} we display the posterior distribution for the overlap table and the overlap table derived from the full estimate, along with the overlap table derived from the ground truth hand labelling. We can see that $n_{111}$, the number of homicides recorded in all three files, and $n_{100}$, the number of homicides recorded in just DANE, are overestimated, and the remaining cells of the overlap table (and $n$) are underestimated. 

\begin{table}[h]
 \centering
  \begin{minipage}[b]{1\textwidth}
 \def\~{\hphantom{0}}
  \caption{Posterior distribution of the overlap table for the Colombian record systems, under the informative prior specification. Black lines indicate the ground truth, dotted lines indicate quantities derived from the full estimate of the tripartite matching.} \label{tab:post_over}
  \begin{tabular*}{\columnwidth}{c@{\extracolsep{\fill}}c@{\extracolsep{\fill}}c@{\extracolsep{\fill}}c@{\extracolsep{\fill}}c@{\extracolsep{\fill}}c@{\extracolsep{\fill}}c@{\extracolsep{\fill}}}
\hline\\ [-22pt]
   &  & \multicolumn{2}{c}{In PN}  & & \multicolumn{2}{c}{Out PN} \\
 \cline{3-7}\\ [-22pt]
 DANE & & In ML & Out ML &  & In ML & Out ML\\
	\cline{1-7} \\ [-10pt]
  In & & \begin{tabular}{c}\includegraphics[width=0.15\columnwidth]{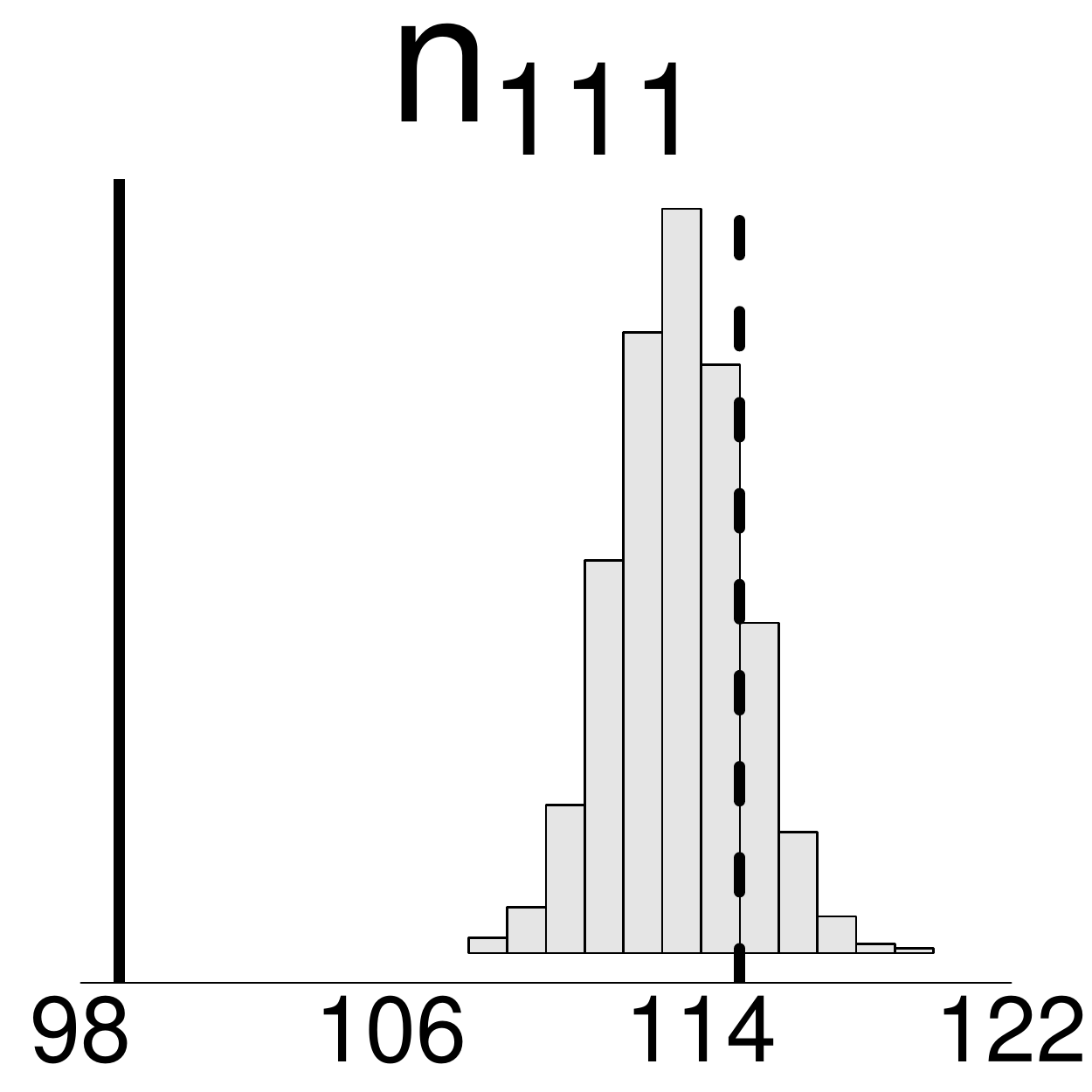}\end{tabular} & \begin{tabular}{c}\includegraphics[width=0.15\columnwidth]{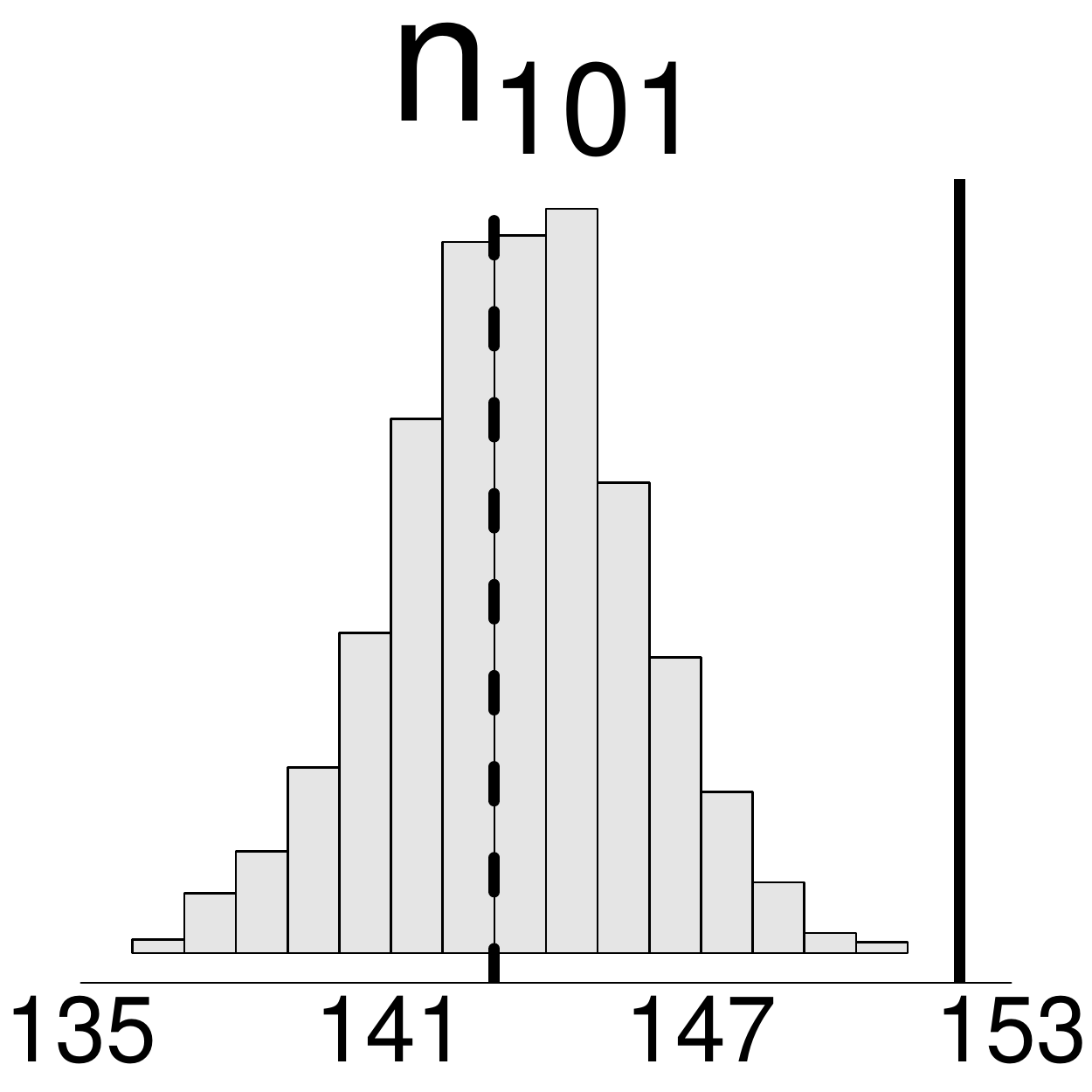}\end{tabular} & & \begin{tabular}{c}\includegraphics[width=0.15\columnwidth]{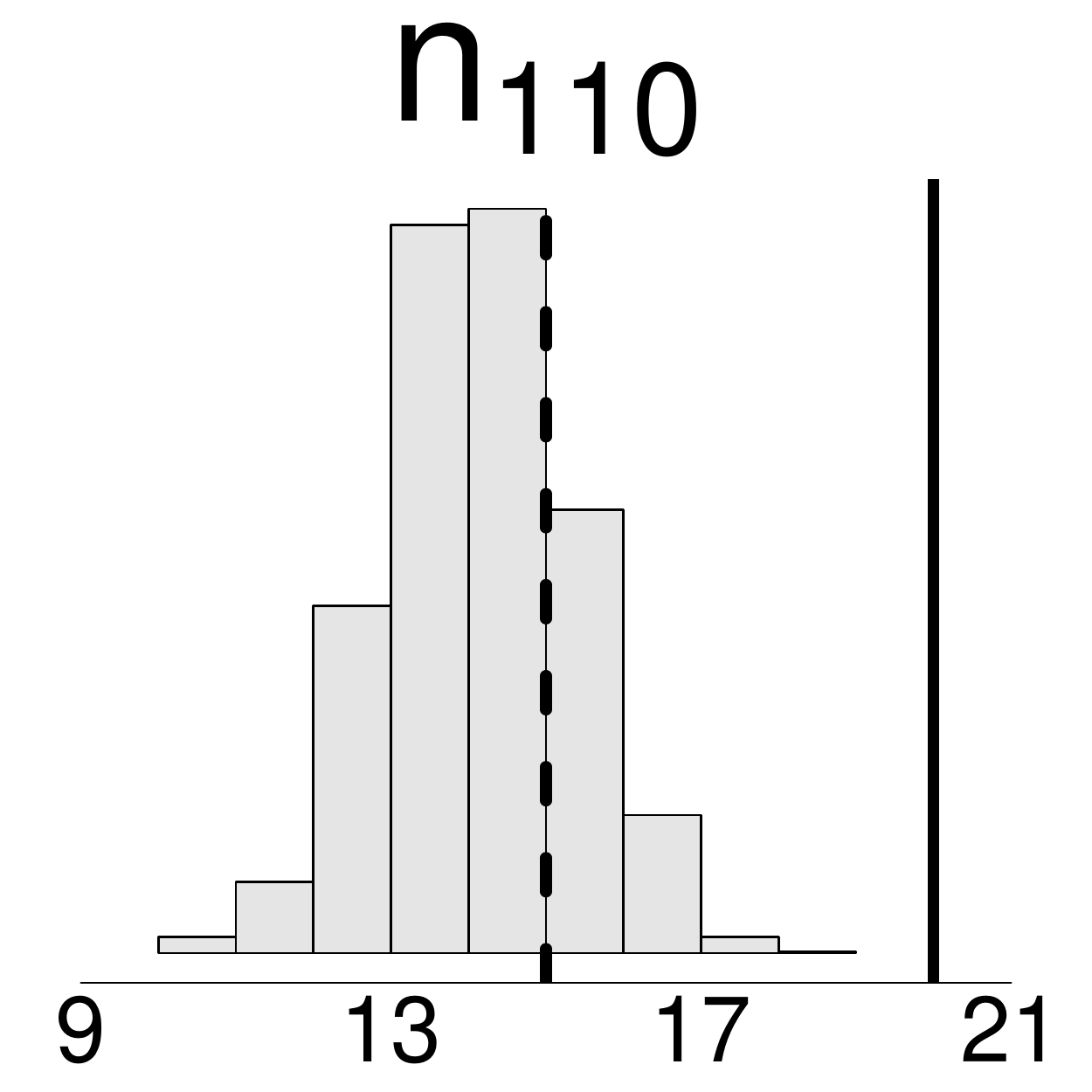}\end{tabular} & \begin{tabular}{c}\includegraphics[width=0.15\columnwidth]{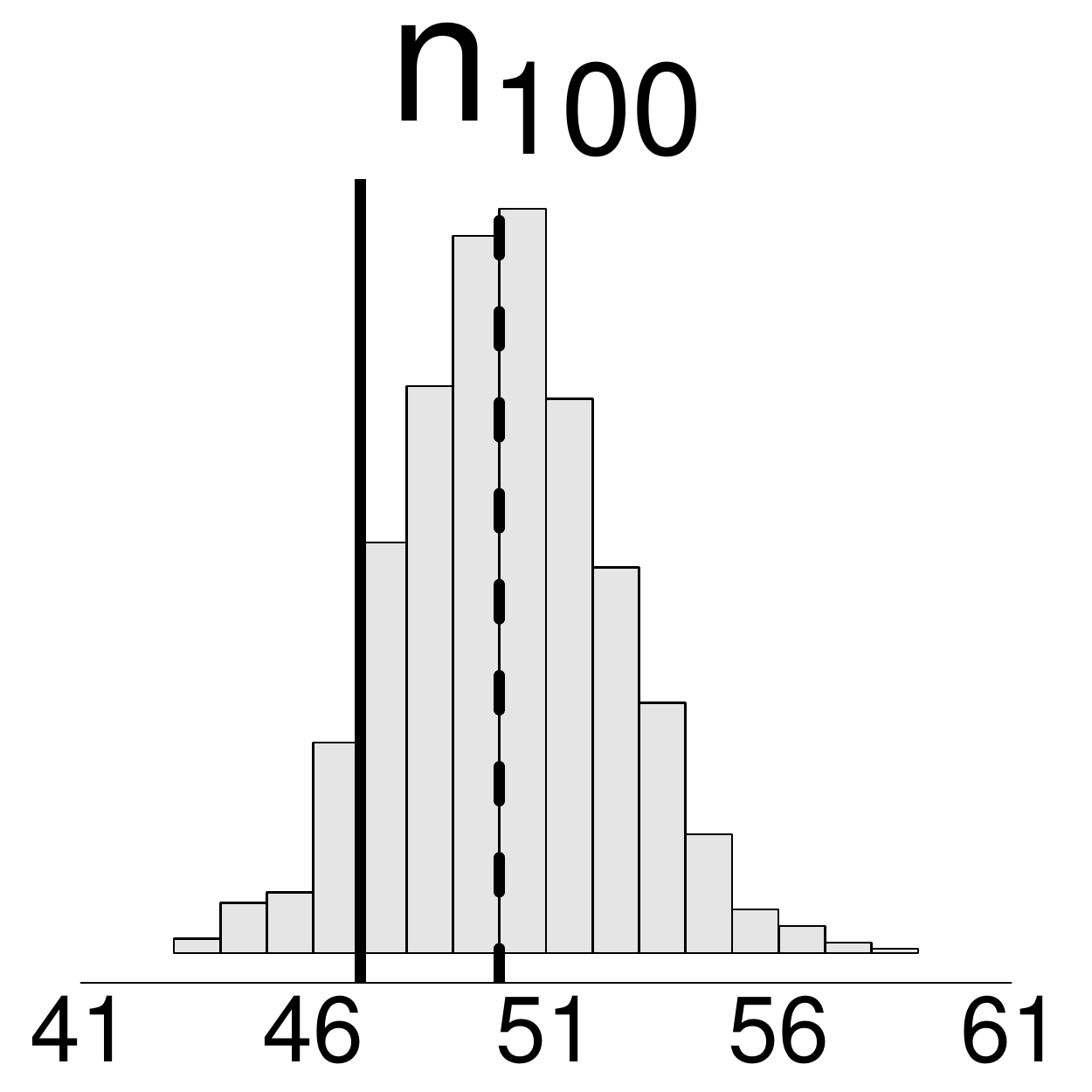}\end{tabular} \\
  Out & & \begin{tabular}{c}\includegraphics[width=0.15\columnwidth]{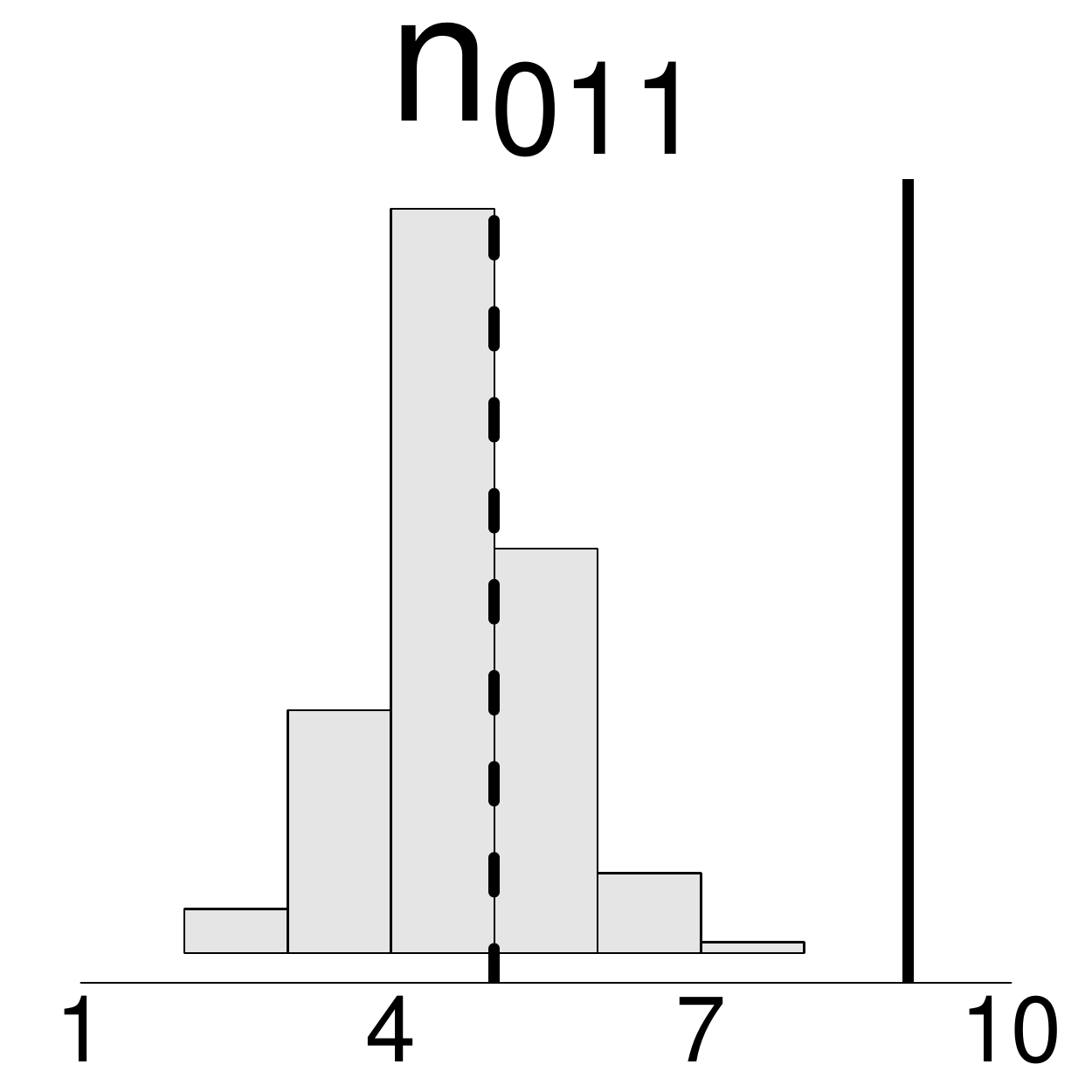}\end{tabular} & \begin{tabular}{c}\includegraphics[width=0.15\columnwidth]{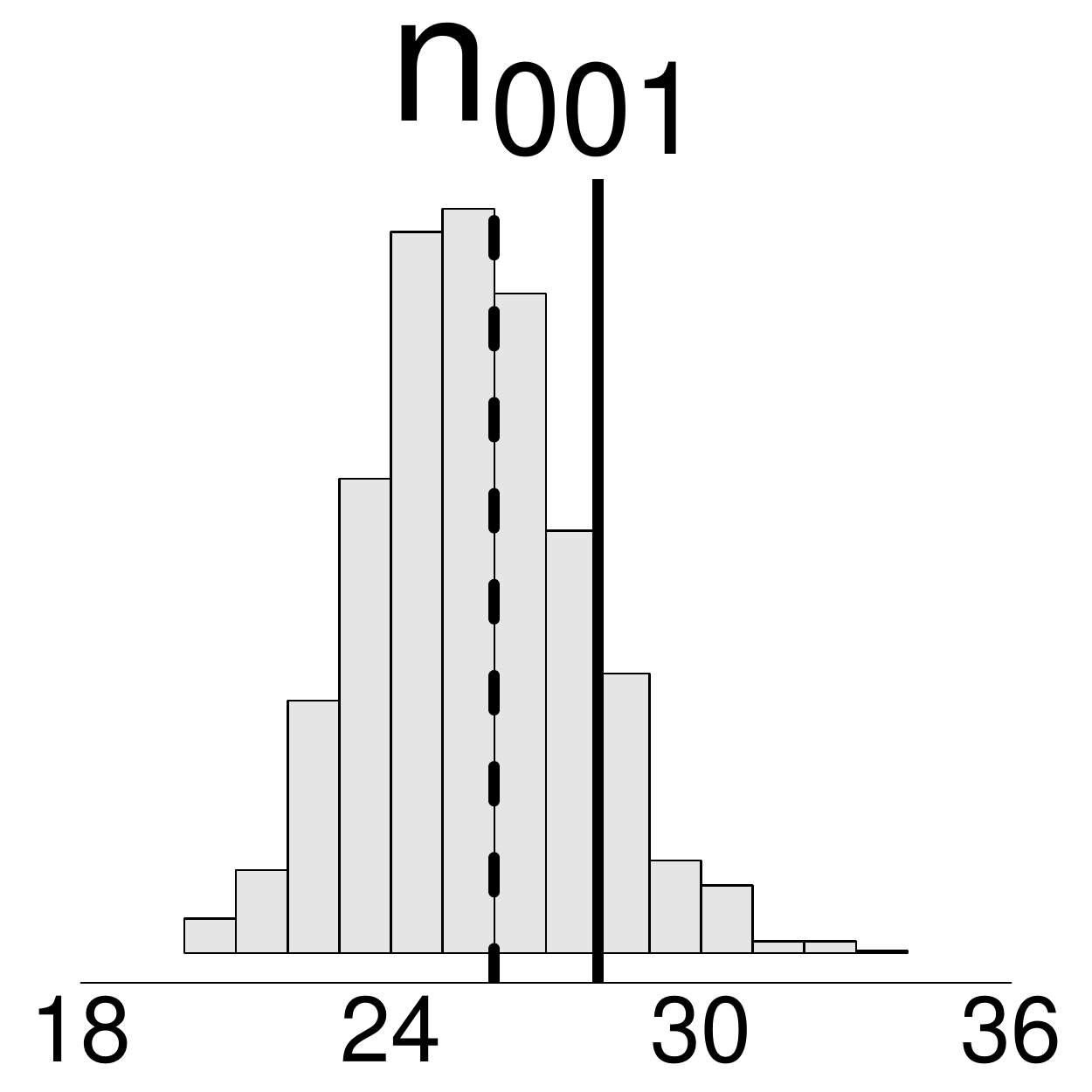}\end{tabular} & & \begin{tabular}{c}\includegraphics[width=0.15\columnwidth]{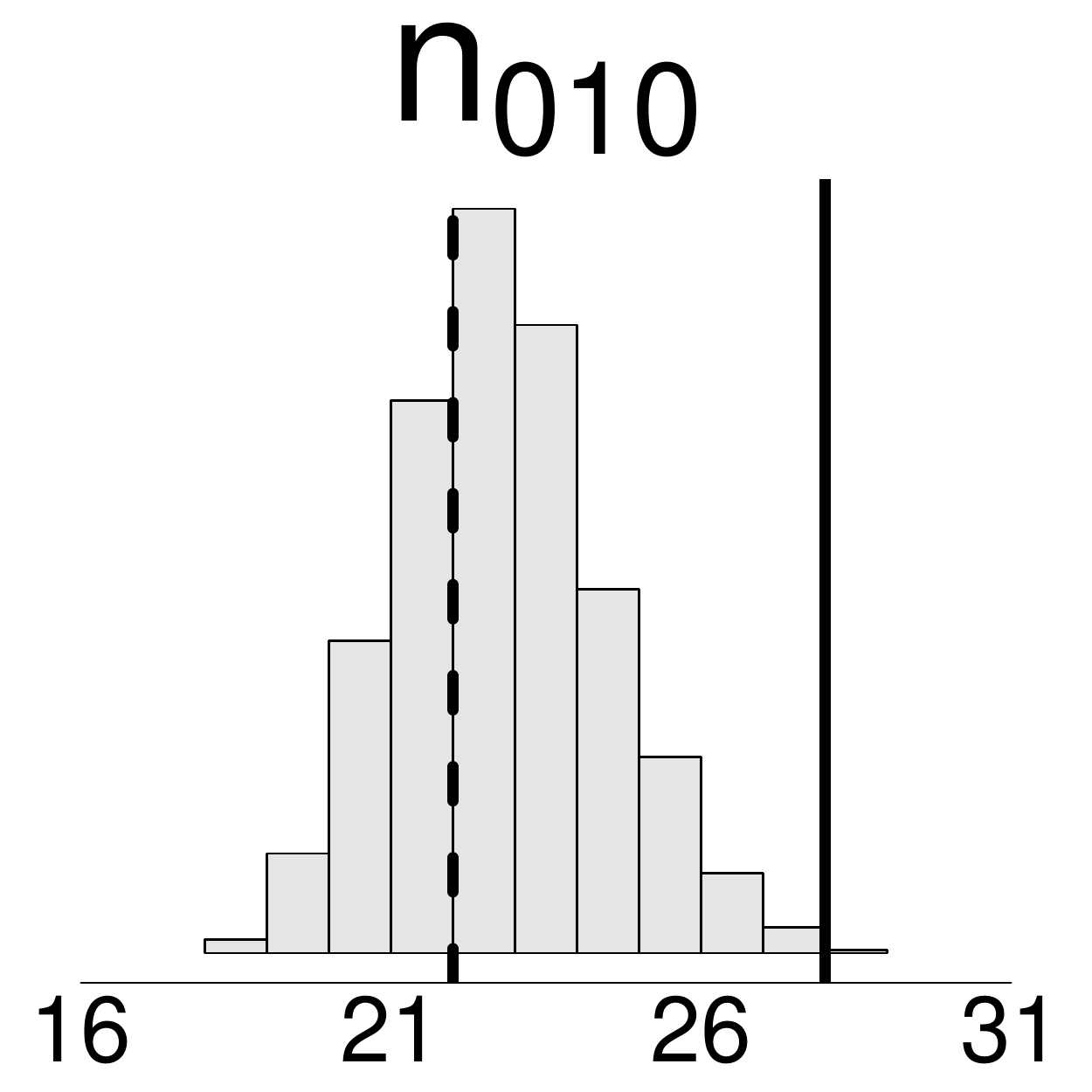}\end{tabular} & $-$ \\[-1pt]\hline
\end{tabular*}\vskip18pt
\end{minipage}
\end{table}

While the performance of the estimated matching is fairly good, with precision of both the full and partial estimate (before clerical review) above $0.9$, the overestimation of $n_{111}$ and the underestimation of $n$ is indicative of over-matching. We believe this over-matching occurrs due to the low amount of discriminative information provided by the fields. In particular, the only fields we believe can be fully trusted are municipality of the homicide (of which there are $12$) and sex of the victim (which is coded as binary), which can only partition the records into $22$ blocks of candidate coreferent records (there are no records of female homicide victims in two municipalities). The remaining fields all have some amount of error and do not provide highly discriminative information since they are either low dimensional categorical fields (urban/rural location of the homicide has two categories, marital status has five categories, and educational status has six categories) or numeric fields that are essentially ordinal categorical (date of homicide and age of victim).  Therefore, records of different homicides may look similar based on the comparisons of these fields, causing them to be mistakenly matched. In these low-information settings, clerical review becomes especially important for the record linkage workflow, which our proposed approach of using partial estimates incorporates by design.

\subsection{Results Under Default Prior}
\label{sec:defaultprior_res}
Under the default prior specification a $95\% $ credible interval for the number of unique homicides $n$ is $[376, 389]$, with an estimate based on the full estimate of the tripartite matching of $378$. Thus we see that $n$ is better estimated under the default prior compared to the informative prior (though the point estimate is still an underestimate). In Table \ref{tab:post_over_default} we display the posterior distribution for the overlap table and the overlap table derived from the full estimate, along with the overlap table derived from the ground truth hand labelling. We can see that, as with the informative prior specification, $n_{111}$ and $n_{100}$ are overestimated, and the remaining cells of the overlap table (and $n$) are underestimated. 

\begin{table}[h]
 \centering
  \begin{minipage}[b]{1\textwidth}
 \def\~{\hphantom{0}}
  \caption{Posterior distribution of the overlap table for the Colombian record systems, under the default prior specification. Black lines indicate the ground truth, dotted lines indicate quantities derived from the full estimate of the tripartite matching.} 
  \label{tab:post_over_default}
  \begin{tabular*}{\columnwidth}{c@{\extracolsep{\fill}}c@{\extracolsep{\fill}}c@{\extracolsep{\fill}}c@{\extracolsep{\fill}}c@{\extracolsep{\fill}}c@{\extracolsep{\fill}}c@{\extracolsep{\fill}}}
\hline\\ [-20pt]
   &  & \multicolumn{2}{c}{In PN}  & & \multicolumn{2}{c}{Out PN} \\
 \cline{3-7}\\ [-20pt]
 DANE & & In ML & Out ML &  & In ML & Out ML\\
	\cline{1-7} \\ [-10pt]
  In & & \begin{tabular}{c}\includegraphics[width=0.15\columnwidth]{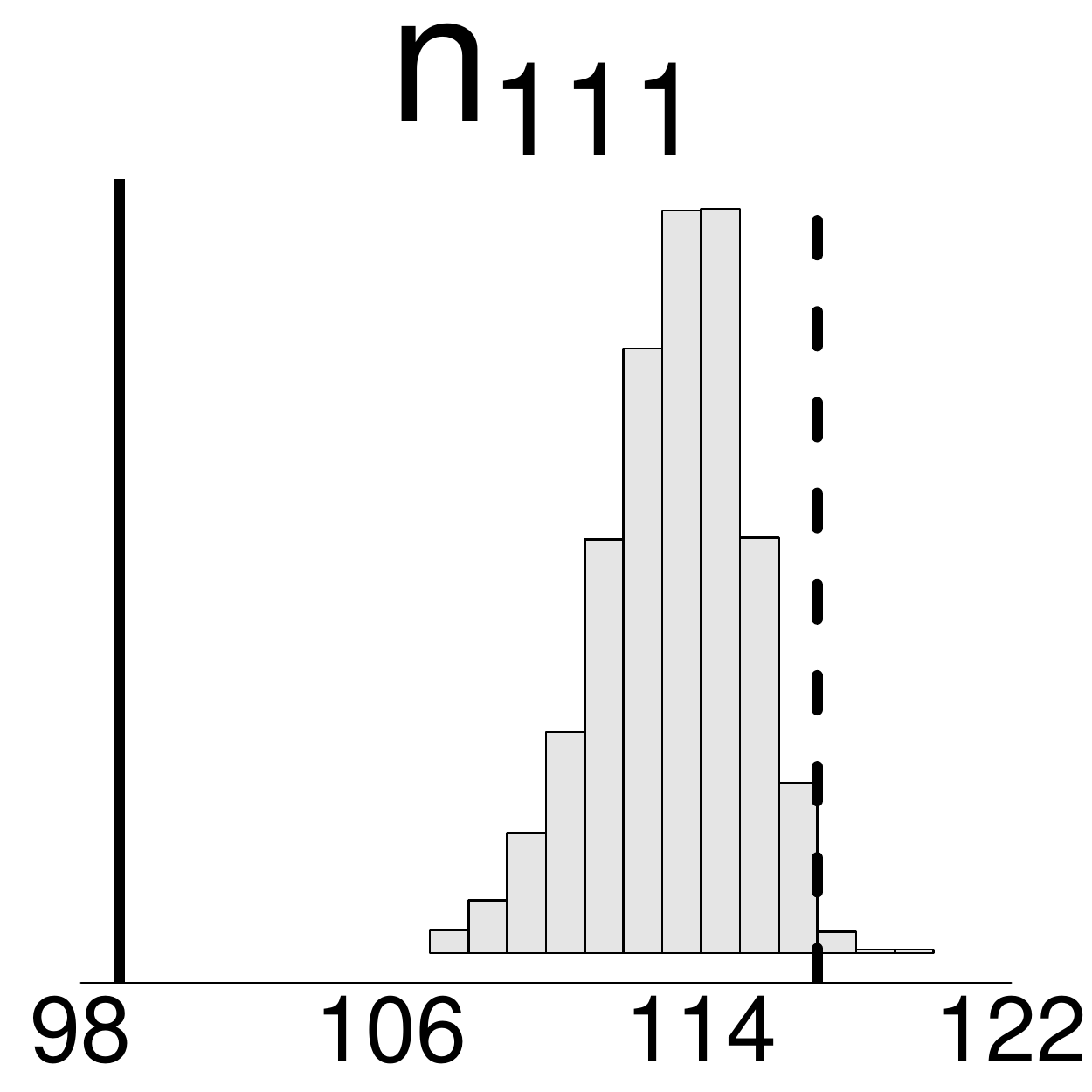}\end{tabular} & \begin{tabular}{c}\includegraphics[width=0.15\columnwidth]{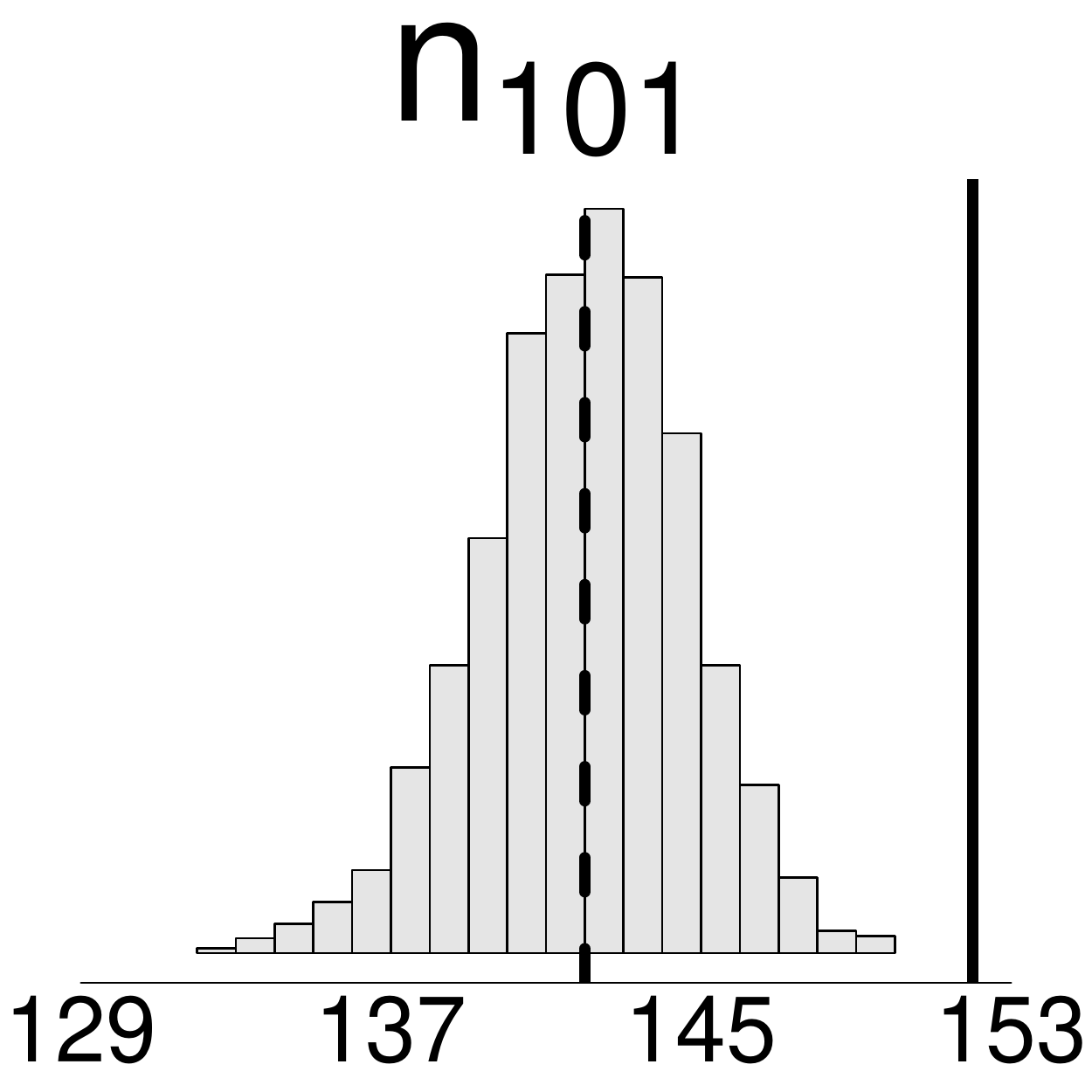}\end{tabular} & & \begin{tabular}{c}\includegraphics[width=0.15\columnwidth]{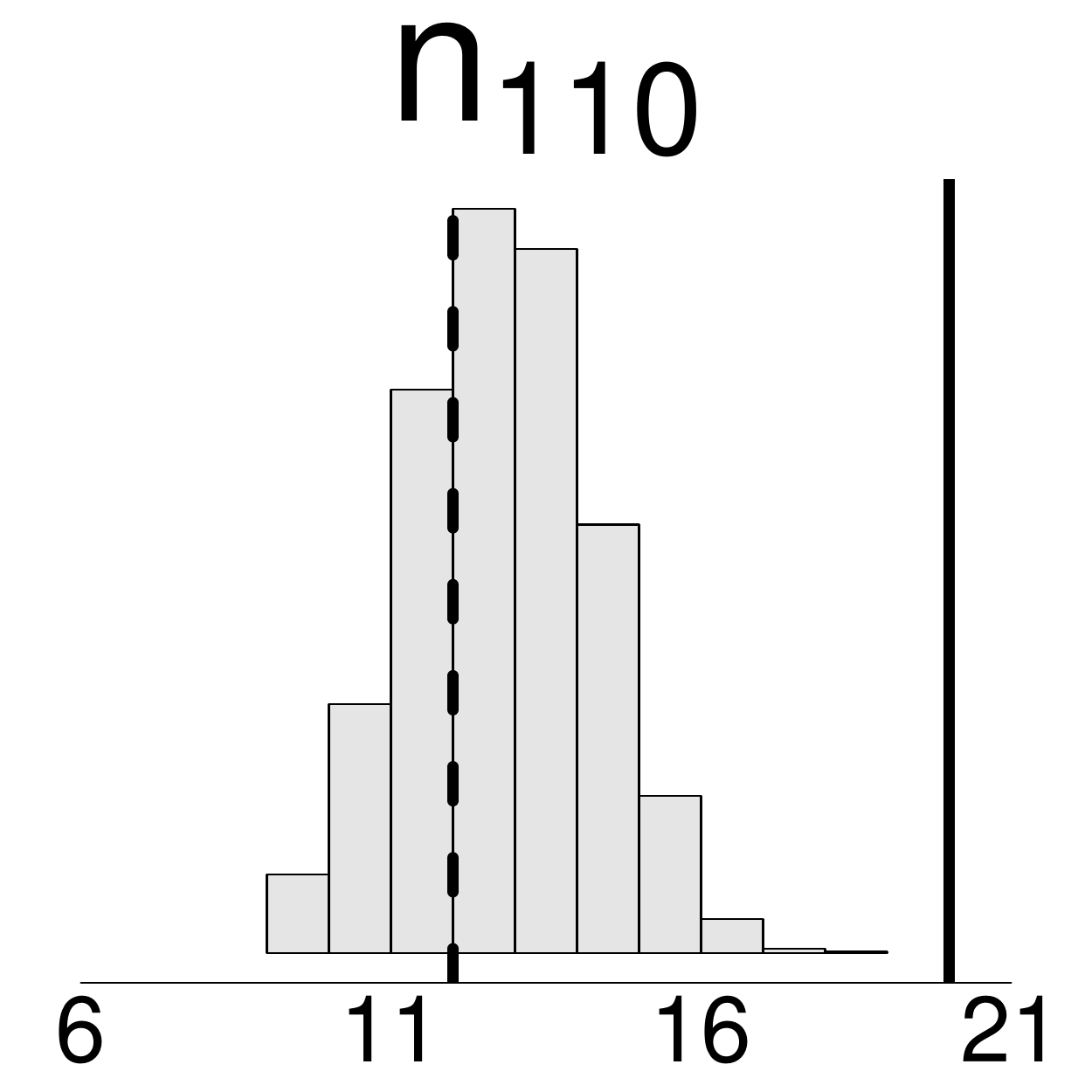}\end{tabular} & \begin{tabular}{c}\includegraphics[width=0.15\columnwidth]{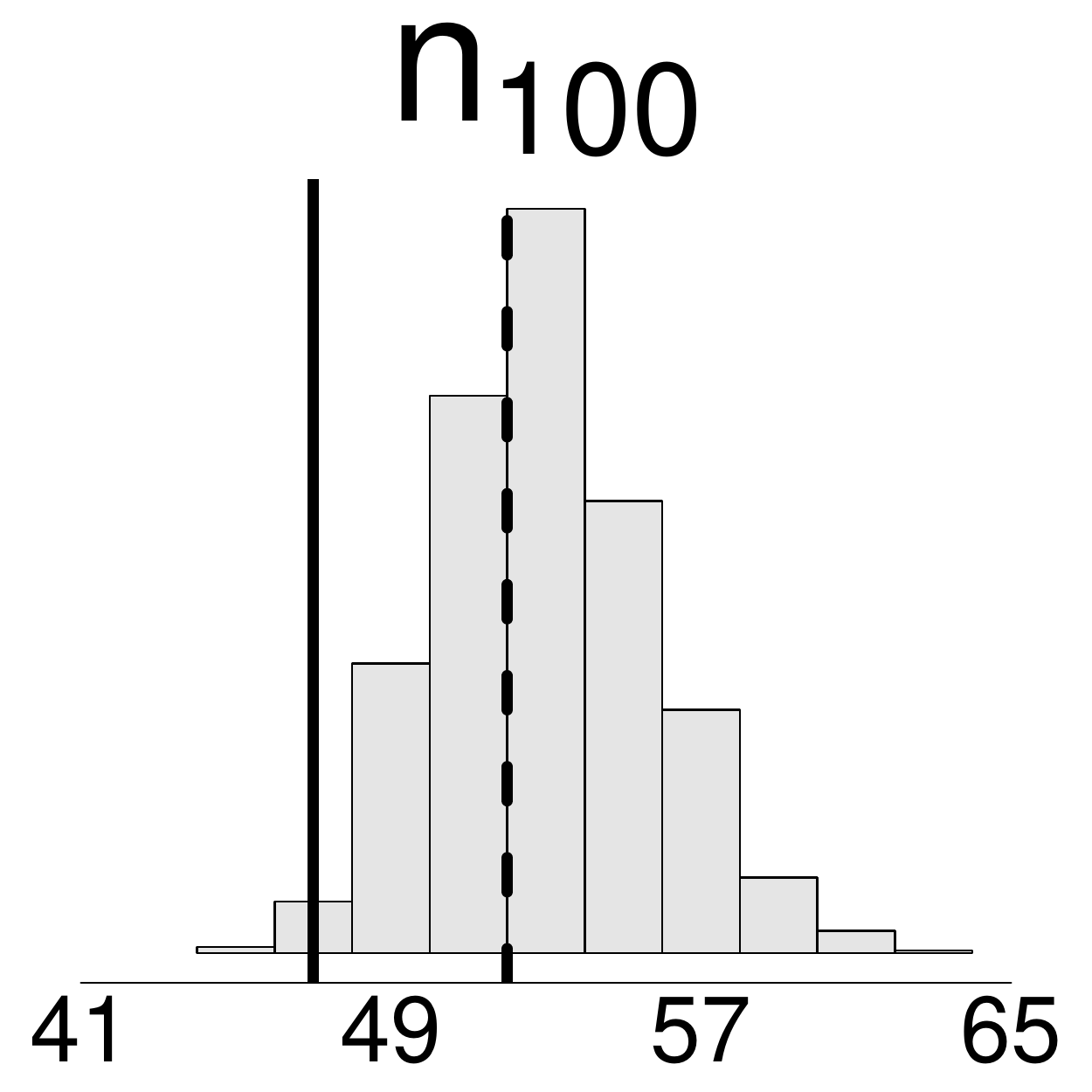}\end{tabular} \\
  Out & & \begin{tabular}{c}\includegraphics[width=0.15\columnwidth]{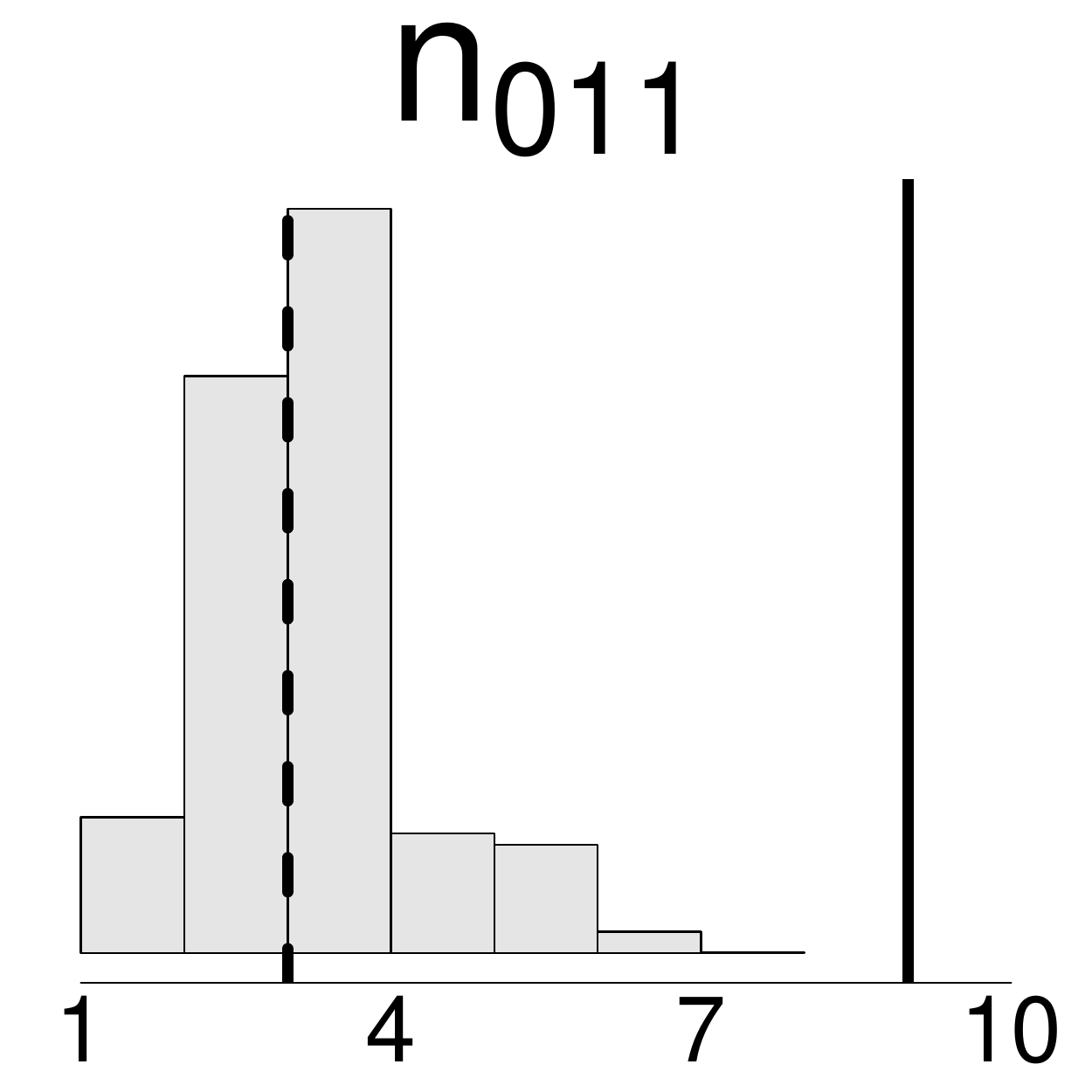}\end{tabular} & \begin{tabular}{c}\includegraphics[width=0.15\columnwidth]{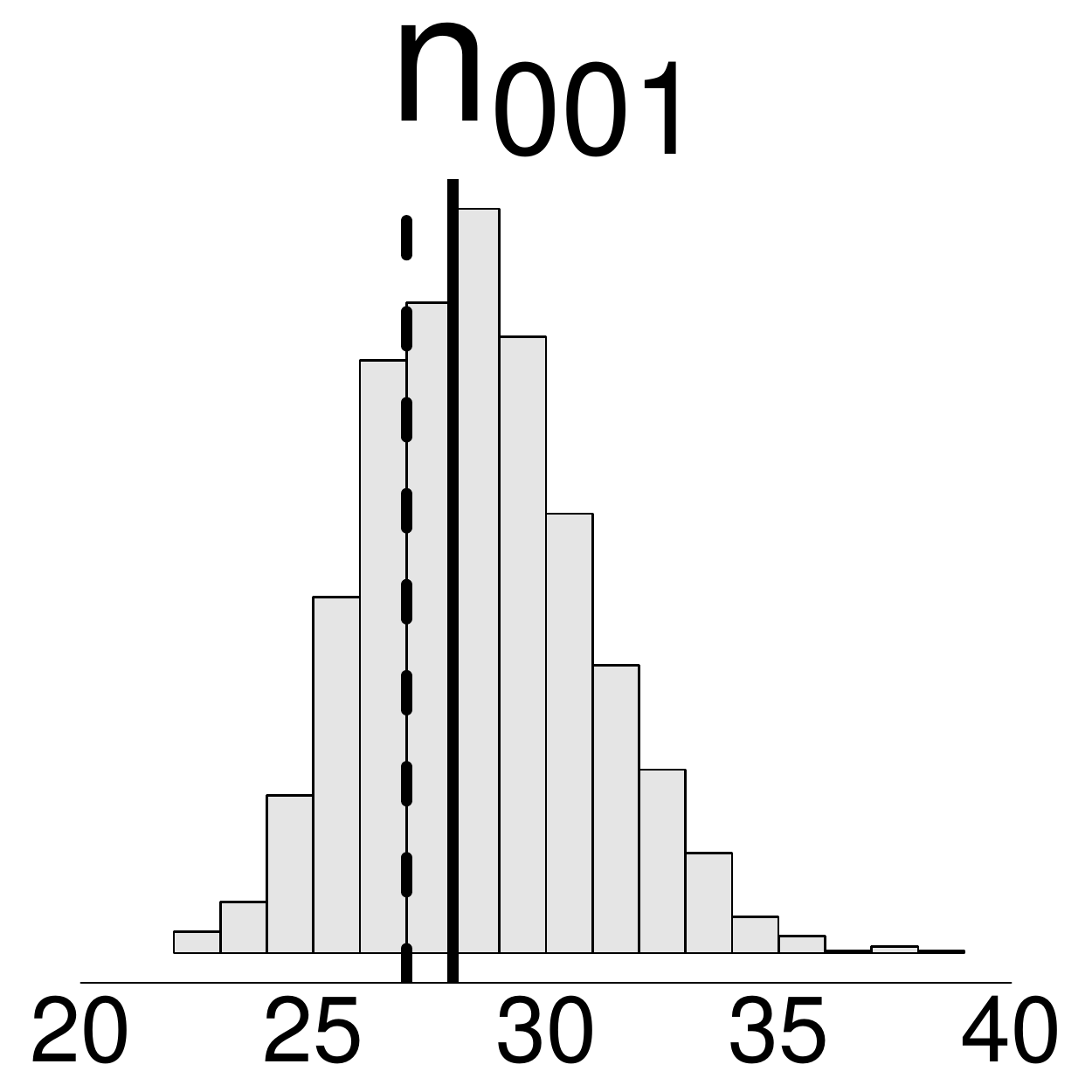}\end{tabular} & & \begin{tabular}{c}\includegraphics[width=0.15\columnwidth]{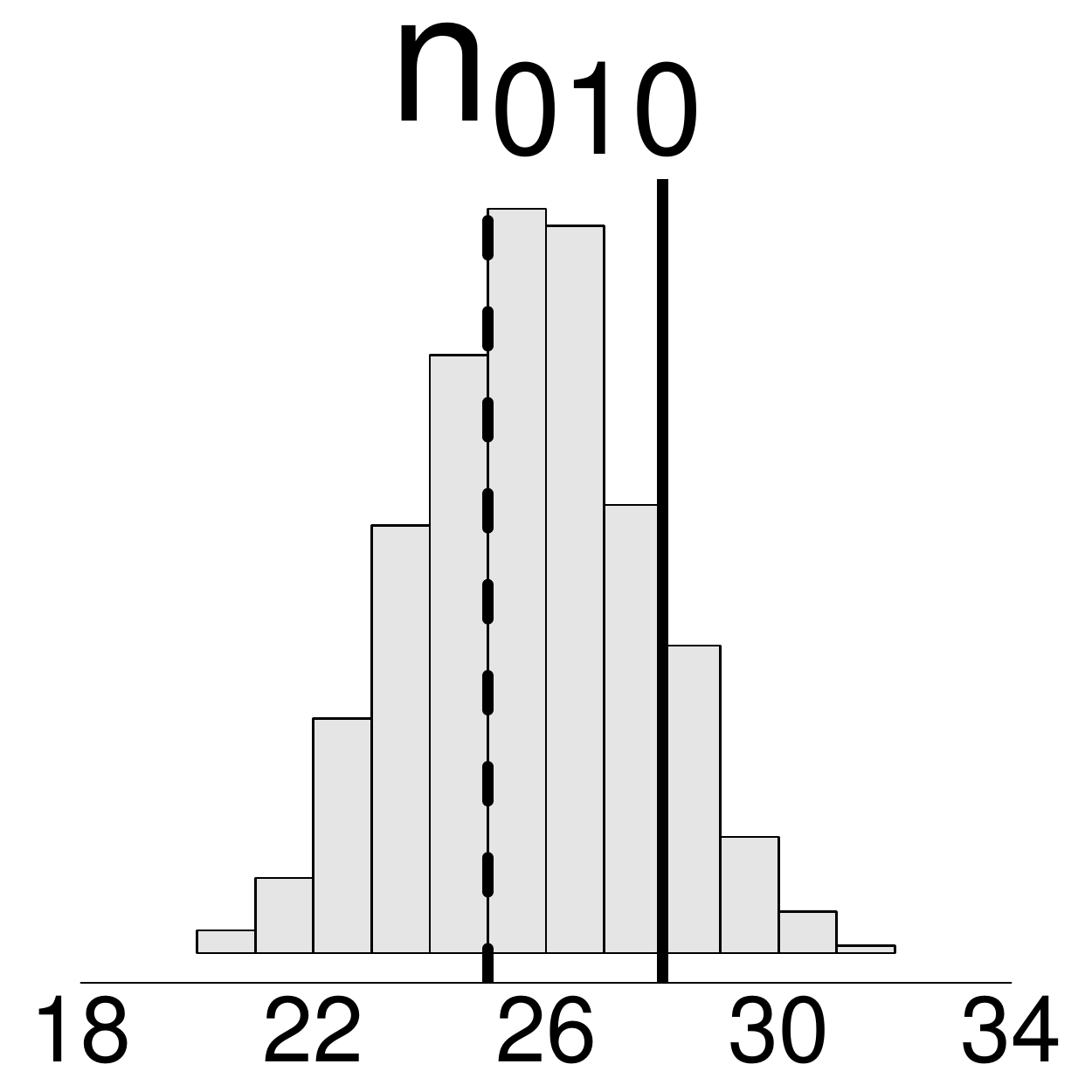}\end{tabular} & $-$ \\[-1pt]\hline
\end{tabular*}\vskip18pt
\end{minipage}
\end{table}

\subsection{Convergence Diagnostics}
\label{sec:trace}
\begin{figure}[!h]
\centering
\includegraphics[width=0.75\linewidth]{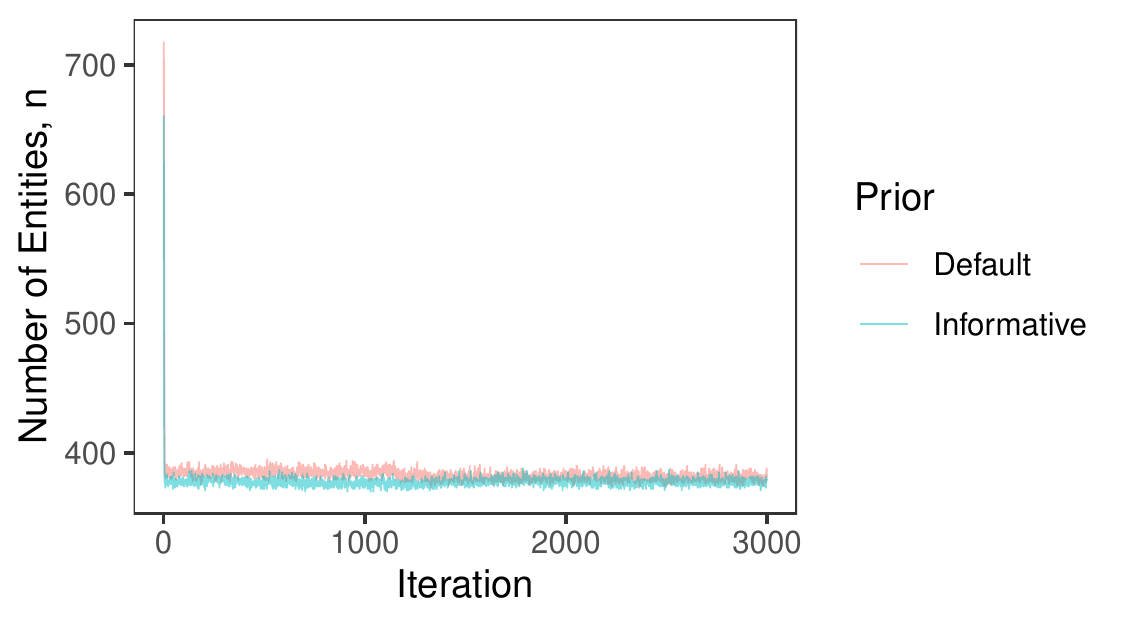}
\begin{minipage}[b]{0.95\textwidth}
    \caption{Trace plots for $n$ in Colombia application.}
    \label{fig:colombia_trace_plots}
\end{minipage} 
\end{figure}

In the application, we ran the the Gibbs sampler described in Appendix \ref{sec:gibbs} for $3,000$ iterations, discarding the first $1,000$ samples as burn-in, under a default and an informative prior specification. In Figure \ref{fig:colombia_trace_plots} we present the the trace plots for the number of entities, $n$, under each of these prior specifications. The chains for $n$ appear to converge quickly based on these trace plots. For each of these chains we computed Geweke's convergence diagnostic as implemented in the \texttt{R} package \texttt{coda} \citep{Plummer_2006}. The Geweke's Z-scores indicated it was reasonable to treat these chains as drawn from their stationary distributions.

\bibliographystyle{agsm}
\bibliography{JASA}

\end{document}